\documentstyle[aps,preprint,tighten,graphicx]{revtex}

\begin{document}

\def\Journal#1#2#3#4{{#1} {\bf #2}, #3 (#4)}

\def\NP{{\em Nucl. Phys.} }
\def\NC{{\em Nuovo Cim.} }
\def\NCL{{\em Nuovo Cim. Lett.} }
\def\PL{{\em Phys. Lett.} }
\def\PR{{\em Phys. Rev.} }
\def\PRL{{\em Phys. Rev. Lett.} }
\def\PREP{{\em Phys.  Rep.} }
\def\AP{{\em Ann. of Phys.} }
\def\ZP{{\em Z. Phys.} }
\def\SJNP{{\em Sov. J. Nucl. Phys.} }
\def\JETPL{{\em Jrn. of Exp. and Theo. Phys. Lett.} }
\def\JPSJ{{\em Jrn. Phys. Soc. Jpn.} }

\newcommand{\mathbold}[1]{\mbox{\protect\boldmath $\displaystyle #1$}}

\def\rme{{\rm e}}
\def\rmi{{\rm i}}

\def\intm1p1{{\int\limits_{-1}^{1}}}
\def\sumlUNE{{\sum\limits_{l=0}^{\infty}}}

\newcommand{\einh}{\frac{1}{2}}
\newcommand{\dreih}{\frac{3}{2}}
\newcommand{\be}{\begin{equation}}
\newcommand{\ee}{\end{equation}}
\newcommand{\bea}{\begin{eqnarray}}
\newcommand{\eea}{\end{eqnarray}}

\newcommand{\nres}[4]{#1_{#2#3}(#4)}

\newcommand{\lmnr}[2]{{\cal L}_{\varphi N #1}^{#2}}

\newcommand{\real}{{\rm Re}}
\newcommand{\imag}{{\rm Im}}

\newcommand{\bsl}[1]{#1 \!\!\! /}

\newcommand{\abs}[1]{\left | #1 \right |}

\newcommand{\dop}[1]{#1^{\prime}}
\newcommand{\dops}[1]{#1^{\prime \, 2}}
\newcommand{\dopp}[1]{#1^{\prime \prime}}
\newcommand{\hatdop}[1]{\hat #1^{\, \prime}}

\newcommand{\Iba}{b a}
\newcommand{\Ipipi}{\pi \pi}
\newcommand{\Ipigam}{\pi \gamma}
\newcommand{\Ibgam}{b \gamma}
\newcommand{\Iphigam}{\varphi \gamma}
\newcommand{\Iphipgam}{\dop \varphi \gamma}
\newcommand{\Iphiphip}{\varphi \dop \varphi}
\newcommand{\Igamgam}{\gamma \gamma}

\newcommand{\solid}{\protect\rule[1mm]{6mm}{.1mm}}
\newcommand{\dash}{\protect\rule[1mm]{2.5mm}{.1mm}\hspace{1mm}
\protect\rule[1mm]{2.5mm}{.1mm}}
\newcommand{\shortdash}{\protect\rule[1mm]{1mm}{.1mm}\hspace{4mm}
\protect\rule[1mm]{1mm}{.1mm}}
\newcommand{\dashdot}{\protect\rule[1mm]{2.5mm}{.1mm}\hspace{1mm}$\cdot\cdot$}
\newcommand{\dashsdot}{\protect\rule[1mm]{2.5mm}{.1mm}\hspace{1mm}$\cdot$}
\newcommand{\dotdot}{\protect$\cdot\!\cdot\!\cdot$}

\newcommand{\FullBox}{\protect\rule[0.5mm]{2.5mm}{2.5mm}}

\pagestyle{plain} \pagenumbering{arabic}

\begin{center}
  {\LARGE Photon- and meson-induced reactions on the nucleon\footnote{Work
      supported by BMBF, GSI Darmstadt}\footnote
    {This paper forms part of the dissertation of T. Feuster}}

  \bigskip \bigskip

  {T. Feuster$^1$\footnote
    {e-mail:Thomas.Feuster@theo.physik.uni-giessen.de} and U.
    Mosel$^1$}

  \bigskip \bigskip

  {\it $^1$Institut f\"ur Theoretische Physik, Universit\"at Giessen\\
    D--35392 Giessen, Germany}\\[3mm]
  UGI-98-16
\end{center}

\bigskip \bigskip

%
%

\section*{Abstract}
Starting from an unitary effective Lagrangian model for the
meson-nucleon scattering developed in \cite{fm98}, we come to a
unified description of both meson scattering and photon-induced
reactions on the nucleon. To this end the photon is added
perturbatively, yielding both Compton scattering and meson
photoproduction amplitudes. In a simultaneous fit to all available
data the parameters of the nucleon resonances are extracted. We find
that a global fit to the data of the various channels involving the
final states $\gamma N$, $\pi N$, $\pi\pi N$, $\eta N$ and $K \Lambda$
is possible. Especially in the eta photoproduction a readjustment of
the masses and widths found in the fits to hadronic reactions alone
\cite{fm98} is necessary to describe the data. Only for the $\nres
D13{1520}$ we find a possible disagreement for the helicity couplings
extracted using the combined dataset and pion photoproduction
multipoles alone. The model-dependence introduced by the restoration
of gauge invariance is discussed and found to be significant mainly
for resonances with small helicity couplings.

{\it PACS}: 14.20.Gk, 11.80Gw, 13.30.Eg, 13.75.Gx, 13.60.Le, 25.20.Lj\\
{\it Keywords}: baryon resonances; unitarity; partial wave analysis;
meson photoproduction; Compton scattering; coupled channels

\maketitle

%
%

\section{Introduction}

Photon- and meson-induced reactions on the nucleon are the main source
of information about the nucleon resonance spectrum. From the
knowledge of the possible excitations of the nucleon and their
properties one hopes to extract information about the structure of the
nucleon. To this end one needs models that allow to determine the
masses and partial decay widths of the resonances. Because of the
constraint of unitarity the different reaction channels can in
principle not be treated separately, but have to be taken into account
simultaneously.

For the purely hadronic reactions models based on unitarity and
analyticity have been widely used
\cite{cfhk79,ka84,pj91,ms92,sg93,sm95}. Using an ansatz developed by
the Carnegie-Mellon Berkeley or CMB group, Dytman et al.\ for example
have recently extracted resonance parameters from a fit to the $\pi N
\to \pi N, \pi\pi N, \eta N$ data \cite{dvl97}.

On the other hand, in calculations of meson photoproduction, effective
Lagrangian models are the main tool for investigations
\cite{nbl90,dmw91,bmz95,sau96}. In these models the important
constraint of gauge invariance can be easily implemented on the
operator level. However, unitarity has only been fulfilled in a few
calculations of pion photoproduction in the $\Delta$-region
\cite{nbl90,dmw91,sg96,sl96} and eta photoproduction \cite{sau96}.
Lately, also a description of Compton scattering in this framework has
been put forward \cite{skpn96}.

With the availability of high-precision data from various accelerator
facilities like MAMI, ELSA, GRAAL and TJNAF, there is also an urgent
need to improve the models for meson photoproduction. The most
important ingredient for these improved models is the dynamical
treatment of the meson rescattering in the same framework as the
initial photoproduction reaction. To this end a description of the
purely hadronic reactions within the effective Lagrangian approach is
necessary.

As a first step, we formulated a model for these hadronic reactions
employing the $K$-matrix approximation \cite{fm98}. The interaction
potential is described solely in terms of $s$-, $u$- and $t$-channel
Feynman diagrams. We have shown that a reliable extraction of the
resonance parameters is possible in this model and that the effective
Lagrangian approach reduces the number of free parameters
considerably, since the non-resonant background is generated by a few
couplings only.

Analyticity is not guaranteed in this approach, but there are
estimates about the quality of the $K$-matrix calculation as compared
to other approximations \cite{pj91,sg93}. These indicate that the final
resonance parameters extracted in different approximations are very
similar.

The aim of this paper is now, to extend this model to include also
photon-induced reactions. This would allow to benefit from the very
accurate data for such reactions that already are or will be available
in the near future. To this end we will at first shortly discuss the
model used and the results of the fits to hadronic data. The extension
to photon-induced reactions and the database available for the various
reactions will then be addressed. After that, the results of the fits
to the combined data are presented and discussed.

An important new feature of this analysis is the extraction of
electromagnetic coupling constants from a combined fit to meson
photoproduction and Compton scattering data. Apart from the
$t$-channel contributions, the latter is determined only by the
electromagnetic couplings and not by a product of strong and
electromagnetic couplings as in meson photoproduction. Compton
scattering data should thus provide an important constraint on the
extraction of electromagnetic coupling constants. We will also use
a comparison with a dispersion theoretical analysis of these data to obtain
information about the quality of the $K$-matrix approximation.

%
%

\section{Description of the model and its application to the purely
  hadronic reactions}
\label{hadronic}

For easier reference we review in this section briefly the treatment
of the hadronic channels in \cite{fm98}. This then forms the basis of
our inclusion of photon-induced reaction channels.

\subsection{The $K$-matrix approximation}
\label{kmat}

The Bethe-Salpeter equations encountered in meson-nucleon scattering
\be
T = V + V G_{BS} T
\label{bseqn}
\ee
can always be decomposed into two coupled equations \cite{pj91,gw64}:
\bea
K &=& V + V \real (G_{BS}) K \nonumber \\
T &=& K - \rmi K \imag (G_{BS}) T .
\label{kteqn}
\eea
Here $G_{BS}$ is the Bethe-Salpeter propagator that describes the
intermediate propagation of the meson and the nucleon.

In (\ref{kteqn}) the second equation can easily be solved, since the
imaginary part of $G_{BS}$ is always proportional to
$\delta$-functions that place the meson and nucleon on its mass-shell.
The real part of $G_{BS}$ in the first equation in (\ref{kteqn})
amounts to a principle-value integral and is, therefore, much harder to
treat.

The $K$-matrix approximation now consists of neglecting this
principle-value integral in the first equation and thus using $K = V$ in
the second one. This leads to
\be             \label{K-matrix}
\left [ T_K \right ] = \left [ V + \rmi V T_K \right ] = 
\left [ \frac{V}{1 - \rmi V} \right ] ,
\label{kmatapprox}
\ee
where the brackets $[ \cdots ]$ indicate that $V$ and $T_K$ are $n
\times n$-matrices ($n$ being the number of asymptotic channels taken
into account) and that (\ref{kmatapprox}) is a matrix equation.

Obviously, $T_K$ as calculated from (\ref{kmatapprox}) does not
necessarily fulfill a dispersion relation, as does the full $T$ from
(\ref{kteqn}). Therefore, the $K$-matrix approximation does not
guarantee analyticity by construction. However, Pearce and Jennings
\cite{pj91} have shown that in $\pi N$-scattering the contributions
from $\real (G_{BS})$ are of minor importance, since the corresponding
principle-value integral is reduced by a very soft cutoff needed to
regularize (\ref{kteqn}). Furthermore, Surya and Gross
  \cite{sg93} estimated that in this process the error made by putting
  the pion onshell is of the order of $(E_{\pi} - m_{\pi})
  E_{\pi}^3/m_N^4$ and can therefore be neglected for small pion
  energies. Based on these studies, it seems that the main
contributions in the Bethe-Salpeter equation (\ref{bseqn}) come from
the rescattering with both intermediate particles onshell. This part
is correctly taken into account in the $K$-matrix approximation.

\subsection{Contributions to the potential $V$}
\label{vhadr}

As asymptotic states $\pi N$, $\pi\pi N$, $\eta N$ and $K \Lambda$
have been taken into account. A detailed description of the database
used in the fits is given in \cite{fm98}. Neglected so far are final
states such as $K \Sigma$ and $\omega N$. This has been done because
either the coupling of resonances to this channel is known to
be small ($K \Sigma$, \cite{pdg96}), or because only one resonance is
known that might have a significant decay into this channel ($\omega
N$, \cite{ms92}).

Also the $\rho N$-channel is not contained in our analysis. To make at
least partly up for this deficiency, we parameterize the $\pi\pi
N$-decay by the coupling to a scalar, isovector $\zeta$-meson
\cite{fm98,sau96,sg96,bdssnl97} with mass $m_{\zeta} = 2 m_{\pi}$. This
ensures that the important $\pi\pi N$-decay is taken into account, but
at the same time the model is kept as simple as possible. A
description of the $\pi\pi N$ final state in terms of two particle
intermediate states like $\pi \Delta$, $\pi \nres P11{1440}$ and $\rho
N$ is currently under way.

The potential $V$ is now calculated from the interaction Lagrangians
collected in App.\ \ref{hadrcoupl}. Taken into account are
contributions from the nucleon Born terms, $t$-channel exchanges of
$\rho$, $a_0$ and $K^*$ and resonance contributions in the $s$- and
$u$-channels.

In this work we limit ourselves to partial waves with spin $\einh$ and
$\dreih$. This is because only for these the Lagrangians can be given
in an unambiguous way \cite{nb80,n81}. Furthermore, we restrict the
energy range to $\sqrt s \le$ 1.9 GeV because for higher energies
additional decay channels might be important.

\subsection{Form factors}

In order to investigate the dependence of the resonance parameters on
the specific choice of the hadronic form factors, in \cite{fm98} fits
using combinations of the three basic forms
\bea
F_p(q^2) &=& \frac{\Lambda^4}{\Lambda^4 + (q^2 - m^2)^2} \nonumber \\
F_e(q^2) &=& \exp (- \frac{(q^2 - m^2)^2}{\Lambda^4}) \nonumber \\
F_t(q^2) &=& \frac{\Lambda^4 + (q_t^2 - m^2/2)^2}{\Lambda^4 + (q^2 -
  (q_t^2 + m^2/2))^2}
\label{ourforms}
\eea
have been carried out ($m$ denotes the mass of the propagating
particle, $q$ its four-momentum and $q_t^2$ is the value of $q^2$ at
the kinematical threshold in the $t$-channel). The following
restrictions were imposed to limit the number of free cutoff
parameters:
\begin{itemize}
\item the same functional form $F$ and cutoff $\Lambda_N$ is used in
  all vertices $\pi NN$, $\eta NN$ and $K N\Lambda$,
\item for all resonances we take the same $F$ as for the nucleon, but
  different values $\Lambda_{1/2}$ and $\Lambda_{3/2}$ for the
  cutoffs for spin-$\einh$ and spin-$\dreih$ resonances,
\item in all $t$-channel diagrams the same $F$ and $\Lambda_t$ are
  used.
\end{itemize}

\subsection{Results for the hadronic reaction channels}

In this section we briefly recapitulate the main results of
the fits to the hadronic reactions $\pi N \to \pi N$, $\pi N \to
\pi\pi N$, $\pi N \to \eta N$ and $\pi N \to K \Lambda$. The reader is
referred to \cite{fm98} for a more detailed discussion.
\begin{itemize}
\item A qualitative and quantitative description of all hadronic data
  is achieved.
\item The resonance parameters we find are in good
  agreement with the values obtained by other groups
  \cite{cfhk79,ka84,ms92,sm95,bdssnl97} and with the PDG-values
  \cite{pdg96}.
\item The $\pi^- p \to \eta n$ data are not good enough to determine
  the $\eta N$-branching fractions of the nucleon resonances very
  accurately. Nevertheless, we find non-vanishing couplings for a few
  resonances ($\nres S11{1535}$, $\nres P11{1710}$, $\nres P13{1720}$
  and $\nres D13{1520}$).
\item Only two resonances ($\nres S11{1650}$ and $\nres P11{1710}$)
  exhibit sizeable couplings to the $K \Lambda$-channel. For higher
  energies the $\pi^- p \to K^0 \Lambda$-reaction is dominated by the
  $t$-channel $K^*$-exchange.
\item There are deviations from the $\pi N$ data for the $P_{I3}$- and
  $D_{I3}$-partial waves. Below the resonance positions we
  underestimate the amplitudes, whereas for energies above the
  resonance the fits give too large amplitudes. This indicates that a
  fit with form factors that are asymmetric around the resonance
  position $m_R$ (instead of our choices $F_{p,e}$) might lead to a
  better description of the data.
\item Already for energies above 1.6 GeV we find that the $t$-channel
  contributions are not adequately described by the corresponding
  Feynman diagrams anymore. It seems that stronger modifications than
  those from any of the form factors $F_{p,e,t}$ are needed. A smooth
  transition to a Regge-like behavior in this energy range would
  probably improve the quality of the fits for the highest energies.

  Because of this, in the fits the $\pi NN$- and $\rho
  NN$-coupling constants are driven below their usual values. This in
  turn also leads to a smaller $\nres P33{1232}$ mass (1.229 GeV) and
  width (110 - 113 MeV).
\end{itemize}

  A comparison with the resonance parameters obtained from other
  unitary and analytic analyses, as for example those listed in
  \cite{pdg96}, shows that the hadronic reactions can be described
  rather well in this effective Lagrangian approach employing the
  $K$-matrix approximation.

%
%

\section{Extension to photon-induced reactions}
\label{photonic}

With the inclusion of the $\gamma N$ final state a model can now be
constructed that combines the advantages of describing the
electromagnetic interactions using effective Lagrangians with a
dynamical treatment of the rescattering. In principle this extension
is straightforward, it mainly consists of enlarging the matrices $[V]$
and $[T]$ to take into account the new final state.
Eq.\ (\ref{kmatapprox}) then gives the unitarized amplitudes for meson
photoproduction and Compton scattering.
However, there are four
important points that make the combined treatment of all possible
reaction channels technically more involved:
\begin{enumerate}
\item The photon can induce electric and magnetic transitions.
  Therefore, in the case of spin-$\dreih$ resonances two amplitudes
  (called multipoles in photoproduction, instead of partial waves as in
  the hadronic reactions) have to be taken into account \cite{gw64}.
\item Furthermore, the interaction of the photon with the nucleon can
  be split up into an isoscalar and an isovector part. For e.g.\ pion
  photoproduction this leads to three different isospin amplitudes
  ($T^0$, $T^{1/2}$ and $T^{3/2}$), instead of two ($T^{1/2}$
  and $T^{3/2}$) as in $\pi N$-scattering \cite{gw64}.
\item In the multipole-decompositions of photoproduction data any
  influence of Compton scattering is usually neglected \cite{bm92}.
\item The Compton amplitudes cannot be fully isospin-decomposed using
  experimental data alone. Since only two physical processes ($\gamma
  p$ and $\gamma n$) exist, only two of the four amplitudes can be
  extracted. Therefore, the rescattering contributions are normally
  calculated in a basis using physical states (e.g.\ $\pi^- p$, $\pi^0
  p$, $\pi^+ n$, $\pi^0 n$) and not pure isospin states \cite{bm92}.
\end{enumerate}
The first point can easily be taken into account by introducing {\it
  two} new final states $(\gamma N)_E$ and $(\gamma N)_M$, where the
index denotes the type of electromagnetic transition. The
isoscalar/isovector nature of the photon can be treated in a similar
way. The usual isospin decomposition of pion photoproduction, for
example, is given by \cite{sau96,gw64}:
\be
\langle \pi_j | T_{\Ipigam} | \gamma \rangle =
\tau_j T_{\Ipigam}^{0} +
\frac{1}{3} \tau_j \tau_3 T_{\Ipigam}^{1/2} +
(\delta_{j3} - \frac{1}{3} \tau_j \tau_3) T_{\Ipigam}^{3/2} .
\label{isogw}
\ee
Here $T_{\Ipigam}^{0}$ contains the amplitude for isoscalar photons,
whereas $T_{\Ipigam}^{1/2,3/2}$ are the amplitudes for total
isospin $I\!=\!\einh,\dreih$ induced by isovector photons. The important
point to note is, that rescattering of $T_{\Ipigam}^{0}$ and
$T_{\Ipigam}^{1/2}$ takes place via the $I\!=\!\einh$ part of the
hadronic reaction channels (e.g.\ $T_{\Ipipi}^{1/2}$), whereas for
$T_{\Ipigam}^{3/2}$ only $T_{\Ipipi}^{3/2}$ contributes. The
$I\!=\!\dreih$ amplitudes can, therefore, be treated in the usual way. For
the $I\!=\!\einh$ sector a further splitting of the final states
into $(\gamma N)_E^{0}$, $(\gamma N)_E^{1/2}$, $(\gamma N)_M^0$ and
$(\gamma N)_M^{1/2}$ needs to be introduced.

The third and the fourth point amount to neglecting
photon-rescattering contributions when extracting the photoproduction
amplitudes from the data. Only a direct term $\sim V_{\Igamgam}$ is
then present in Eq.\ (\ref{kmatapprox}). In our analysis of those
amplitudes this can only be taken into account by using the form $T =
V + \rmi T V$ and not $T = V/(1 - \rmi V)$, since in the latter case
direct and rescattering terms cannot be disentangled. For the
photoproduction of a scalar meson $\varphi$ this schematically leads
to
\be
T_{\Iphigam}^{I_{\gamma}} =
V_{\Iphigam}^{I_{\gamma}} +
\rmi \sum\limits_{\dop \varphi} T_{\Iphiphip}^{I_{\varphi}}
V_{\Iphipgam}^{I_{\gamma}} ,
\qquad I_{\gamma} = 0, \einh, \dreih, I_{\varphi} = \einh, \dreih .
\label{unitarpipho}
\ee
For Compton scattering we end up with:
\be
T_{\Igamgam}^{p,n} = V_{\Igamgam}^{p,n} +
\rmi \sum\limits_{c} T_{\gamma c} V_{c \gamma} .
\label{unitarcompton}
\ee
Here the sum runs over all physical intermediate states $c$ (e.g.\
$\pi^0 p$, $\pi^+ n$, ... for $\gamma p \to \gamma p$). Neglected are
in both cases contributions from the photon rescattering, since they
are suppressed by an additional factor $e^2$. Because of the same
suppression we also do not take electromagnetic corrections to the
hadronic channel into account. We have checked this approximation by
also calculating the photon-rescattering contributions in the
$\Delta$-region and found them to be negligible.

\subsection{Background contributions and resonance couplings}

The contributions to the potential $V$ in the case of photon-induced
reactions come from Bremsstrahlung of asymptotic particles ($N$,
$\pi$, $K$) and electromagnetic decays (e.g.\ nucleon resonances and
vector mesons). The Bremsstrahlung leads to Born diagrams in the $s$-,
$u$- and $t$-channels; decays of nucleon resonances give contributions
in the $s$- and $u$-channels, whereas meson decays enter as
$t$-channel diagrams.

To calculate the different contributions, we first of all need to
specify the couplings to the photon. Since the Lagrangians have
already been discussed many times \cite{dmw91,nbl90,bmz95,sau96,fm97},
we limit ourselves to a short summary.

For the nucleon (N, $\Lambda$ and $\Sigma$) and the scalar mesons
(denoted by $\varphi$) we have:
\bea
{\cal L}_{\gamma NN} &=&
- e \bar N \left [ \frac{(1 + \tau_3)}{2} \gamma_{\mu} A^{\mu} -
(\kappa^s + \kappa^v \tau_3) \frac{\sigma_{\mu \nu}}{4 m_N}
F^{\mu \nu} \right ] N  \\
{\cal L}_{\gamma N \Lambda,\Sigma} &=&
e \bar N \kappa_{\Lambda, \Sigma} \frac{\sigma_{\mu \nu}}{4 m_N}
F^{\mu \nu} N \\
{\cal L}_{\gamma \varphi\varphi} &=&
- e \left [ \mathbold \varphi \times (\partial_{\mu} \mathbold\varphi)
\right ]_3 A^{\mu} \\
{\cal L}_{\gamma \varphi NN} &=&
- e \frac{g_{\varphi NN}}{2 m_N}
\bar N \gamma_5 \gamma_{\mu} \left [ \mathbold \tau \times \mathbold
  \varphi \right ]_3 A^{\mu} .
\eea
The magnetic transitions moments are given by $\kappa_{\Lambda} =
-0.613$ and $\kappa_{\Sigma} = 1.610$ \cite{pdg96,as90}. In the case
of pseudovector $\varphi NN$-coupling, ${\cal L}_{\gamma \varphi NN}$
gives rise to the so-called seagull- or 4-point-diagram.

For the decay of scalar ($\varphi$) and vector ($v$) mesons we use
($F^{\mu \nu} = \partial^{\nu} A^{\mu} - \partial^{\mu} A^{\nu}$,
$v^{\lambda \sigma} = \partial^{\sigma} v^{\lambda} -
\partial^{\lambda} v^{\sigma}$):
\bea
{\cal L}_{\gamma\gamma \varphi} &=&
e^2 \frac{g_{\gamma\gamma \varphi}}{2 m_{\varphi}}
\varepsilon_{\mu \nu \lambda \sigma}
F^{\mu \nu} (\partial^{\lambda} \varphi) A^{\sigma} \nonumber \\
{\cal L}_{\gamma \varphi v} &=&
e \frac{g_{\gamma \varphi v}}{4 m_{\varphi}}
\varepsilon_{\mu \nu \lambda \sigma}
F^{\mu \nu} v^{\lambda \sigma} \varphi ,
\eea
with the couplings extracted from the corresponding decay widths \cite{pdg96}:
\bea
g_{\gamma \gamma \pi^0} &=& -0.044, \qquad
g_{\gamma \gamma \eta} = 0.167 \nonumber \\
g_{\gamma \pi^0 \rho^0} &=& 0.131, \qquad
g_{\gamma \pi^{\pm} \rho^{\pm}} = 0.103, \qquad
g_{\gamma \eta \rho^0} = 1.020 \nonumber \\
g_{\gamma \pi^0 \omega} &=& 0.313, \qquad
g_{\gamma \eta \omega} = 0.329 \nonumber \\
g_{\gamma \pi^0 K^{*, 0}} &=& 0.631, \qquad
g_{\gamma \pi^{\pm} K^{*, \pm}} = 0.415 .
\label{mesondecays}
\eea

For the spin-$\einh$ resonances only a magnetic transition is
possible:
\be
{\cal L}_{\gamma N R_{1/2}} =
e \bar R g_1
\frac{\Gamma_{\mu \nu}}{4 m_N} N F^{\mu \nu} + h.c.  ,
\label{lrng12}
\ee
where $R$ is the resonance spinor. The operator $\Gamma_{\mu \nu}$ is
given by $\gamma_{5} \sigma_{\mu \nu}$ for odd-parity resonances and
by $\sigma_{\mu \nu}$ otherwise. In the spin-$\dreih$ case two
couplings can contribute:
\bea
{\cal L}_{\gamma N R_{3/2}} &=&
- \frac{\rmi e g_1}{2 m_N} \bar R^{\alpha} \Theta_{\alpha \mu} (z_1)
\gamma_{\nu}
\Gamma N F^{\nu \mu} + \nonumber \\
& & - \frac{e g_2}{4 m_N^2} \bar R^{\alpha} \Theta_{\alpha \mu} (z_2)
\Gamma (\partial_{\nu} N) F^{\nu \mu} + h.c.  ,
\label{lrng32} \\
\Theta_{\alpha \mu} (z) &=&
g_{\alpha \mu} - \frac{1}{2} (1 + 2 z) \gamma_{\alpha} \gamma_{\mu} \nonumber .
\eea
Here, the operator $\Gamma$ is either $1$ or $\gamma_5$ for odd and
even parity resonances, respectively.

The initial values for the electromagnetic couplings of the resonances
are calculated from the helicity couplings given by the
Particle-Data-Group \cite{pdg96}. The connections between the
$g_{1,2}$ used here and the $A_{1/2, 3/2}$ are as follows
\cite{wspr90,p97}:
\bea
\mbox{Spin-$\einh$ : } & &
A_{1/2} = \mp \frac{e \xi_R}{2 m_N} \sqrt {\frac{m_R^2 - m_N^2}{2 m_N}} g_1
\nonumber \\
\mbox{Spin-$\dreih$ : } & &
A_{1/2} = - \frac{e \xi_R}{4 m_R} \sqrt {\frac{m_R^2 - m_N^2}{3 m_N}}
(\pm g_1 + \frac{m_R}{4 m_N^2}(m_R \mp m_N) g_2) \nonumber \\
& &
A_{3/2} = \mp \frac{e \xi_R}{4 m_N} \sqrt {\frac{m_R^2 - m_N^2}{m_N}}
(g_1 - \frac{1}{4 m_N}(m_R \mp m_N) g_2) .
\label{ganda}
\eea
Here $\xi_R$ denotes the phase at the $\pi NR$-vertex. These relations
can easily be inverted to extract the $g_{1,2}$.

In all these couplings we have to account for the isoscalar/isovector
nature of the photon. This is done in the case of $I\!=\!\einh$ resonances
by using $g_{1,2} = g^s_{1,2} + \tau_3 g^v_{1,2}$ in (\ref{lrng12})
and (\ref{lrng32}). Note that $I\!=\!\dreih$ resonances only couple to
isovector photons and therefore have the same coupling $g_{1,2} T_3$ to
$\gamma p$ and $\gamma n$.

The formulae needed for the multipole decomposition of photoproduction
and Compton scattering are somewhat lengthy and have been collected in
App.\ \ref{multipoles}. There we also list the connections of the
amplitudes to the various observables.

\subsection{Form factors and gauge invariance}
\label{ffandgauge}

As we have already stated in the introduction, an effective Lagrangian
model allows to address the question of gauge invariance on a
fundamental level. Unfortunately, there is no unique way to restore
gauge invariance once hadronic form factors have been introduced. This
ambiguity affects only the Born terms, since all transition vertices
fulfill gauge invariance by construction \cite{nbl90}. That is, having
a vertex function $\Gamma^{\mu}_{RN \gamma}$ constructed from the
corresponding Lagrangian, we always have
\be
\Gamma^{\mu}_{RN \gamma} k_{\mu} = 0 ,
\label{reswti}
\ee
where $k_{\mu}$ is the photon-momentum. The main reason for this is
that a $R \to N \gamma$ transition cannot be derived from the
hadronic Lagrangians by minimal substitution, but has to be
constructed by hand. Therefore, all corresponding vertex functions
have to fulfill (\ref{reswti}) separately, in order to preserve gauge
invariance. Because of this, we have also introduced independent form
factors at the electromagnetic vertices of the nucleon resonances.
These form factors are taken to have the same analytic form as the
hadronic ones (cf. Eq.\ (\ref{ourforms})), only the cutoffs $\Lambda$
are chosen independently.

For vertices derived through minimal substitution, only the sum of all
contributing ($s$-, $u$- and $t$-channel and 4-point) diagrams needs
to fulfill a constraint similar to (\ref{reswti}):
\be
\sum\limits_{i=s,u,t,4} {\cal M}_{\gamma \varphi, i}^{\mu} k_{\mu} = 0 .
\label{bornwti}
\ee
Here ${\cal M}_{\gamma \varphi, i}^{\mu}$ denotes the contribution of
the $i$'th Born diagram to the photoproduction amplitude.

In the case of pion photoproduction the vertex functions are
$\Gamma^{\mu}_{\gamma NN}$, $\Gamma^{\mu}_{\gamma \pi\pi}$ and
$\Gamma^{\mu}_{\gamma \pi NN}$ and lead to four possible diagrams:
$s$-, $u$- and $t$-channel and 4-point. Now the momenta at the
hadronic vertices are different in all diagrams. Since,
correspondingly, the form factors have different values for each
diagram, all terms in (\ref{bornwti}) are weighted differently once
hadronic form factors are introduced. Hence, the inclusion of any of
the factors $F_{p,e,t}$ will results in a violation of gauge
invariance.

The first model that tries to solve this problem was proposed by Ohta
\cite{o89}. In his ansatz it is assumed that the form factor $F$ can
be separately Taylor-expanded with respect to the momenta $p$ $\dop p$
and $q$. After minimal substitution in all three momenta, the
resulting expressions can then be resummed in closed form.

To illustrate this in some detail, we focus on the $s$-channel
Born diagram. For our choice of hadronic form factors the amplitude
${\cal M}_{\gamma \pi, s}^{\mu}$ is given by (suppressing factors of
$\rmi$):
\be
{\cal M}_{\gamma \pi, s}^{\mu} =
{\bar u} (\dop p) g_{\pi NN} F(s)
\gamma_5 \bsl q
\frac{(\bsl p + \bsl k) + m_N}{(p + k)^2 - m_N^2}
e \left [
\gamma^{\mu} + \frac{\rmi \sigma^{\mu \nu}}{2 m_N} \kappa_N k_{\nu}
\right ] u(p) .
\label{borns}
\ee
Using now Ohta's prescription we obtain an additional counter term
$\left . {\widetilde {\cal M}}_{\gamma \pi, s}^{\mu} \right |_{Ohta}$.
After some rearrangements the sum of both can be expressed as:
\bea
\left . {\cal M}_{\gamma \pi, s}^{\mu} \right |_{Ohta} &=& 
{\cal M}_{\gamma \pi, s}^{\mu} +
\left . {\widetilde {\cal M}}_{\gamma \pi, s}^{\mu} \right |_{Ohta} = 
\nonumber \\
&&
{\bar u} (\dop p) g_{\pi NN} \gamma_5 \bsl q
\frac{(\bsl p + \bsl k) + m_N}{(p + k)^2 - m_N^2}
e \gamma^{\mu}
u(p) +
\mbox{Terms } \sim \sigma^{\mu \nu} k_{\nu} .
\label{bornohta}
\eea
From this it is clear that the influence of the form factor $F$ on the
coupling to the charge of the nucleon has been fully removed by the
Ohta-prescription. Only the coupling to the magnetic moment (which is
gauge invariant by itself) is affected by the introduction of $F$ at
the hadronic vertex. The same also holds for the $u$- and $t$-channel
diagrams.

Haberzettl \cite{hbmf97} has argued against this procedure because
four-momentum conservation connects $p$, $\dop p$ and $q$ and the form
factor $F$ is needed at an unphysical point $F(m_N^2, m_N^2,
m_{\varphi}^2)$. To incorporate both points, he has developed an
alternative method that leads to different expressions for the final
amplitude.

If we choose, for example, $p$ and $\dop p$ as the independent
variables, we can remove the $q$-dependence by using $q = \dop p - p$.
After this, we get one counter-term less than in Ohta's method. The
net result for the electric $s$-channel contribution is in this case
given by:
\be
\left . {\cal M}_{\gamma \pi, s}^{\mu} \right |_{Haberzettl} =
\left . {\cal M}_{\gamma \pi, s}^{\mu} \right |_{Ohta} 
\cdot F(t) .
\label{bornhaber}
\ee
Instead of removing the influence of $F$ altogether, we now have
replaced the form factor used in the bare Born diagram with the one
from the $t$-channel exchange. Since the same happens for the other
diagrams as well, we have an overall factor $F(t)$ multiplying the
charge contributions from all four Born diagrams. Of course, the
choice of independent variables is arbitrary, so that in its most
general form the amplitude (\ref{bornhaber}) contains, instead of
$F(t)$, a common form factor ${\widetilde F} = {\widetilde F}
(s,u,t)$. This degree of freedom has for example been used in the
calculation of Nozawa et al.\ \cite{nbl90}, where a form factor
${\widetilde F} = F(s)$ was taken into account. Following Haberzettl
\cite{hbmf97}, we use here an ansatz of the form
\be
{\widetilde F} (s,u,t) =
a_1 F(s) + a_2 F(u) + a_3 F(t) ,
\label{haberform}
\ee
with the additonal constraint $a_1 + a_2 + a_3 = 1$; in \cite{hbmf97}
the coefficients $a_i$ were fitted in kaon photoproduction.

To investigate the dependence of the electromagnetic couplings of the
resonances on the gauge procedure, we have performed fits using both
Ohta's method and the prescription of Haberzettl. Since this is mainly
an exploratory study we have in the latter case adopted the
`democratic' choice and taken the $a_i$'s to be equal. For pion
photoproduction this leads to the ansatz ${\widetilde F} (s,u,t) = 1/3
(F(s) + F(u) + F(t))$ (\ref{haberform}). In the case of eta and kaon
photoproduction, however, we use a slightly different form factor
${\widetilde F}$ because in both cases the Born terms consist of a
smaller number of diagrams. For the eta case we don't have a
$t$-channel contribution, whereas for the kaon photoproduction there
is no $u$-channel diagram involving a coupling to the charge of the
particles. Therefore, in our calculation the ansatz
\be
\begin{array}{lclclcl}
{\widetilde F}_{\pi} (s,u,t) & = & 
\frac{1}{3} F(s) & + & \frac{1}{3} F(u) & + & \frac{1}{3} F(t) \\
{\widetilde F}_{\eta} (s,u) & = & 
\einh F(s) & + & \einh F(u) \\
{\widetilde F}_{K} (s,t) & = & 
\einh F(s) & + & & & \einh F(t) ,
\label{haberformreal}
\end{array}
\ee
was used for the different form factors. This prescription has been
chosen to avoid additional free parameters. But we want to stress
again that (\ref{haberform}) and (\ref{haberformreal}) can only be
motivated by the structure of the hadronic vertices present in the
photoproduction diagrams. Only in a microscopic model for the
electromagnetic vertex the form factor could be determined
unambiguously.\\

Finally, one could think about introducing form factors at the
electromagnetic vertices of the nucleon and the pion and kaon as well.
For the $\gamma NN$ case this would lead to the following vertex:
\be 
\Gamma_{\gamma NN}^{\mu} = 
{\bar u} (\dop p) e \left [
F_1 (p^2, \dops p, k^2) \gamma^{\mu} + 
F_2 (p^2, \dops p, k^2) \frac{\rmi \sigma^{\mu \nu}}{2 m_N} 
k_{\nu} \right ] u(p) 
\ee
For real photons gauge invariance demands $F_1 (p^2, \dops p, 0) = 1$.
In addition, for $F_2$ we have $F_2 (m_N^2, m_N^2, 0) = \kappa_N$.
Besides this, we have no further constraint on $F_2$ for the case of
meson photoproduction. In the case of Compton scattering it can be
shown \cite{n91}, however, that additional contributions are needed to
restore gauge invariance, once a form factor $F_2$ has been
introduced. Unfortunately, also these contributions cannot be
determined in an unambiguous way. To avoid this additional
uncertainty, we have chosen not to include electromagnetic form
factors for the nucleon and scalar mesons in our calculation.

\subsection{Calculation of photon-induced reactions}

The calculation of photon-induced reactions is now carried out as
follows: i) calculate the potential $[V]$ from all contributing
Feynman diagrams; ii) invert the hadronic submatrix to give the
$T$-matrix $[T_{hadr}] = [V_{hadr}/(1 - \rmi V_{hadr})]$; iii)
unitarize the meson photoproduction with the help of
(\ref{unitarpipho}); iv) finally, calculate the $T$-matrix for Compton
scattering using (\ref{unitarcompton}).

The potential $[V_{hadr}]$ is calculated using the results of
\cite{fm98}, with only one exception: at the $\eta NN$-vertex we now
use pseudoscalar (PS) instead of pseudovector (PV) coupling. From eta
photoproduction \cite{tbk94} it is known that PS coupling leads to a
much better description of the data. In purely hadronic reactions this
PS $\leftrightarrow$ PV difference is hardly visible because of the
suppresion of the Born terms due to the hadronic form factor. But
since, at least for the Ohta-prescription, the influence of the form
factor is removed by restoring gauge invariance, in photoproduction
the contribution of the Born terms is enhanced as compared to others.
For kaon photoproduction the situation is not so clear \cite{bmk97},
so we will use both PS and PV coupling at the $K N
\Lambda,\Sigma$-vertices in the fits.

\subsection{Summary of the model}

Our model thus consists of a $K$-matrix treatment of the following
{\em asymptotic} channels: $\pi N$, $\pi \pi N$, $\eta N$, $K \Lambda$,
and $\gamma N$. The $K$-matrix elements are calculated from a Lagrangian
given in \cite{fm98} for the hadronic couplings and in the present paper for
the electromagnetic couplings. Possible {\em intermediate} states are the
nucleon and all three-star nucleon resonances with spins 1/2 and 3/2 up to an
invariant mass of $\sqrt{1.9}$ GeV; thus no nucleon resonances appear in
the final states of a Feynman diagram. As a consequence of the $K$-matrix
approximation all intermediate particles propagate only on-shell.

With these ingredients all Feynman-amplitudes are calculated taking
into account the $s$-, $u$- and $t$-channel diagrams. The latter are
used to consistently generate the background amplitudes thus
eliminating the need to introduce separate ad-hoc background
parametrizations. The iteration of the amplitude $V$ contained in
Eq.\ (\ref{K-matrix}) then leads to a consistent description both of
resonance decay widths and rescattering in a hadronic reaction. For
example, in a reaction with asymptotic channels $\gamma N \rightarrow
\eta N$ a whole chain of rescatterings, like e.g., $\gamma N
\rightarrow N^* \rightarrow \pi N \rightarrow N^{**} \rightarrow \eta
N$, where $N^*$ and $N^**$ are different nucleon resonances, is
automatically generated. While the resonances in this way all acquire
width, and are thus `dressed', higher order contributions to the
vertices are not taken into account and the extracted couplings
therefore are `bare' ones.

%
%

\section{Results of the fits}
\label{results}

The parameters of the model can be grouped into resonant and
non-resonant ones. The first group is given by the masses and the
hadronic and electromagnetic couplings of the resonances. Furthermore,
it also contains the $z$-parameters of the spin-$\dreih$ resonances
and the cutoffs $\Lambda$ of the form factors at all vertices. In the
case of the non-resonant background we have the couplings of the
scalar and vector mesons to the nucleon and the cutoffs at those
vertices. The vector meson decays are calculated with the couplings
fixed to (\ref{mesondecays}). Only $g_{\gamma \eta \rho^0}$ and
$g_{\gamma \eta \omega}$ have been allowed to vary in the fits because
the corresponding decay widths are only known to about 25\%.

As has been explained in Sec.\ \ref{vhadr}, only resonances with spin
$\einh$ and $\dreih$ have been taken into account. These are: $\nres
S11{1535}$, $\nres S11{1650}$, $\nres P11{1440}$, $\nres P11{1710}$,
$\nres D13{1520}$, $\nres D13{1700}$, $\nres S31{1620}$, $\nres
P33{1232}$, $\nres P33{1600}$ and $\nres D33{1700}$. Furthermore, the
potential contains $t$-channel contributions from the hadronic and
electromagnetic decays of the vector mesons $\rho$, $\omega$, $a_0$
and $K^*$. For the coupling of the $\omega$ to the nucleon, the values
\be g_{\omega NN} = 7.98 , \qquad \kappa_{\omega NN} = -0.12 \ee have
been used \cite{nbl90,fm97}. The dependence of the
  electromagnetic couplings of the $\nres P33{1232}$ on the $\omega
  NN$-couplings have been investigated in \cite{dmw91,sg96,sl96}.
  Mainly due to the $\omega$-exchange contributions to the
  $M_{1+}^{3/2}$-multipole both couplings $g_1$ and $g_2$ of the
  $\nres P33{1232}$ are influenced by variations in $g_{\omega NN}$
  and $\kappa_{\omega NN}$. Therefore, also the extracted $E/M$-ratio
  is somewhat sensitive to the values of those couplings. In principle
  one could go ahead and also determine $g_{\omega NN}$ and
  $\kappa_{\omega NN}$ through fits to the data. However, we have not
  chosen to do so, because we feel that the pion photoproduction data
  alone do not offer enough sensitivity to reliably determine the
  $\omega NN$-couplings. In view of the aim of this study to find a
  simultaneous description of all included channels the so induced
  error in the $\nres P33{1232}$-couplings is acceptable.

\subsection{Reaction channels and database}

The hadronic database has been described in \cite{fm98} and consists
of PWA results for both $\pi N \to \pi N, \pi\pi N$
\cite{ka84,ms92,sm95} and cross section and polarization data for
$\pi^- p \to \eta n, K^0 \Lambda$. Because of the much larger database
used in the SM95-PWA as compared to the KA84-solution, we only use the
SM95 data and the corresponding parameter set of \cite{fm98} in this
calculation.

\begin{itemize}
\item \mathbold {\gamma p \to \gamma p}: Differential cross section
  data from various measurements \cite{gptogp} have been used in the
  fits. Furthermore, we include the LEGS data on photon polarizations
  \cite{b96}, the old data on the recoil polarization from Wada et al.\
  \cite{gptogp} have not been used here. Since, from the helicity
  couplings given in \cite{pdg96}, we expect sizable contributions
  from spin-$\frac{5}{2}$ resonances ($\nres D15{1675}$ and $\nres
  F15{1680}$) not taken into account here, we fit the Compton data
  only for energies $<$ 1.6 GeV. Only in this energy range we can be
  sure that all resonance contributions are taken into account.
\item \mathbold {\gamma N \to \pi N}: In this channel we use the
  single-energy data of the multipole analysis SP97 \cite{sm95}. In
  principle, also another analysis (MA97 \cite{ma97}) is available,
  based mainly on the latest MAMI and ELSA data. This analysis,
  however, only covers the energy region $<$ 1.35 GeV. Because of
  this restriction, we do not use it in our fits. Nevertheless,
  differences between the two solutions SP97 and MA97 and our fits are
  shown in some of the plots.  In a later stage of the investigation
  it would be very interesting to perform restricted fits using the
  MA97 data. This could allow to investigate the dependence of the
  $\nres P33{1232}$ couplings on the multipole analysis used.
  
  Unfortunately, the spreading of the single-energy data SP97 is much
  larger than for the $\pi N$-analysis SM95. In order to further
  restrict the parameters, we also use the speeds calculated from the
  energy-dependent solution SP97 in our fits. However, some
  precautions are necessary when incorporating these data, in
  particular near the resonance positions of the $\nres P11{1710}$ and
  the $\nres D13{1700}$. Even though a resonant structure cannot be
  ruled out from the SP97 single-energy solution, the SP97 speed data
  are smooth in the vicinity of both resonances. In our calculation,
  however, we always find resonant structures, even for resonances
  coupling only weakly to $\gamma N$. For energies near these
  resonances, therefore, no fit to the SP97 speeds is possible.  For
  this reason the speed data for the resonant multipoles were not used
  near the $\nres P11{1710}$ and the $\nres D13{1700}$ in our fit.
\item \mathbold {\gamma N \to \eta N}: In the energy range below 1.54
  GeV we only use the very precise data of Krusche et al.\
  \cite{k95} for the differential and total cross section. For
  energies above that only sparse data from different groups are
  available \cite{gptoep}. For the total cross section there are also
  data from ELSA \cite{w93} for eta electroproduction at very small
  $k^2$ (= $-$0.056 GeV$^2$), but no differential cross sections.  From
  measurements on deuterium targets also neutron to proton ratios have
  been extracted \cite{hr97}. Furthermore, a few target asymmetry data
  are available \cite{bock97}. The latest measurements at GRAAL on
  photon asymmetries, however, are not yet published.
\item \mathbold {\gamma p \to K^+ \Lambda}: Here, the best data come
  from the SAPHIR experiment \cite{barth97}. The older measurements of
  differential cross sections and $\Lambda$-polarizations \cite{gptokl}
  have been carefully investigated by Adelseck and Saghai
  \cite{as90}.  Because of systematic deviations of certain datasets,
  the errorbars of these data have been enlarged. In our fits we also
  use these newly assigned errors.
\end{itemize}

The fits are labeled (as in \cite{fm98}) by the $\pi N$-PWA used to
determine the hadronic parameters and the type of form factors for the
$s$- and $t$-channel resonances. An additional number indicates the
method used to gauge the Born contributions: 1 - Ohta's method
\cite{o89} with fixed hadronic parameters, 2 - Ohta's method with all
parameters fitted and 3 - Haberzettl's method \cite{hbmf97} with fixed
$a_i$'s (cf. Eq.\ (\ref{haberformreal})). Furthermore,
from the fits performed for the hadronic channels, it is obvious that
the exponential form $F_e$ leads to larger values of $\chi^2$ as
compared to the other form factors $F_p$ and $F_t$. Therefore, we do
not use the parameterization $F_e$ for the case of photon-induced
reactions. Finally, since the fits performed in \cite{fm98} using
$F_p$ and $F_t$ lead to very similar descriptions of the data, we
limit ourselves to fits starting from the parameter set SM95-pt
(given in \cite{fm98} and also in the Tables\ \ref{mesparmSM95} -
\ref{reszparmSM95}). Here one must keep in mind that the use of
different parameterizations for the form factors introduces a source
of systematic error that can be of comparable size to the statistical
uncertainty induced by the error of the data.

  In summary, the fit results are then labeled by SM95-pt-1,
  SM95-pt-2 and SM95-pt-3, standing for the SM95 partial wave analysis
  of $\pi N$ scattering of the Virginia group \cite{sm95}, all using
  the form factor $F_p$ from Eq.\ (\ref{ourforms}) for all vertices
  with propagating hadrons and $F_t$ from the same equation for the
  $t$-channel diagrams. They employ Ohta's gauge fixing method with
  hadronic parameters determined from a fit to hadronic channels
  alone, Ohta's gauge fixing method with all parameters refitted to
  all channels, including the photonuclear ones, and Haberzettl's
  gauge fixing method, respectively.

\subsection{Fit with fixed hadronic parameters}
\label{fitphoto}
\setcounter{paragraph}{0}

In a first fit, we allowed only the electromagnetic couplings to vary.
All other values of masses and decay widths were taken from the
parameter set SM95-pt of \cite{fm98}. Using the procedure of Ohta to
restore gauge invariance, the parameters have been determined by a
simultaneous fit to the full database as described above.

In Figs.\ \ref{gpipSM} - \ref{ggSM95} we show the results of this fit
as dotted lines, together with the other fits described below. From
the plots it is clear that the experimental data in all channels can
be reproduced rather well. The improvement over our old non-unitary
calculation \cite{fm97} using the $T$-matrix approximation (also shown in
Figs.\ \ref{gpipSM} - \ref{gpi32SM}) is obvious.

\paragraph{Comparison to a $T$-matrix calculation}

One of the most noticeable differences between the calculations
performed here and those in Ref.\ \cite{fm97} is the improvement in
the description of the $E_{1+}^{3/2}$- and $M_{2-}^{n}$-multipoles. In
the case of $E_{1+}^{3/2}$, it is well known that only a correct
treatment of the rescattering allows a quantitative description of
this channel. The reason for this can best be seen in Fig.\ 
\ref{p33d13unitar}, where a calculation with both $\nres P33{1232} N
\gamma$-couplings set to zero is shown. Even with no direct coupling
to the resonance, the structure of the data in the
$E_{1+}^{3/2}$-multipole can already be reproduced quite well. This
shows that the rescattering is responsible for the shape of this
multipole and not the direct excitation of the $\Delta$. From this it
is obvious, why our old model \cite{fm97} failed in describing this
multipole.

In \cite{fm97} we have speculated that the same might be true for the
$M_{2-}^{n}$-multipole, but Fig.\ \ref{p33d13unitar} shows that this
is not the case. The direct coupling of the resonance is essential to
describe the data for both $E_{2-}^{n}$- and $M_{2-}^{n}$-multipoles.
Therefore, the poor fit in the old calculation was obviously driven by
the contributions of the $\nres D13{1520}$ to some other multipole.
These were most likely the offshell contributions that are not treated
correctly in the $T$-matrix approximation. One important deficiency in
this approximation is the appearance of spurious resonance-like
structures (e.g.\ $E_{1+}^{p,n}$, $M_{1+}^{p,n}$, $E_{0+}^{3/2}$ and
$M_{1-}^{3/2}$). These are induced by the offshell contributions of
the spin-$\dreih$ resonances. As has been demonstrated in \cite{fm98},
these structures are an artifact of the $T$-matrix approximation and
do not appear in a unitary calculation. Therefore, the investigation
of offshell contributions of spin-$\dreih$ resonances and the
corresponding $z$-parameters, as has been done recently by Mizutani et
al.\ \cite{mfls97}, is not very meaningful in a $T$-matrix calculation.
Without dynamical rescattering, the $z$-parameters are mainly adjusted
to minimize the induced offshell-structures and reveal only little
about the nature of the spin-$\dreih$ resonances. From what we have
said about the $M_{2-}^{n}$-multipole, it furthermore seems that also
some resonance couplings are influenced by this effect.

Also, the dynamical generation of the imaginary parts of the
amplitudes leads to an improved fit. Especially in cases where the
Born terms dominate the amplitude, the old calculation did not
generate the correct imaginary part ($E_{0+}$, $E_{1+}^{p,n}$
and $M_{1+}^{p,n}$). This is also easily understood, since in the
$T$-matrix approach the Born terms are purely real.

\paragraph{Compton scattering}

This is the one of the first attempts to calculate Compton scattering
in a dynamical model beyond the $\Delta$ resonance. Therefore, we are
for the first time able to check if the data on Compton scattering and
meson photoproduction can be described using the same helicity
couplings for the various resonances. As can be seen from Fig.\ 
\ref{ggSM95}, we are able to fit the available data on Compton
scattering very well. Both the differential cross section and the photon
asymmetry $\Sigma$ are reproduced over the whole energy range. From
this we conclude that the Compton data are indeed compatible to the
experimental results for the photoproduction channels.

The main contributions to the cross section come from the Born $s$-
and $u$-channel diagrams and the resonances $\nres P33{1232}$ and
$\nres D13{1520}$. This can be seen from Fig.\ \ref{ggcontrib}, where
a decomposition of the differential cross section into the individual
contributions is shown. It is also obvious that the $\pi^0$ and $\eta$
$t$-channel diagrams have a small influence under backward angles
only. For energies below 1.6 GeV all other resonance contributions could
be safely neglected, none of them exceeds 5 nb/sr.

Furthermore, in our calculation there is no need for an additional
attenuation factor for the Born terms, as introduced by Ishii et al.\
\cite{gptogp} ($x = \cos \theta$):
\be
{\widetilde A}_{Born} = A_{Born} \rme^{-C (1 - x)} ,
\ee
with a free parameter C fitted to the Compton data. The strong
backward peaking of the Born contributions is an artifact of the
$T$-matrix approximation employed by Ishii et al.\ and does not persist
once the amplitudes are properly unitarized. To illustrate the
difference between the $K$- and $T$-matrix calculation, we show in
Fig.\ \ref{ggborn} the Born $s$- and $u$-channel contribution to
the differential cross section employing both $K$- and $T$-matrix
approximation. The inclusion of mainly $\pi N$ rescattering leads to
an enhancement of the cross section under forward angles and to the
abovementioned reduction under backward angles. Therefore, we are able
to fit the Compton data without an additional factor $\rme^{-C (1 -
  x)}$.

For the photon asymmetry $\Sigma$ we also show the results of the
isobar model of Wada et al.\ \cite{gptogp} in the lower part of Fig.\ 
\ref{ggSM95}. Obviously, this ansatz is not able to describe the data.
Both magnitude and shape are in disagreement with the experimental
results. The dispersion relation calculations of L'vov \cite{lvov81},
on the other hand, can reproduce the polarization data very well. In
this region they practically coincide with our results, whereas for
energies $\sqrt s \approx$ 1.08 GeV both approaches differ by a factor
of two.

This observation allows us to investigate the validity of the
$K$-matrix approximation in some detail. The main difference to the
dispersion relation calculation performed in \cite{lvov81} is the
onshell-approximation for $G_{BS}$ in Eq.\ (\ref{kteqn}). Therefore,
in our calculation the $\pi N$ intermediate state does not contribute
below $m_N + m_{\pi} \approx$ 1.08 GeV. Taking also the
offshell-propagation into account (as the dispersion relation
implicitly does), the Compton scattering `sees' the $\pi N$-channel
already for lower energies. This is the main reason for the different
results in $\Sigma$ within about 50 MeV around the $\pi N$-threshold.
We thus conclude that the offshell `tails' of the propagator $G_{BS}$
do not extend much further than $\approx$ 50 MeV. This observation is in
agreement with the results of Pearce and Jennings \cite{pj91}, who
found that a rather soft offshell-cutoff in $G_{BS}$ is needed
($\Lambda \approx$ 300 MeV) to describe the $\pi N$-phasehifts. Thus,
it seems that the $K$-matrix approximation yields reliable results for
energies not too close to a meson-production threshold.

\paragraph{Eta photoproduction}

Looking at the overall result from our fit, we find that all major
structures of the data are also visible in the calculation. Only for a
few channels significant deviations from the data can be seen. The
most prominent of these can be found in eta photoproduction below 1.6
GeV (cf. Figs.\ \ref{gekSM95} and \ref{geSM95}). As we have pointed
out earlier, especially in this region high precision measurements are
already available \cite{k95}. Since the forthcoming experiments should
yield data with comparable quality, the eta photoproduction can be
seen as a `testing ground' for all models that try to describe
photon-induced reactions. Only the full description of this data in
all details might allow the unambiguous extraction of the $\nres
S11{1535}$ resonance parameters.

From the differential cross section it is clear that mainly the
absolute magnitude is too small for energies below 1.5 GeV, whereas
the isotropy is well reproduced. Therefore, an increase of the $\nres
S11{1535} p\gamma$-coupling alone would not improve the overall fit,
since it would lead to a drastic over-prediction of the data for
energies above 1.5 GeV. From this we conclude that a change in the
{\it energy dependence} of the resonance contribution is needed for a
better fit in this channel. Such a change can only result from a
variation of the hadronic masses and couplings; the coupling to the
photon mainly influence the magnitude and not the shape of the $\nres
S11{1535}$ contribution.

This observation coincides with the fact that the poor $\pi^- p \to
\eta n$ data were responsible for the spread of the $\nres S11{1535}$
parameters between the different fits carried out in \cite{fm98}. Also
the smaller $\eta N$-couplings of the other resonances could not be
extracted reliably. In the moment, it seems that the eta
photoproduction imposes much stricter constraints on the resonance
parameters, as the purely hadronic data does. This clearly shows that
new precise photoproduction measurements need to be accompanied by
improved hadronic data as well. Otherwise, extractions of resonance
parameters will always be handicapped by the quality of the hadronic
database.

\paragraph{Kaon photoproduction}

In the other reaction channels this problem does not show up, mainly
because of the lack of high precision data. Only in the case of kaon
photoproduction for energies around 1.75 GeV we have indications of
systematic deviations in backward directions. Here the cross section
is dominated by the Born contributions, since $g_{K \Lambda N}$ is
rather large ($\approx$ $-$6, e.g.\ compared to $g_{\eta NN} \approx$
1-2). In the hadronic channels only the product of coupling constant
and hadronic form factor enters, which is much smaller ($\approx$
$-$2.5 in the case of $g_{K \Lambda N}$). Additionally, $\pi^- p \to
K^0 \Lambda$ is for higher energies dominated by the $K^*$-exchange in
the $t$-channel. So the Born contribution, and thus $g_{K \Lambda N}$,
is not well determined by the hadronic data.

In the kaon photoproduction, however, $g_{K \Lambda N}$ plays a
dominant role, since in the Ohta prescription the influence of the
hadronic form factor is completely removed. Furthermore, the
contribution from the charge of the proton is not cancelled by a
similar $u$-channel contribution, as it is the case in $\pi^0$
production. Because of the hidden strangeness in the $K \Lambda$ final
state, we have a $\Lambda$ (or $\Sigma$) propagating in the crossed
diagram that only exhibits magnetic coupling (as in $\pi^+$
production). Since all major contributions are therefore fixed from
the outset, the fit could only be improved by reducing the photon
coupling of the $\nres S11{1650}$. The resulting value ($A_{1/2}^p =
31\!\times\!10^{-3}$ GeV$^{-1/2}$) is significantly smaller than the
number deduced from pion photoproduction ($A_{1/2}^p =
69\!\times\!10^{-3}$ GeV$^{-1/2}$, \cite{sm95}). In our fit to the
combined data of both channels the $E_{0+}^p$ data on pion
photoproduction obviously do not play such an important role as the
kaon photoproduction data because of the large uncertainty of the
former in the region of the $\nres S11{1650}$.\\

Already with fixed hadronic parameters we obtain a good overall fit.
From the observed deviations it is clear that a further improvement
can only be achieved by simultaneously varying some of the hadronic
parameters as well. Before we show the results for such combined fits,
we want to stress again that already with fixed hadronic parameters a
reasonable description of all data is possible. Especially due to the
dynamical rescattering, the main shortcomings of a $T$-matrix
calculation have been resolved.

\subsection{Global fit using Ohta's prescription}
\label{fitohta}
\setcounter{paragraph}{0}

In this section we now discuss the results of a global fit to all
hadronic and photon-induced channels in which also the hadronic
parameters are allowed to vary. Ohta's prescription was used to gauge
the hadronic form factors. The results are also shown in Figs.\
\ref{gpipSM} - \ref{ggSM95}.

\paragraph{Compton scattering}

Looking at the plots for the different reaction channels, we in
general find only a slight improvement using SM95-pt-2. In the case of
Compton scattering, the fit with fixed hadronic parameters already
describes the data rather well, so that the new fit leads only to a
relatively small decrease in $\chi^2_{\gamma \gamma}$ (7.15 $\to$
5.20). Since the main contributions here come from the Born terms and
the $\nres P33{1232}$ and $\nres D13{1520}$ resonance, the changes in
the differential cross sections can easily be explained by the
slightly different helicity couplings found in both fits. For the
$\nres P33{1232}$ both $A_{1/2}$ and $A_{3/2}$ are reduced and lead to
the observed reduction for energies up to 1.3 GeV. In contrast to
this, the increase in the $\nres D13{1520}$ helicities increases the
interference with the other contributions to Compton scattering.
Therefore, the cross section is reduced slightly in this case as well.

\paragraph{Pion photoproduction}

For the pion photoproduction the reduction of $\chi^2_{\gamma \pi}/DF$
is due to a the better fit of the $M_{1+}^{3/2}$-multipole. The increase
in $\chi^2_{\pi \pi}/DF$ comes mainly from the $S_{11}$-channel, since
the $\nres S11{1535}$-parameters exhibit the largest changes as
compared to the SM95-pt-1 values. Except from this we note only minor
changes, mainly for channels where the background is dominated by the
Born contribution (e.g.\ $M_{1-}$, and $E_{1+}^{p,n}$). Accordingly,
the values for the helicity couplings we extract are very similar for
both fits SM95-pt-1 and SM95-pt-2. In general, the agreement to the
PDG-values is quite good. Serious discrepancies we find for the $\nres
S11{1650}$ (for the reasons discussed in the last section) and for
both the $\nres P13{1720}$ and $\nres D33{1700}$. That we find no
agreement in the case of the $\nres D13{1700}$ comes as no surprise,
keeping in mind that this state is not well established and is found
at rather different energies in the different analyses. Furthermore,
the helicity couplings are known to be small and have very large
errorbars.

For both the $\nres P13{1720}$ and $\nres D33{1700}$ the background is
mainly due to the Born terms. As can be seen from $\imag (E_{1+}^p)$,
$\imag (M_{1+}^n)$ and $\real (E_{2-}^{3/2})$, this background is too
large for higher energies. Accordingly, the helicity couplings of
both resonances are adjusted to compensate this contribution.
Especially for the $\nres D33{1700}$ it is obvious that no good fit to
the multipole data is possible with this large background.

\paragraph{Eta photoproduction}

The drastic reduction of $\chi^2_{\gamma \eta}/DF$ (6.09 $\to$ 3.00)
is accompanied by only a small increase of $\chi^2_{\pi \eta}/DF$
(1.73 $\to$ 1.95), so that we have an overall decrease for both
channels (4.25 $\to$ 2.56). Furthermore, the dramatic increase of the
$\nres D13{1520} N \eta$-decay width again shows the importance of a
global fit to the full dataset. Fits to the hadronic data alone always
yield very small values for this decay ($\approx$ 10 keV), whereas the
combined fits are much closer ($\approx$ 50 keV) to the values found
elsewhere (e.g.\ $\approx$ 130 keV in \cite{bdssnl97}), if one takes
into account the lower mass of the $\nres D13{1520}$ found here. In
addition also the $\eta N$-decay width of the $\nres S11{1535}$
significantly increases. As can be seen from Figs.\ \ref{gekSM95} and
\ref{geSM95}, this increase is driven by the better fit to the cross
sections close to threshold. 

Since we now have a reasonable agreement with the precise data of
Krusche et al.\ \cite{k95}, we can turn to the extracted polarization
observables. The results for the polarized photon asymmetry $\Sigma$,
the recoil nucleon polarization ${\cal P}$ and the polarized target
asymmetry ${\cal T}$ are shown in Fig.\ \ref{gepolSM95}, together with
the few datapoints available and the calculations of Kn\"ochlein et
al.\ \cite{kdt95} for the photon asymmetries. The agreement with their
calculation up to 1.6 GeV is obvious. Since $\Sigma$ is dominated by
the $\nres D13{1520}$ contribution it seems that the $\eta N$-coupling
of this resonance is already well determined by the differential cross
section. In contrast to this, we are not able to reproduce the target
asymmetries for the low energies. At threshold none of our fits shows
the measured forward-backward asymmetry. This is in agreement with the
results of \cite{kdt95} that practically coincide with ours in this
energy region. In a recent analysis Tiator and Kn\"ochlein \cite{tk98}
have investigated the target asymmetry in a more phenomenological
approach and have shown that a reasonable description of all data is
only possible if one assumes a rather large, energy-dependent phase
between the $S_{11}$- and $D_{13}$-contributions; such a phase is
obviously not present in our results. For the higher energies we find
no consistent results for all three polarization observables. Above
the $\nres S11{1535}$-resonance the small contributions of the various
resonances coupling to $\eta N$ interfere strongly with each other and
with the background contributions. Further detailed investigations
have to show, if the sensitivity of $\Sigma$, ${\cal P}$ and ${\cal
  T}$ to small contributions can be used to uniquely disentangle the
different resonances. In the moment we can only observe that the
polarization observables are not well determined from the fit to the
differential cross section alone.

In addition to $\Sigma$, ${\cal P}$ and ${\cal T}$ we also show our
results for the neutron/proton ratios of the differential cross
sections. From Fig.\ \ref{genprSM95} it can be seen that the few data
points do not put strong constraints on the fits because of their
large error bars. Nevertheless, it seems as if the helicity coupling
of the neutral $\nres S11{1535}$ deduced from the $E_{0+}^n$-multipole
from pion photoproduction is too small to yield the measured
neutron/proton ratios. 

We find large variations of $d \sigma_n / d \sigma_p$ for the higher
energies. From the differential cross section it can be seen that $d
\sigma_p / d \Omega$ is rather small for forward and backward angles.
Therefore, we are extremely sensitive to the exact numbers obtained in
the calculation in this regions. This indicates that $d \sigma_n / d
\sigma_p$ is not a good quantity to investigate if one of the cross
sections is close to zero. For example, even if there would be data
for both channels under $\theta = 180^o$ with an accuracy comparable
to the results of Krusche et al.\ ($\approx$ 0.01 $\mu b /sr$), we
could still vary $d \sigma_n / d \sigma_p$ by an order of magnitude
without loosing the fit to the differential cross sections.
Therefore, we define an isospin asymmetry ${\cal I}$ similar to the
polarizations:
\be
{\cal I} = \frac{d \sigma_p - d \sigma_n}{d \sigma_p + d \sigma_n} ,
\ee
which is limited to $-1 \le {\cal I} \le 1$. Besides this more
technical advantage it also has a simple interpretation in terms of
isoscalar/isovector couplings, provided one contribution to the
amplitudes is dominant:
\be
{\cal I} = 
\frac{(g_s + g_v)^2 - (g_s - g_v)^2}{(g_s + g_v)^2 + (g_s - g_v)^2} .
\label{isoasym}
\ee
Obviously, ${\cal I}$ vanishes only if either $g_s$ or $g_v$ vanish.
If the coupling to the proton $(g_s + g_v)$ or to the neutron $(g_s -
g_v)$ vanish, ${\cal I}$ takes on its maximum values $\mp 1$.
Furthermore, in the case of one dominant amplitude ${\cal I}$ should
be rather isotropic.

The results for ${\cal I}$ are shown in Fig.\ \ref{genprSM95}. It can
be clearly seen that two different production mechanisms for forward
and backward angles develop above 1.6 GeV. Below this energy the
amplitude is dominated by the $\nres S11{1535}$-contribution. The
positive value for the higher energies in forward directions can be
understood from the then dominant $\rho$- and $\omega$-contributions
to eta photoproduction. Since both add up for the proton case and have
opposite sign for the production on the neutron, we would clearly
expect ${\cal I} > 0$. Even the magnitude can be explained in this
simple picture: taking $g_{\gamma \eta \rho^0} = 1$ and $g_{\gamma
  \eta \omega} = 0.3$ we readily obtain ${\cal I} = 0.55$ from Eq.\ 
(\ref{isoasym}). Under backward angles the situation is more
complicated, since we there do not have one dominant contribution to
the cross section. Here the Born terms, which are determined by $e_N$
and $\kappa_N$, play an important role. Already these two couplings
alone have different decompositions into isoscalar and isovector
couplings. The same is true for the small contributions of the nucleon
resonances. Therefore, one would not expect a simple explanation for
the extracted values of ${\cal I}$ in this region.\\

In summary, we find that the quality of the fits naturally improves,
once we allow the hadronic parameters to readjust. The improvement is
most significant in the eta photoproduction. Mainly the resonances
$\nres S11{1535}$ and $\nres D13{1520}$ are affected by such a
readjustment. For the other hadronic parameters, the masses of the
resonances and the branching fractions do not change in a global fit
using Ohta's method (except for $\nres S11{1535}$ and $\nres
D13{1520}$, as explained above). The partial decay widths vary, but
only in some cases ($\nres P33{1600}$, $\Gamma_{\zeta N}$ 346 MeV
$\to$ 494 MeV and $\nres D33{1700}$, $\Gamma_{\zeta N}$ 477 MeV $\to$
337 MeV) the changes are significant. Furthermore, we have found that
the fits cannot be improved in channels that are dominated by Born
contributions. Since we have used Ohta's prescription to restore gauge
invariance for both fits, the contributions of the Born terms are
nearly unchanged. From this we conclude that a further improvement is
only possible, if the couplings to the charge of the nucleon, the pion
and kaon are also changed by the inclusion of a form factor.

\subsection{Global fit using Haberzettl's prescription}
\label{fithaber}
\setcounter{paragraph}{0}

From the $\chi^2$-values (Table\ \ref{chi2comp}) it is obvious that
already a fit with fixed $a_i$'s leads to a further improvement. We
find that mainly the better fit to the Compton scattering and the pion
production data are responsible for this, whereas the $\chi^2$-values
for the other channels remain fairly constant.

\paragraph{Compton scattering}

The main improvement can be found in the fits to the differential
cross sections at higher energies (cf. Fig.\ \ref{ggSM95}). Looking at
the photon asymmetry $\Sigma$, one is tempted to conclude that the fit
SM95-pt-3 is worse than SM95-pt-{1,2}. The solution to this puzzle is
that, in the $\chi^2$-analysis, the $\Sigma$ data around 1.2 GeV are
more important than the other points, since their errorbars are much
smaller. Since these three points are reproduced better in SM95-pt-3,
the $\chi^2_{\gamma\gamma}$ does not increase, even though the slope
of the data seems to be described better using the other two fits.

\paragraph{Pion photoproduction}

In the pion photoproduction the most significant changes can be seen
in the $E_{1+}$,$M_{1+}$- and $E_{2-}$,$M_{2-}$-mutipoles. This can be
understood from the ansatz ${\widetilde F}_{\pi} = 1/3(F(s) + F(u) +
F(t))$ (\ref{haberformreal}). The Born contributions to the s-wave
multipoles $E_{0+}$ for example are mainly affected by $F(s)$, since
$F(u)$ and $F(t)$ induce angular-dependent modifications. Therefore,
the changes in the s-wave contributions are not very large for
energies $<$ 1.5 GeV. Also the larger changes in the helicity
amplitudes of the $P_{I3}$- and $D_{I3}$ resonances (cf. Table\ 
\ref{resheli12SM95}) show that the $S_{I1}$- and $P_{I1}$-channels are
affected only for higher energies (e.g.\ for the $\nres P11{1710}$).

An interesting effect can be see for the changes in the
$E_{2-}^{3/2}$- and $M_{2-}^{3/2}$-multipoles as compared to
$M_{2-}^n$. For the first two we have already concluded in Sec.\ 
\ref{fitohta} that an improved description might only be found by
changing the Born contributions. This can now be confirmed using
Haberzettl's method. Also, the helicity couplings of the $\nres
D33{1700}$ are now in somewhat better agreement with the PDG-values,
as can be seen in Table\ \ref{resheli32SM95}. Obviously, ${\widetilde
  F}_{\pi}$ led to a significant reduction of the non-resonant
background (cf. Fig.\ \ref{gpi32SM}). Since ${\widetilde F}_{\pi}$
does not depend on isospin, we expect to have a similar reduction for
$E_{2-}^{p,n}$ and $M_{2-}^{p,n}$. That this is indeed the case can be
seen in Figs.\ \ref{gpipSM} and \ref{gpinSM}. In all four multipoles
the agreement to the data for energies $<$ 1.5 GeV is reduced due to
the smaller background. The readjustment of the $\nres
D13{1520}$-parameters in fit SM95-pt-3 then results in some
deviations, mainly in the $M_{2-}^n$-multipole, since there the
errorbars are largest. From this we conclude that SM95-pt-3 represents
a compromise between the improvement in the $\nres D33{1700}$-case and
the larger deviations for the multipoles containing the $\nres
D13{1520}$.

Additionally, in all three fits we find that we overestimate the
$E_{0+}^{p,n}$-multipoles for energies around 1.3 GeV (Figs.\ 
\ref{gpipSM} and \ref{gpinSM}). Only part of this is due to the $\nres
S11{1535}$, as can be seen from Fig.\ \ref{s11contrib}, where the
results for $\real (E_{0+}^p)$ are shown with and without this
resonance. From this it seems that the background is too large in this
energy region. It is interesting to note that a similar discrepancy
was also found in the $K$-matrix calculation of Deutsch-Sauermann et
al.\ \cite{sau96}, where an even larger non-resonant contribution was
found (cf. Fig.\ \ref{s11contrib}). Since the background for this
energies is dominated by the Born terms, it might be possible to find
a better description of the data if the parameters $a_i$ in
(\ref{haberformreal}) were also allowed to vary.

\paragraph{Eta photoproduction}

Since $g_{\eta NN}$ is small compared to $g_{\pi NN}$, we find only
minor changes in the case of eta photoproduction. The differential
cross section in backward directions is larger for energies $<$ 1.7
GeV; above this energy we have a reduction as compared to
SM95-pt-{1,2}. This is due to the changes in the interference of the
Born terms with the $\rho$- and $\omega$-contributions. Since we have
no dominant resonance in this energy range, these changes can be
easily observed.

\paragraph{Kaon photoproduction}

For the kaon channels, both $\chi^2_{\gamma K}$ (3.91 $\to$ 4.09) and
$\chi^2_{\pi K}$ (3.77 $\to$ 4.21) increase slightly as compared to
SM95-pt-2. Obviously, in the case of kaon photoproduction the effect
of ${\widetilde F}_K$ is largely compensated by the increase of $g_{KN
  \Lambda}$ ($-$6.25 $\to$ $-$8.65, cf. Table\ \ref{mesparmSM95}).
Since ${\widetilde F}_K = 1/2 (F(s) + F(t))$, this compensation cannot
be complete. Because of the $t$-dependence of ${\widetilde F}_K$ a
small reduction under backward angles and a similar increase in
forward directions is expected. The calculated differential cross
sections indeed show this behavior (Fig.\ \ref{gkSM95}). It can be
seen that the use of ${\widetilde F}_K$ does not lead to an improved
description of the cross sections.

In contrast to this, the $\Lambda$-polarizations can clearly be
reproduced better using SM95-pt-3. Especially close to threshold we
find the polarization to have the right sign and magnitude, in
contrast to the other fits SM95-pt-{1,2}. Responsible for this
improvement are not the changes in either Born or $\nres
S11{1650}$-parameters, but mainly the $\nres P11{1710} p
\gamma$-coupling that increased by a factor of two in SM95-pt-3 (cf.
Table\ \ref{resheli12SM95}).\\

In summary, we find that Haberzettl's method, aside from its
theoretical appeal in that it incorporates the physical constraints on
all momenta, also leads to a better fit, mainly in the Compton scattering
and photo-pion channels. In a calculation, in which also the parameters $a_i$
in eq.\ \ref{haberform} are allowed to vary, even more significant
improvements can be obtained \cite{hbmf97}.

%
%

\section{Parameters and couplings}

After the more phenomenological discussion in the last section we now
want to focus on the extracted parameters. To this end we first
investigate the non-resonant couplings and after that the resonance
parameters as found in the different fits.

\subsection{Background parameters}
\label{nonrescoupl}

In Table\ \ref{mesparmSM95} we list the final values for the couplings
of the mesons to the $N$ and $\Lambda$ and also those for $g_{\gamma
  \eta \rho^0}$ and $g_{\gamma \eta \omega}$. The other photon-decay
couplings of the mesons have been kept at the values given in
(\ref{mesondecays}). For these the errors deduced from the
uncertainties in the decay width are of the order $\sim$ 5\%, whereas
for $g_{\gamma \eta \rho^0,\omega}$ we have $\sim$ 25\%. Furthermore,
only in the eta photoproduction we have some sensitivity on the
background couplings also at higher energies because there is no
dominant resonance contribution. From previous studies
\cite{tbk94,bmz95,sau96} it is also know that in this channel we have
a large cancellation between the Born contributions and the $\rho$ and
$\omega$ $t$-channel exchanges. This also enhances the sensitivity of
the fits onto the parameters $g_{\gamma \eta \rho^0,\omega}$.

For $g_{\pi NN}$ and $g_{\eta NN}$ obviously all fits yield very
similar results, comparable to the SM95-pt-1 values, which have been
deduced in \cite{fm98} from the hadronic data alone. For $g_{\pi NN}$
this comes as no surprise, since this value has already been extracted
many times consistently from hadronic and photon-induced reactions. 

In the case of $g_{\eta NN}$ other groups find somewhat larger values
than the ones deduced here: 2.24 from eta photoproduction
\cite{tbk94}, 6-9 from NN potentials \cite{bm90}. In our analysis the
main sensitivity comes from the data under backward angles for both
$\pi^- p \to \eta n$ and $\gamma p \to \eta p$. Interestingly, the
values we find are even smaller than those from other fits to eta
photoproduction: Benmerrouche et al.\ find $g_{\eta NN} \sim$ 5 in a
$T$-matrix calculation using effective Lagrangians \cite{bmz95}. In
contrast to that Tiator et al.\ \cite{tbk94} deduce 2.24 from a model
that tries to incorporate unitarity by the use of energy dependent
phases at the $RN\gamma$-vertices. Since we also find somewhat smaller
cross sections ($\sim$ 10 \%) using the $T$-matrix approximation, our
even smaller values $g_{\eta NN} \sim$ 1.0 might be due to
rescattering effects. Especially, since we have the above mentioned
cancellations between different non-resonant contributions, the
extracted values for the background couplings are rather sensitive to
the approximation used.

For the $g_{KN \Lambda}$ we have a totally different situation. Here,
the fits using Haberzettl's prescription find a much larger value than
those using no form factor at the charge contributions. However, as we
have already pointed out in Sec.\ \ref{fithaber}, the {\it effective}
coupling $g \cdot \widetilde F$ is of the same order for all fits,
e.g.\ $g \cdot \widetilde F$ = $-$3.79 for SM95-pt-3. This coupling is
therefore a very good example that it does not make sense to compare
the bare couplings deduced from different ansatzes/reactions/models
once form factors have been taken into account.

Since we are now able to describe both the meson- and photon-induced
production of $K\Lambda$ using rather small values $g_{KN \Lambda}$
($\sim$ $-$6), it seems that the discrepancy to the SU(3) predictions
($\sim$ $-$(10.3 - 16.7)) cannot be removed easily. But as can be seen
from SM95-pt-3, only the effective coupling is determined by the fits.
Therefore, the SU(3)-values can of course be used, as long as one
introduces a suitable form factor $\widetilde F$. This is clearly not
very satisfactory, since it renders the whole procedure of determining
$g_{KN \Lambda}$ using SU(3) questionable. Only from a microscopic
model for the form factor one could judge, if a bare coupling $g_{KN
  \Lambda}$ compatible with SU(3) would lead to reasonable fits to
the data.

One other solution to this problem was sought in the use of PS-
instead of PV-coupling at the $KN \Lambda$-vertex. The investigation
of kaon photoproduction in non-unitary models (see \cite{bmk97} for a
detailed discussion) suggested that PS-coupling leads to larger values
for $g_{KN \Lambda}$, in better agreement with SU(3). To check whether
this conclusion also holds in a multichannel calculation, we performed
a fit starting from SM95-pt-2 employing PS-coupling. We only show the
most important results of this fit in Table\ \ref{PSvsPV}. From the
$\chi^2$-values we can see that both PS and PV yield fits of similar
quality. However, we do not find a significant increase of $g_{KN
  \Lambda}$ ($-$6.8 instead of $-$6.2). Already the $\pi^- p \to K^0
\Lambda$ data alone are not compatible with $| g_{KN \Lambda} | >$ 10,
even though the contribution of the Born terms is suppressed by the
hadronic form factors \cite{fm98}.

\subsection{Resonance parameters}
\label{rescoupl}

Finally, we discuss the parameters of the nucleon resonances as found
in the fits. These are collected in Tables\ \ref{rescouplSM95} -
\ref{resheli32SM95}, where also the PDG-values and the results of
Arndt et al.\ \cite{sm95} for the various helicity couplings are
listed. We do not show the results of other models for the purely
hadronic parameters. These, and a detailed discussion, can be found in
\cite{fm98}. As can be seen from Tables\ \ref{resheli12SM95} -
\ref{resheli32SM95}, the agreement of the helicity couplings deduced
in this work to the PDG-values and the values given by Arndt et al.\ is
quite reasonable. The most important deviations ($A_{1/2}^p$ of the
$\nres S11{1535}$, $\nres S11{1650}$ and the $\nres D13{1520}$ and
also for $A_{1/2}$ of the $\nres D33{1700}$) have already been
addressed in the previous sections. They have been related to the use
of additional data besides the pion photoproduction multipoles and to
the influence of the gauge prescription used in the fits.

\mathbold {S_{11}}: For the $\nres S11{1535}$ all three fits lead to
helicity couplings much larger than the PDG-values. A similar
discrepancy was also found in the extraction of $A_{1/2}^p$ from eta
photoproduction \cite{kmzb97}. There resonance parameters very similar
to ours have been found:
\bea
m_R &=& 1.544 \mbox{ GeV} \nonumber \\
\Gamma_{tot} &=& 212 \pm 20 \mbox{ MeV} \nonumber \\
A_{1/2}^p &=& (120 \pm 26)\!\times\!10^{-3} \mbox{ GeV}^{-1/2} .
\eea
Furthermore, it has been demonstrated that larger values for $m_R$
also lead to a larger width. This trend can clearly be found in our
results as well. We therefore confirm the findings of \cite{kmzb97}
that the eta photoproduction data can only be explained using helicity
couplings larger than the PDG-value. 

Since the PDG-value has been extracted using pion photoproduction data
alone, one might think of an inconsistency between both datasets. Our
results show that this is not the case and that both reactions can be
described using a large $A_{1/2}^p$. This observation was first made by
Deutsch-Sauermann et al.\ \cite{sau96}. There the value of $A_{1/2}^p =
102\!\times\!10^{-3}$ GeV$^{-1/2}$ was extracted by a combined fit to
pion and eta photoproduction. For this conclusion the treatment of the
rescattering seems to be important, since in a $T$-matrix calculation
using effective Lagrangians a smaller helicity coupling was deduced
($A_{1/2}^p = 87\!\times\!10^{-3}$ GeV$^{-1/2}$,
\cite{fm97}) from the pion data.

The values for $A_{1/2}^p$ of the $\nres S11{1650}$ are found to be
smaller than the PDG-values. It is mainly determined by the $K
\Lambda$-channel, as has been discussed in the last section. Here
indeed it seems that this small value is in contradiction to the pion
photoproduction. Unfortunately, the data on the $E_{0+}^p$-multipole
are not very good in this energy range. Therefore, they do not
constraint $A_{1/2}^p$ very much. Especially a better determination of
$\imag (E_{0+}^p)$ would help to clarify this
situation.\\

\mathbold {P_{11}}: In the case of the $\nres P11{1440}$ the fits
SM95-pt-{1,2} agree very well with the values obtained elsewhere. Only
for SM95-pt-3 we find a somewhat larger coupling ($A_{1/2}^p = -84\!
\times\!10^{-3}$ GeV$^{-1/2}$). Nevertheless, the fit to the
$M_{1-}^p$-multipole is still as good as for the other two fits. The
change in $A_{1/2}^p$ is therefore be due to the use of Haberzettl's
prescription, which leads to a reduction of the
Born contributions.

For the second resonance we also find larger couplings once we allow
for a residual form factor $\widetilde F$. Unfortunately, the data for
pion photoproduction are not good enough to constrain the fit. As we
have discussed in Sec.\ \ref{fithaber}, the increase of $A_{1/2}^p$
in SM95-pt-3 is mainly driven by the data on the
$\Lambda$-polarization in kaon photoproduction.\\

\mathbold {P_{13}}: In this channel we are not able to describe the
imaginary part of the multipoles. There are indications for a second
state with these quantum numbers ($\nres P13{1879}$, \cite{ms92}),
which was not taken into account here. Maybe the inclusion of an
additional resonance would lead to a better fit. Accordingly, the
couplings that we find are not well determined and vary between the
different fits.\\

\mathbold {D_{13}}: For the $\nres D13{1520}$ the most obvious
deviation from the PDG-values is for $A_{1/2}^p$ (-6 in SM95-pt-1 as
compared to $-$24 $\pm$ 9). In the multipoles this shows up in the
$M_{2-}^p$-channel, where we miss the imaginary part by roughly 20\%.
This disagreement is clearly driven by the simultaneous description of
Compton scattering. This is illustrated in Fig.\ \ref{ggd13} where we
show the Compton result using the PDG helicity couplings for the
$\nres D13{1520}$. With these values the Compton data are
overestimated by a factor of 2-3, while with our couplings a good fit
to the data is obtained. Also shown in Fig.\ \ref{ggd13} is the cross
section using {\it only} the $\nres P33{1232}$. It can be seen that
this resonance gives a divergent contribution for energies $>$ 1.8
GeV. This is due to the rather large cutoff ($\Lambda_{3/2}^e \approx$
4.0 GeV), see Table\ \ref{cutoffsSM95}) obtained in all fits, which 
use the Compton data only up to
1.6 GeV. As a consequence, the $\nres P33{1232}$ gives a large
background already at lower energies, accounting for $\sim$ 50\% of
the cross section in the $\nres D13{1520}$-region. In this energy
range the $s$- and $u$-channel diagram of the $\nres P33{1232}$ lead
to a comparable cross section. Therefore, the small value of
$A_{1/2}^p$ might be forced by the interference with an unreasonably
large background.

To investigate this question in more detail, we have performed a fit
starting from SM95-pt-2 with a fixed cutoff $\Lambda_{3/2}^e$ = 1.1
GeV; the Compton scattering cross section obtained with this value is
also shown in Fig.\ \ref{ggd13}. Mainly because of the data in the
pion $M_{1+}^{3/2}$-multipole the total $\chi^2$ increases to 8.70 and
the values for the Compton channel and the pion photoproduction are
found to be $\chi^2_{\gamma \gamma}$ = 12.44, $\chi^2_{\gamma \pi}$ =
10.88 (as compared to $\chi^2_{\gamma \gamma}$ = 3.40, $\chi^2_{\gamma
  \pi}$ = 6.69 in SM95-pt-3). For the $\nres D13{1520}$ the resulting
value of $A_{1/2}^p$ changed from 3 to $-$6 and is therefore not much
closer to the PDG-value $-$24 $\pm$ 9.

Interestingly, in their old isobar model analysis of Compton
scattering, Wada et al.\ \cite{gptogp} found $A_{1/2}^p$ close to our
values, namely $A_{1/2}^p = -12.1\!\times\!10^{-3}$ GeV$^{-1/2}$. Thus
it seems that this value is not dependent on the details of the model,
but is forced by the Compton data; here clearly further investigations
are necessary.

The changes in the couplings for fit SM95-pt-3 have already been
discussed in Sec.\ \ref{fithaber}. There it was shown that the use
of a residual form factor at the Born terms results in large changes
of the background amplitudes. Especially for $M_{2-}^n$ no good fit
can be found in this case.

As in the purely hadronic fits, we find the second $D_{13}$-resonance
at rather high energies $>$ 1.9 GeV. Since we limited our fits to this
energy, we cannot meaningfully extract the parameters of this
resonance. Therefore, the values given in Tables\ \ref{rescouplSM95}
and \ref{resheli12SM95} should only be seen as an indication that a
second resonance exists in this energy range.\\

\mathbold {S_{31}}: Even without a direct photon coupling of the
$\nres S31{1620}$ we can describe the multipole data quite reasonably.
As in the hadronic case, the background in this channel is dominated
by the Born terms and the $\nres P33{1232}$ contribution. Therefore,
the resonance parameters of the $\nres S31{1620}$ are very sensitive
to the $z$-parameters of the $\nres P33{1232}$. This can be seen from
fit SM95-pt-3, where the sign change in $A_{1/2}$ is mainly driven by
the change of $z_2$.\\

\mathbold {P_{31}}: We do not include a resonance in this channel.
Nevertheless, we are able to reproduce the $M_{1-}^{3/2}$-multipole up
to energies $\sim$ 1.7 GeV. Above that, in the imaginary part clearly
the contribution of a higher lying state becomes visible. Therefore,
we use the data only up to this energy.\\

\mathbold {P_{33}}: In the global fits the hadronic parameters $\nres
P33{1232}$ tend to decrease even further. The numbers found are all at
the lower end of the allowed region. An unconstrained fit would lead
to even smaller values ($m_R$ = 1.226 GeV, $\Gamma_{\pi N}$ = 105
MeV), but the improvement in $\chi^2$ would be minimal. Since these
small numbers can be understood in terms of the $\rho$ contribution to
$\pi N$ scattering at higher energies (as has been discussed in
\cite{fm98}) we have chosen to limit the parameter range for the
$\nres P33{1232}$. The numbers in Table\ \ref{rescouplSM95}, which are
all at the lower bounds, therefore indicate that the hadronic
parameters of the $\nres P33{1232}$ are still too small, even in a
global fit to all reactions.

All three fits yield somewhat smaller electromagnetic couplings for
the $\nres P33{1232}$ than the PDG-values. As in the case of the
$\nres D13{1520}$, this is due to the inclusion of the Compton data. A
fit without this channel would lead to somewhat larger couplings.
Nevertheless, these changes are small. In this energy range both
reactions can therefore be described by a single set of parameters
that is in agreement with the values deduced from photoproduction
alone.

The $E/M$-ratio deduced in our fits is \cite{nbl90}:
\bea
R_{E/M} &=& 
- (m_R - m_N)
\frac{g_1 - g_2 m_R / (2 m_N)}
{g_1 (3 m_R + m_N) - g_2 m_R (m_R - m_N) / (2 m_N)} \nonumber \\
&=& -2.1(3) \% ,
\eea
where the number in brackets denotes the error in the last digit.
There has been some debate about the exact value of this quantity, but
it has been demonstrated that the different values extracted were due
to the use of different datasets (see \cite{asw97} for details). Using
the SM97-PWA, Arndt et al.\ deduced $R_{E/M}$ = $-$1.5(5) \%, the use of
the MA97-PWA led to $R_{E/M}$ = $-$2.4(4) \%. Wilbois et al.\ have
pointed out that the $E/M$-ratio extracted from dynamical models
depends on the unitarization method used \cite{wwa98}. From the
different ansatzes they found $R_{E/M}$ = $-$(0.7 - 5.7) \%. In order to
have a model-independent quantity it was proposed \cite{hdt96} to use
a speed plot analysis to determine the $E/M$-ratio at the pole of the
$\nres P33{1232}$ (for a detailed discussion see \cite{wwa98}). From
this one finds \cite{asw97}:
\be
\begin{array}{lcl}
\mbox{SP97} & : & R_{E/M} = -0.034(5) - 0.055(5) \rmi , \\
\mbox{MA97} & : & R_{E/M} = -0.035 - 0.046 \rmi .
\label{remlit}
\end{array}
\ee
Using the same technique here for the numerical results of our fits,
we extract an $E/M$-ratio of:
\be
R_{E/M} = -0.022(3) - 0.002(2) \rmi .
\ee
The obvious disagreement to the values given in (\ref{remlit}) can be
traced back to the imaginary part of the $E_{1+}^{3/2}$-multipole. As
can be seen from Figs.\ \ref{gpi32SM} and \ref{p33d13unitar}, it does
not rise steeply enough in our fits.

For the second $P_{33}$-resonance we find rather large values for
$A_{3/2}$. Furthermore, in the fits SM95-pt-{2,3} also the mass and
width increase as compared to the purely hadronic fits. Both effects
are driven by the fit to the $M_{1+}^{3/2}$-multipole. As can be seen
from Fig.\ \ref{gpi32SM}, for the higher energies only a change in the
mass and width leads to a good description of the data. In the
$E_{1+}^{3/2}$ case only SM95-pt-3 yields a reasonable fit for
energies $>$ 1.5 GeV. However, also in this channel the data are not
good enough to extract the helicity couplings of the $\nres P33{1600}$
unambiguously.\\

\mathbold {D_{33}}: As discussed in Sec.\ \ref{fithaber}, a
satisfactory fit could only be found using a residual form factor
$\widetilde F_{\pi}$. This again shows up in the extracted helicity
couplings. Only for SM95-pt-3 we find good agreement with the values
obtained by Arndt et al.\ \cite{sm95}.\\

\mathbold z{\bf -parameters}: We confirm the finding of \cite{fm98}
that only the $z$-parameters of the $\nres P33{1232}$ and the $\nres
D13{1520}$ can be extracted reliably. Since these two resonances give
large contributions to the background in the $S$- and $P$-waves, their
magnitude can be determined independently form the resonances in these
channels.

Clearly visible in all cases is a dependence of the $z$'s on the
gauge prescription used. This can be understood because the residual
form factor $\widetilde F$ changes the non-resonant Born terms, so
that the background determined by the $z$-parameters needs to readjust
during the fits. Mainly $z_2$ is effected by this, whereas
$z_{\varphi}$ and $z_1$ remain rather stable. The values for the
$\nres P33{1232}$ are found to be:
\be
z_{\pi} = - (0.31 - 0.35) , \qquad
z_1 = - (0.27 - 0.53) , \qquad
z_2 = - 0.66 - 1.37 ,
\ee
and can be compared to the extractions by Davidson et al.\ \cite{dmw91}
and by Olson and Osypowski \cite{oo78} who, however, do not use the
second coupling of the $\nres P33{1232}$ to $\gamma N$:
\be
\begin{array}{lrlrlrl}
\mbox{Davidson et al.:} \qquad & z_{\pi} = & -0.24 , & \qquad
z_1 = & -0.53 , & \qquad
z_2 = & 2.39 , \\
\mbox{Olson and Osypowski:} \qquad & z_{\pi} = & -0.29(10) , & \qquad
z_1 = & 0.78(30) .
\end{array}
\ee
The values found here are obviously in good agreement with both
extractions.

For the $\nres D13{1520}$ our results are given by:
\be
\begin{array}{rlrlrl}
z_{\pi} = & 0.31 - 0.35 , & \qquad
z_{\zeta} = & -(0.17 - 0.66) , & \qquad
z_{\eta} = & 0.57 - 0.82, \\
z_1 = & -(0.26 - 0.32) , & \qquad
z_2 = & -(0.48 - 1.36) .
\end{array}
\ee 
Here no other systematic investigation is available. Fits to the
eta photoproduction alone found no sensitivity to these parameters
\cite{bmz95}.

%
%

\section{Summary and conclusions}

The aim of this paper was to extend the $K$-matrix calculation of
\cite{fm98} to include photon-induced reactions. To this end we have
introduced the final state $\gamma N$ into our model, leading to the
new reaction channels $\gamma N \to \gamma N, \pi N, \eta N, K
\Lambda$. From a fit to the combined database of hadronic and
photon-induced reactions three parametersets have been extracted.

It was shown that keeping the hadronic parameters fixed to the
values obtained in \cite{fm98} no satisfactory fit to the eta
photoproduction data could be obtained. Therefore, at least in this
channel, the photoproduction measurements are needed as additional
input for the extraction of the resonance masses and width.

The use of hadronic form factors at the $\varphi NN$-vertices violates
gauge invariance so that counter terms have to be introduced to
restore it. Unfortunately, these counter terms cannot be constructed
in an unambiguous way. To investigate this systematic uncertainty, we
have performed fits using both Ohta's and Haberzettl's method.

We find that only a few couplings are sensitive to the gauge
prescription used. Mainly the parameters determining the non-resonant
contributions (e.g.\ the $z$-parameters of the spin-$\dreih$
resonances) change between the fits. Furthermore, some partial waves
are affected more than others. The reason for this is the additional
angular dependence introduced by the residual form factor $\widetilde
F (s,u,t)$ found in Haberzettl's ansatz. Even though we find it to be
small, this model-dependence has to be kept in mind when extracting
helicity couplings using effective Lagrangians. Only in a microscopic
model for the hadronic form factors this problem might be resolved.

We have shown that the Compton data provide an important additional
source of information for the extraction of the electromagnetic
coupling constant. The combined use of the Compton date and the pion
photoproduction multipoles shows only one possible candidate for an
inconsistency between both data sets: the helicity couplings of the
$\nres D13{1520}$ are found to be smaller as compared to values
extracted from pion multipoles alone. Here further investigations are
necessary, since we also find a rather large background in the Compton
amplitudes. For the $\nres P33{1232}$ a combined fit yields couplings
$\sim$ 5\% smaller than the PDG-values.

In the case of eta photoproduction the cross sections can be fitted
rather well, but we are not able to reproduce the measured target
asymmetry close to threshold. For the higher energies we find that
especially the polarization observable are sensitive to the tails of
the $\nres S11{1535}$ and $\nres D13{1520}$. The interference of
weekly coupling resonances with this background leads to large
fluctuations between the different fits. This effect might be used to
look for these resonances, since they are otherwise hard to detect
from the differential cross section alone.

In all fits we find a $KN \Lambda$-coupling of half the size of the
SU(3)-value. Since this results also persist using PS-coupling at the
$KN \Lambda$-vertex, we conclude that with $g_{KN \Lambda}$ taken from
SU(3) no fit to the combined hadronic and photoproduction data is
possible.

  The basis for all these results is the $K$-matrix method used in
  the present study. The main approximation in this method is that all
  intermediate state particles are put onshell; this violates
  causality since the connection between real and imaginary part of
  the propagator is lost.  This approximation is clearly unphysical
  for bound states, but for scattering states it seems to be quite
  reliable. We conclude this from the cited studies of Pearce and
  Jennings \cite{pj91} and Surya and Gross \cite{sg93}. Further, more
  heuristically, the agreement of the resonance parameters extracted
  in the present study with those obtained from other unitary analyses
  (for smaller channel spaces and limited energy regimes) also shows
  that the $K$-matrix approximation is quite reliable.
  
  We have tried to obtain a more quantitative estimate for the
  validity of the $K$-matrix approximation from a comparison with
  dispersion relation calculations for Compton scattering. There we
  have found that both calculations agree very well except for
  energies $\pm$ 50 MeV around the pion threshold. This is quite
  understandable in view of the discussion above: just above threshold
  the bound state behavior can still be felt. We thus expect the
  $K$-matrix approximation to be resonable also for higher energies,
  except close to particle production thresholds. Any remaining
  shortcomings of this method also have to be seen in comparison to
  its inherent and practical advantages. First, because of the absence
  of off-shell propagators no regularization of amplitudes is needed.
  There is therefore no need to renormalize, for example, coupling
  constants etc. Second, after a partial wave expansion of the
  scattering amplitude has been made, the original integral equation
  reduces to an algebraic equation which can
  more easily be solved.
  
  In the future it is clearly desirable to combine the multi-channel
  calculation performed here with approximations beyond the $K$-matrix
  ansatz, as they have been developed e.g. by Surya and Gross
  \cite{sg93} and Sato and Lee \cite{sl96}. Work along these lines is
  currently under way.\\

In summary, we were able to describe the combined dataset of hadronic
and photon-induced reaction using the same set of parameters. This
model can therefore be used for numerous detailed investigations that
have not been possible before.

%
%

\section{Acknowledgments}

The authors would like to thank H. Haberzettl for fruitful discussions
regarding the problem of gauge invariance. Furthermore, we are obliged
to C. Bennhold, A. Bock, O. Hanstein, P. Hoffmann-Rothe, B. Krusche,
T. Mart, V. Pascalutsa and L. Tiator for making their experimental and
theoretical results available to us.

\clearpage

%
%

\appendix

\section{Hadronic couplings}
\label{hadrcoupl}

In this section we list the hadronic couplings used in our model. For
a more detailed account, see \cite{fm98}. Throughout this paper we use
the notation of Bjorken and Drell \cite{bd66}. Four-momenta are
denoted by $x$. ${\mathrm x}$ is the absolute value of the
corresponding three-momentum \mathbold x. Furthermore, $\mathbold
{\hat x}$ is a unit-vector in the direction of \mathbold x: $\mathbold
{\hat x} = \mathbold x / \mathrm x$.

For the nucleon, the following couplings have been used:
\bea
{\cal L}_{NR} = &-& \frac{g_{\varphi NN}}{2 m_N} \bar N \gamma_5
\gamma_{\mu} (\partial^{\mu} \varphi) N
- g_{sNN} s (\bar N N)
- g_{s\varphi \varphi} s (\varphi^* \varphi) \nonumber \\
&-& g_{vNN} \bar N \left ( \gamma_{\mu} v^{\mu} - \kappa_v
  \frac{\sigma_{\mu \nu}}{4 m_N} v^{\mu \nu} \right ) N -
g_{v\varphi\varphi} \left [ \varphi \times (\partial_{\mu} \varphi)
\right ] v^{\mu}.
\label{backcoupl}
\eea
Here $\varphi$ denotes the asymptotic mesons $\pi$, $\eta$ and $K$, a
coupling to the $\zeta$-meson is not taken into account. $s$ and $v$
are the intermediate scalar and vector mesons ($a_0$, $\rho$ and
$K^*$) and $v^{\mu \nu} = \partial^{\nu} v^{\mu} - \partial^{\mu}
v^{\nu}$ is the field tensor of the vector mesons; $N$ is either a
nucleon or a $\Lambda$ spinor. For the $I\!=\!1$-mesons ($\pi$ and
$\rho$) $\varphi$ and $v^{\mu}$ need to be replaced by \mathbold {\tau
  \cdot \varphi} and $\mathbold {\tau \cdot v}^{\mu}$ in the $\varphi,
v NN$-couplings and by \mathbold {\varphi} and $\mathbold v^{\mu}$
otherwise.

For the $S_{I1}$ resonances we employ (pseudo-)scalar coupling to
mesons and nucleons in accordance with \cite{sau96}. The
(pseudo-)vector coupling is used in the case of nucleon and all
$P_{I1}$ resonances. This is done in order to reproduce the $\pi
N$-scattering lengths without an additional sigma meson. For the
spin-$\dreih$ resonances offshell parameters $z_i$ are taken into
account at all vertices and fitted to the available data.

For the $S_{11}$ and $S_{31}$ we therefore use:
\be
\lmnr {R_{1/2}}{PS} = - g_{\varphi NR} \bar R \: \Gamma \varphi N
+ h.c.,
\label{lnp12PS} \\
\ee
and in the case of $P_{11}$ and $P_{31}$ the couplings are given by
\be
\lmnr {R_{1/2}}{PV} = - \frac{g_{\varphi NR}}{m_R \pm m_N} \bar R
\Gamma_{\mu} (\partial^{\mu} \varphi) N + h.c. ,
\label{lnp12PV} \\
\ee
with the upper sign for positive parity. The vertex-operators $\Gamma$
and $\Gamma_{\mu}$ depend on the parity of the particles involved. For
a meson with negative intrinsic parity coupling to two baryons with
positive parity (e.g.\ $\pi NN$) they are given by $\Gamma = \rmi
\gamma_5$ and $\Gamma_{\mu} = \gamma_5 \gamma_{\mu}$, otherwise (e.g.\
$\pi N \nres S11{1535}$) we have $\Gamma = \rmi$ and $\Gamma_{\mu} =
\gamma_{\mu}$.

For the spin-$\dreih$ resonances the following coupling is used:
\bea 
\lmnr {R_{3/2}}{} &=& \frac{g_{\varphi NR}}{m_{\pi}} \bar
R^{\alpha} \Theta_{\alpha \mu} (z_{\varphi}) \Gamma
(\partial^{\mu} \varphi) N + h.c. \nonumber \\
\Theta_{\alpha \mu} (z) &=& g_{\alpha \mu} - \frac{1}{2} (1 + 2 z)
\gamma_{\alpha} \gamma_{\mu} ,
\label{lnp32}
\eea
again with a vertex-operator $\Gamma$ that is $1$ for a particle with
negative intrinsic parity and $\gamma_5$ otherwise.

For the isovector-mesons $\pi$ and $\zeta$, $\varphi$ in
(\ref{lnp12PS}) - (\ref{lnp32}) needs to be replaced by $\mathbold
{\tau \cdot \varphi}$ for $I\!=\!\einh$ resonances and by $\mathbold {T
  \cdot \varphi}$ otherwise.\\

The couplings constants can be derived from the decay widths using the
following formulae (${\mathrm p}$ denotes the three-momentum of
the meson and nucleon, $E_N$ and $E_{\varphi}$ the nucleon and meson
energy, respectively):

For spin-$\einh$ resonances we have:
\bea
\mbox {PS-coupling} &:& \nonumber \\
\Gamma_{\pm} &=& 
{\mathrm {ISO}} \; \frac{g_{\varphi NR}^2}{4 \pi} \;
{\mathrm p} \; \frac{E_N \mp m_N}{\sqrt s} \nonumber \\
\mbox {PV-coupling} &:& \nonumber \\
\Gamma_{\pm} &=& 
{\mathrm {ISO}} \; \frac{g_{\varphi NR}^2}{4 \pi (m_R
  \pm m_N)^2} \; {\mathrm p} \; \frac{2 E_{\varphi} (E_N E_{\varphi} +
  {\mathrm p}^2) - m_{\varphi}^2 (E_N \pm m_N)}{\sqrt s} . 
\eea
The upper sign corresponds to decays of resonances into mesons with
opposite parity (e.g.\ $\nres P11{1440} \to \pi N$), the lower sign
holds if both have the same parity (e.g.\ $\nres S11{1535} \to \pi N$).
$\mathrm {ISO}$ is the isospin factor, it is equal to 3 for decays into
mesons with isospin one, 1 otherwise.

Spin-$\dreih$ resonances:
\be 
\Gamma_{\pm} = 
{\mathrm {ISO}} \; \frac{g_{\varphi NR}^2}{12 \pi m_{\pi}^2} \;
{\mathrm p}^3 \; \frac{E_N \pm m_N}{\sqrt s} . 
\ee
Again, the upper sign is used if resonance and meson are of opposite
parity.

%
%

\section{Extraction of partial waves and multipoles}
\label{multipoles}

In general, the extraction of the partial waves in hadronic reactions
and the multipoles in the photon-induced channels is done similarly.
Starting point is always the invariant matrix element ${\cal M}_{fi}$
and its connection to ${\cal F}$, the scattering amplitude:
\be
{\cal M}_{fi} = 
\bar u (\dop p, \dop s) \widehat O u (p, s) = 
\frac{4 \pi \sqrt s}{\sqrt {m \; \dop m}} 
\chi_f^{\dagger} {\cal F} \chi_i .
\ee
Here, $\widehat O$ denotes the transition-operator for a given
reaction, $m$ and $\dop m$ are the masses of the initial and final
baryon, respectively. The decomposition of ${\cal F}$ in terms of
partial waves/multipoles is usually straightforward, but tedious. Once
the relations between the expansions of $\widehat O$ and ${\cal F}$
have been established, any contribution to the invariant matrix
element can therefore be decomposed into partial waves/multipoles. In
the following we want to list the relations $\widehat O \to {\cal F}$
for the three types of reactions that are described in our model.

\subsection{Meson nucleon scattering}

In this case, the invariant matrix elements ${\cal M}_{fi}$ and the
scattering amplitudes ${\cal F}$ are given by:
\bea
{\cal M}_{fi} &=& \bar u (\dop p, \dop s) \Gamma
\left ( A + B \bsl Q \right ) u (p, s) , \nonumber \\
{\cal F}_e &=& ( \widetilde A + \widetilde B \, \mathbold
{\sigma \cdot \hatdop p} \; \mathbold {\sigma \cdot \hat p} ) ,
\nonumber \\
{\cal F}_n &=& ( \widetilde A \; \mathbold {\sigma \cdot
\hatdop p} + \widetilde B \; \mathbold {\sigma \cdot \hat p} ) ,
\label{calfhadr}
\eea
with $\Gamma = 1$ for mesons with equal parity in the intital and
final state, $\Gamma = \gamma_5$ otherwise. $F_{e,n}$ are the
scattering amplitudes for the case of equal/non-equal parity,
respectively. The relation between the amplitudes $A, B$ and their
counterparts $\widetilde A, \widetilde B$ can easily be established 
\cite{fm98,gw64}, and in turn can be used to calculate the partial
waves $T_{l\pm}$ from the invariant matrix element ($x = \cos \theta$):
\bea
\widetilde A &=&
\frac{\sqrt {(\dop E \pm \dop m)(E + m)}}{8 \pi \sqrt s}
(A + B (\sqrt s - \widetilde m)) \nonumber \\
\widetilde B &=& -
\frac{\sqrt {(\dop E \mp \dop m)(E - m)}}{8 \pi \sqrt s}
(A - B (\sqrt s + \widetilde m)) , \label{abviaf} \\
T_{l\pm} &=& \frac{\sqrt {\mathrm {q \dop q}}}{2} \intm1p1 d x 
\; ( \widetilde A P_l(x) + \widetilde B P_{l\pm}(x) ) .
\label{amesonpart}
\eea
In (\ref{abviaf}), the upper sign holds for mesons with equal parity,
$\widetilde m = (\dop m \pm m)/2$. The reader is referred to
\cite{gw64} for a detailed derivation of this relation.

\subsection{Photoproduction}

In this section we list the formulae needed for extracting the
multipole amplitudes from the invariant matrix elements. This is by no
means a complete listing of all the relations between the different
set of amplitudes; a detailed account of this can be found in
\cite{kdt95}.

For the multipole decomposition of the meson photoproduction, we first
need the relation between the invariant matrix elements ${\cal
  M}_{fi}$ and the CGLN invariants ${\cal F}_i$ \cite{cgln57}. To this
end we expand the Feynman amplitude as \cite{gd63}
\be
\rmi {\cal M}_{fi} = \bar u (\dop p, \dop s)
\sum\limits_{j = 1}^{4} A_j M_j u (p, s) ,
\label{mfipipho}
\ee
where
\bea
M_1 &=&
- \Gamma \bsl \varepsilon \bsl k , \nonumber \\
M_2 &=&
2 \Gamma ((\varepsilon p) (k \dop p) -
(k p) (\varepsilon \dop p)) , \nonumber \\
M_3 &=&
\Gamma (\bsl \varepsilon (k p) -
\bsl k (\varepsilon p) ) , \nonumber \\
M_4 &=&
\Gamma (\bsl \varepsilon (k \dop p) -
\bsl k (\varepsilon \dop p) ) .
\label{apipho}
\eea
$\Gamma = \gamma_5$ for the production of mesons with negative parity,
$\Gamma = 1$ otherwise. This decomposition differs from the one
usually used in pion and eta photoproduction
\cite{bmz95,kdt95,cgln57}, to simplify the inclusion of kaon
photoproduction. The relation to the usual CGLN invariants $\widetilde
M_i$ is given by:
\be
\widetilde M_1 = M_1, \quad
\widetilde M_2 = - M_2, \quad
\widetilde M_3 = - (M_3 - M_4), \quad
\widetilde M_4 = - (M_3 + M_4) - 2 m_N M_1 .
\ee
In terms of Pauli spinors the scattering amplitude can be expressed
using the ${\cal F}_i$'s \cite{sau96}:
\bea
{\cal F}_p &=&
\rmi \mathbold {\sigma \cdot \hat q} \;
\mathbold {\sigma \cdot \varepsilon} {\cal F}_1 +
\mathbold {\sigma \cdot} (\mathbold {\hat k \times \varepsilon} )
{\cal F}_2 +
\rmi \mathbold {\sigma \cdot \hat q} \; \mathbold {\sigma \cdot \hat k} \;
\mathbold {\varepsilon \cdot \hat q} {\cal F}_3 +
\rmi \mathbold {\varepsilon \cdot \hat q} {\cal F}_4 , \nonumber \\
{\cal F}_m &=&
\rmi \mathbold {\sigma \cdot \varepsilon} {\cal F}_1 +
\mathbold {\sigma \cdot \hat q} \;
\mathbold {\sigma \cdot} (\mathbold {\hat k \times \varepsilon} )
{\cal F}_2 +
\rmi \mathbold {\sigma \cdot \hat k} \;
\mathbold {\varepsilon \cdot \hat q} {\cal F}_3 +
\rmi \mathbold {\sigma \cdot \hat q} \;
\mathbold {\varepsilon \cdot \hat q} {\cal F}_4 .
\label{calfpipho}
\eea
Here, ${\cal F}_{p,m}$ refer to the parity of the meson that is produced.
$\mathbold k$ and $\mathbold q$ denote the three-momentum of the
photon and the meson, respectively. $\mathbold {\hat x}$ is a
unit-vector in the direction of $\mathbold x$.

For our choice of amplitudes $M_i$, the relations between the CGLN
invariants ${\cal F}_i$ and the $A_j$'s of Eq.\ (\ref{mfipipho}) are
the following \cite{gd63}:
\bea
{\cal F}_1 &=&
\frac{\mathrm k}{4 \pi} \sqrt \frac{\dop E \pm \dop m}{2 \sqrt s}
\left [ A_1 - \frac{\sqrt s + m}{2} A_3 +
\frac{(k \dop p)}{\sqrt s - m} A_4 \right ] , \nonumber \\
{\cal F}_2 &=&
\frac{\mathrm k}{4 \pi} \sqrt \frac{\dop E \mp \dop m}{2 \sqrt s}
\left [ - A_1 - \frac{\sqrt s - m}{2} A_3 +
\frac{(k \dop p)}{\sqrt s + m} A_4 \right ] , \nonumber \\
{\cal F}_3 &=&
\frac{\mathrm k \mathrm q}{4 \pi}
\sqrt \frac{\dop E \pm \dop m}{2 \sqrt s}
\left [ - (\sqrt s - m) A_3 - A_4 \right ] , \nonumber \\
{\cal F}_4 &=&
\frac{\mathrm k \mathrm q}{4 \pi}
\sqrt \frac{\dop E \mp \dop m}{2 \sqrt s}
\left [ (\sqrt s + m) A_3 + A_4 \right ] ,
\eea
where the upper sign holds for mesons with negative parity. $\mathrm
k$ and $\mathrm q$ are the absolute values of the three-momenta
$\mathbold k$ and $\mathbold q$.

In order to unify the formulae for mesons of positive/negative parity,
we introduce a notation different to the usual one, by labeling the
multipoles using initial variables instead of final variables \cite{gd63}.
In Table\ \ref{multinot} we list the relation between both notations.
The ${\cal F}_i$ can now be expressed in terms of these new multipoles
${\cal M}_{l \pm}$ and ${\cal E}_{l \pm}$:
\be
\left (
\begin{array}{c}
{\cal F}_1 \\ {\cal F}_2 \\ {\cal F}_3 \\ {\cal F}_4 \\
\end{array}
\right )
= \sum\limits_l
\left (
\begin{array}{cccc}
 l \dop P_{l+1} & (l+1) \dop P_{l-1} &        \dop P_l &        \dop P_l \\
 (l+1) \dop P_l &         l \dop P_l &               0 &               0 \\
- \dopp P_{l+1} &      \dopp P_{l-1} &       \dopp P_l &       \dopp P_l \\
      \dopp P_l &        - \dopp P_l & - \dopp P_{l+1} & - \dopp P_{l+1} \\
\end{array}
\right )
\left (
\begin{array}{c}
{\cal M}_{l+} \\ {\cal M}_{l-} \\ {\cal E}_{l+} \\ {\cal E}_{l-} \\
\end{array}
\right ) .
\label{fviaem}
\ee
Eq.\ (\ref{fviaem}) can now be inverted to yield the electric and
magnetic multipoles. Using the orthogonality of the Legendre
polynomials $P(x)$ and recurrence relations to relate the derivatives
$\dop P$ and $\dopp P$ to the $P$'s, we obtain
\cite{bmz95,sau96,kdt95}:
\be
\left (
\begin{array}{c}
2 (l+1) {\cal M}_{l+} \\ 2 l {\cal M}_{l-} \\ 2 (l+1) {\cal E}_{l+} \\ 2 l{\cal E}_{l-} \\
\end{array}
\right )
= \intm1p1 dx
\left (
\begin{array}{cccc}
 P_l & - P_{l+1} &  \frac{1}{2l+1}\partial P & 0 \\
-P_l &   P_{l-1} & \frac{-1}{2l+1}\partial P & 0 \\
P_{l+1} & - P_l &  \frac{l+2}{2l+3}(P_{l+2} - P_l) & 
\frac{l+1}{2l+1}\partial P \\
P_{l-1} & - P_l &  \frac{1-l}{2l-1}(P_l - P_{l-2}) & 
\frac{-l}{2l+1}\partial P \\
\end{array}
\right )
\left (
\begin{array}{c}
{\cal F}_1 \\ {\cal F}_2 \\ {\cal F}_3 \\ {\cal F}_4 \\
\end{array}
\right ) ,
\label{emviaf}
\ee
with $\partial P = P_{l+1} - P_{l-1}$. The $T$-matrix amplitudes
${\cal T}_{l \pm}^{M,E}$ can now be obtained by multiplying the ${\cal
  M}_{l \pm}$ and ${\cal E}_{l \pm}$ with $\mp \sqrt {{\mathrm {k q}} \;
  l(l+1)}$ (cf. Table\ \ref{multinot}).

Finally, we want to give the connection of the CGLN amplitudes ${\cal
  F}_i$ to the helicity amplitudes $H_i$ \cite{bmz95}:
\be
\begin{array}{lclcl}
H_1 & = & 
- \rmi e^{-\rmi \phi} \langle -\einh | {\cal F} | 1, -\einh \rangle & = &
- \frac{1}{\sqrt 2} \sin \theta \cos \frac{\theta}{2} 
({\cal F}_3 + {\cal F}_4) \\
H_2 & = & 
- \rmi \langle -\einh | {\cal F} | 1, \einh \rangle & = &
\sqrt 2 \cos \frac{\theta}{2} \left [ 
({\cal F}_2 - {\cal F}_1) + \sin^2 \frac{\theta}{2} ({\cal F}_3 -
{\cal F}_4)
\right ] \\
H_3 & = & 
- \rmi e^{-2 \rmi \phi} \langle \einh | {\cal F} | 1, -\einh \rangle &
= &
\frac{1}{\sqrt 2} \sin \theta \sin \frac{\theta}{2} 
({\cal F}_3 - {\cal F}_4) \\
H_4 &=& 
- \rmi e^{- \rmi \phi} \langle \einh | {\cal F} | 1, \einh \rangle &
= &
\sqrt 2 \cos \frac{\theta}{2} \left [ 
({\cal F}_2 + {\cal F}_1) + \cos^2 \frac{\theta}{2} ({\cal F}_3 +
{\cal F}_4)
\right ] . \\
\end{array}
\label{hviafpipho}
\ee
The numbers in the brackets denote the helicities of the photon and the
initial/final baryon. In the c.m. system, the spins of the nucleon are
opposite to the helicities.

\subsection{Compton scattering}

In the case of Compton scattering, the decompositions of ${\cal F}$
and ${\cal M}_{fi}$ analogous to (\ref{calfhadr}) and
(\ref{calfpipho}) are rather lengthy and will not be given here. They
can be found, for example, in \cite{gpr78}. Its is more convenient to
start from a set of helicity amplitudes, analogous to
(\ref{hviafpipho}) \cite{prs74}:
\be
\begin{array}{lclclcclcl}
\Phi_1 & = & \Phi_{\einh \einh} & = & 
\frac{1}{8 \pi \sqrt s} \langle 1, \einh | T | 1, \einh \rangle ,
\qquad &
\Phi_4 & = & \Phi_{\einh \dreih} & = & 
\frac{1}{8 \pi \sqrt s} \langle 1, -\einh | T | 1, \einh \rangle ,\\
\Phi_2 & = & \Phi_{\einh -\einh} & = & 
\frac{1}{8 \pi \sqrt s} \langle -1, -\einh | T | 1, \einh \rangle ,
\qquad &
\Phi_5 & = & \Phi_{\dreih \dreih} & = & 
\frac{1}{8 \pi \sqrt s} \langle 1, -\einh | T | 1, -\einh \rangle ,\\
\Phi_3 & = & \Phi_{\einh -\dreih} & = & 
\frac{1}{8 \pi \sqrt s} \langle -1, \einh | T | 1, \einh \rangle ,
\qquad &
\Phi_6 & = & \Phi_{\dreih -\dreih} & = & 
\frac{1}{8 \pi \sqrt s} \langle -1, \einh | T | 1, -\einh \rangle .\\
\end{array}
\label{hviatcomp}
\ee
Again, the numbers in the brackets denote the helicities of the
initial and final photon and baryon. Their expansion in terms of the
amplitudes $f_{fi}^{l\pm}$ is given by
\be
\begin{array}{lcll}
\Phi {\scriptstyle {1 \atop 2}} &=& \frac{1}{2} \sum\limits_l 
(l+1) & [ (l+2)^2 (f_{EE}^{(l+1)-} \pm f_{MM}^{(l+1)-}) \\
 & & & 
\pm l^2 (f_{EE}^{l+} \pm f_{MM}^{l+}) 
\mp 2l(l+2) (f_{EM}^{l+} \pm f_{ME}^{l+}) ] 
d_{\einh, \pm \einh}^{\; l+\einh} , \\
\Phi {\scriptstyle {3 \atop 4}} &=& \frac{1}{2} \sum\limits_l 
(l+1) \sqrt{l(l+2)} & [ (l+2) (f_{EE}^{(l+1)-} \mp f_{MM}^{(l+1)-}) \\
 & & & 
\pm l (f_{EE}^{l+} \mp f_{MM}^{l+}) 
\mp 2 (f_{EM}^{l+} \mp f_{ME}^{l+}) ] 
d_{\einh, \mp \dreih}^{\; l+\einh} , \\
\Phi {\scriptstyle {5 \atop 6}} &=& \frac{1}{2} \sum\limits_l
(l+1) l(l+2) & [ (f_{EE}^{(l+1)-} \pm f_{MM}^{(l+1)-}) \\
 & & & 
\pm (f_{EE}^{l+} \pm f_{MM}^{l+}) 
\pm 2 (f_{EM}^{l+} \pm f_{ME}^{l+}) ] 
d_{\dreih, \pm \dreih}^{\; l+\einh} , \\
\end{array}
\label{hviafcomp}
\ee
with $f_{EE}^{0+} = f_{MM}^{0+} = f_{EM}^{0+} = f_{ME}^{0+}
\stackrel{\mathrm def.}{=} 0$. In an abbreviated notation, we can
write (\ref{hviafcomp}) as:
\be
\Phi_{s \dop s} = \sum\limits_J (2J+1) 
\Phi_{s \dop s}^J d_{s \dop s}^{\; J}, 
\qquad \mbox{with} \qquad
\Phi_{s \dop s}^J = \frac{1}{2} \intm1p1 dx \;
\Phi_{s \dop s} (x) d_{s \dop s}^{\; J}(x) .
\ee
The inversion of (\ref{hviafcomp}) can now be shown to be \cite{prs74}:
\be
\begin{array}{lcl}
f_{{EE \atop MM}}^{l+} &=& \frac{1}{(l+1)^2}
\left [ \frac{1}{2} (\Phi_1^{l+\einh} \mp \Phi_1^{l+\einh}) 
\pm \frac{l+2}{\sqrt {l(l+2)}} (\Phi_3^{l+\einh} \mp \Phi_4^{l+\einh})
+ \frac{l+2}{2l} (\Phi_5^{l+\einh} \mp \Phi_6^{l+\einh}) \right ] ,\\
f_{{EE \atop MM}}^{(l+1)-} &=& \frac{1}{(l+1)^2}
\left [ \frac{1}{2} (\Phi_1^{l+\einh} \pm \Phi_1^{l+\einh}) 
\pm \frac{l}{\sqrt {l(l+2)}} (\Phi_3^{l+\einh} \pm \Phi_4^{l+\einh})
+ \frac{l}{2(l+2)} (\Phi_5^{l+\einh} \pm \Phi_6^{l+\einh}) \right ] ,\\
f_{{EM \atop ME}}^{l+} &=& \frac{1}{(l+1)^2}
\left [ -\frac{1}{2} (\Phi_1^{l+\einh} \mp \Phi_1^{l+\einh}) 
\mp \frac{1}{\sqrt {l(l+2)}} (\Phi_3^{l+\einh} \mp \Phi_4^{l+\einh})
+ \frac{l}{2} (\Phi_5^{l+\einh} \mp \Phi_6^{l+\einh}) \right ] , \\
f_{{ME \atop EM}}^{(l+1)-} &=& f_{{EM \atop ME}}^{l+} , \\
\end{array}
\label{fviahcomp}
\ee
where the last line follows from time-reversal symmetry. The
$T$-matrix amplitudes $T_{l \pm}$ are now given by:
\be
T^{{EE \atop MM}}_{l \pm} = 
{\mathrm k} \; l (l+1) \; f_{{EE \atop MM}}^{l \pm} , \qquad
T^{{EM \atop ME}}_{l +} =
{\mathrm k} \; \sqrt {l (l+2)} (l+1) \; f_{{EM \atop ME}}^{l +} .
\ee
%

%
%

\section{Isospin decomposition}
\label{decompose}

\subsection{Meson nucleon scattering}

For the $I\!=\!1$ mesons $\pi$ and $\zeta$ we use the usual projection
operators\cite{ew88} with the matrix elements (a, b = $\pi, \zeta$)
\bea
\langle b_j | P_{1/2} | a_i \rangle &=&
\frac{1}{3} \tau_j \tau_i \nonumber \\
\langle b_j | P_{3/2} | a_i \rangle &=&
\delta_{ji} - \frac{1}{3}\tau_j \tau_i
\eea
in a cartesian basis. With the help of this, all possible reactions can
be written as:
\be
\langle b_j | T_{\Iba} | a_i \rangle = \frac{1}{3} \tau_j
\tau_i T_{\Iba}^{1/2} +
(\delta_{ji} - \frac{1}{3} \tau_j \tau_i) T_{\Iba}^{3/2} .
\label{decompi1}
\ee

For the pure $I\!=\!\einh$ reactions involving $\pi$ and $\zeta$ the
projector is usually taken to be $P_{1/2} = \mathbold \tau$
\cite{ew88}. This choice has the disadvantage that it does not agree
with the Clebsch-Gordan coefficients for the different reactions
channels. Therefore we choose instead ($a = \pi, \zeta$, $b = \eta, k$):
\be
\langle b | T_{\Iba} | a_i \rangle =
-\frac{1}{\sqrt 3} \tau_i T^{1/2}_{\Iba}.
\label{decompi0}
\ee
This has no influence on the calculated quantities, since in the end,
we convert our amplitudes to the normal convention.

\subsection{Photoproduction}

In our calculation we use a decomposition slightly different from the
one normally used in pion photoproduction (\ref{isogw}). For the same
reason as in the hadronic case, we always use $-\frac{1}{\sqrt 3}
\mathbold \tau$ instead of $\mathbold \tau$ as projection operator for
transition were either the photon or the meson have isospin $I\!=\!0$.
For the $I\!=\!1$ mesons this leads to:
\be
\langle b_j | T_{\Ibgam} | \gamma \rangle =
-\frac{1}{\sqrt 3} \tau_j T_{\Ibgam}^{0} +
\frac{1}{3} \tau_j \tau_3 T_{\Ibgam}^{1/2} +
(\delta_{j3} - \frac{1}{3} \tau_j \tau_3) T_{\Ibgam}^{3/2} ,
\label{isopipho1}
\ee
while for the other mesons we have:
\be
\langle b | T_{\Ibgam} | \gamma \rangle =
T_{\Ibgam}^{0} -
\frac{1}{\sqrt 3} \tau_3 T_{\Ibgam}^{1/2} .
\label{isopipho0}
\ee
This choice guarantees that there are no additional factors to be
taken into account in the calculation of the hadronic rescattering.

\subsection{Compton scattering}

For Compton scattering no isospin decomposition is performed. As has
been discussed in Sec.\ \ref{photonic}, in this case the rescattering
takes place through the physical intermediate states ($\pi^0 p$,
$\pi^0 n$, etc.). Therefore, the Compton amplitudes are calculated
for both channels $\gamma p$ and $\gamma n$ directly.

%
%

\section{Observables}
\label{observables}

In this section we want to summarize the relations of the various
amplitudes from App.\ \ref{multipoles} to the observables. Again, more
detailed accounts can be found in various places
\cite{fm98,kdt95,prs74}. In the following, the notation $\widetilde O$
should indicate that the amplitude $O$ for a specific reaction channel
has been constructed from the isospin-decomposed amplitudes:
\be 
\widetilde O = \sum\limits_I p^I O^I .
\ee
The factors $p^I$ are Clebsch-Gordan coefficients and can be
determined from isospin decompositions listed in App.\ \ref{decompose}.

\subsection{Meson nucleon scattering}

The observables consist of the total cross sections $\sigma$, the
differential cross sections $\frac{d \sigma}{d \Omega}$ and the
final-state polarizations ${\cal P}$ \cite{gw64}:
\bea
\sigma &=& \frac{4 \pi}{{\mathrm q}^2} \sum\limits_l
\left [ 
(l+1) {\abs {\widetilde T_{l+}}}^2 + l {\abs {\widetilde T_{l-}}}^2
\right ] , \nonumber \\
f &=& \frac{1}{\mathrm q} \sum\limits_l
\left [ 
(l+1) \widetilde T_{l+} + l \widetilde T_{l-} 
\right ] P_l \nonumber \\
g &=& \frac{1}{\mathrm q} \sin \theta \sum\limits_l
\left [ 
\widetilde T_{l+} - \widetilde T_{l-} \right ] 
\dop P_l \nonumber \\
\frac{d \sigma}{d \Omega} &=& {\abs f}^2 + {\abs g}^2 , \quad \frac{d
  \sigma}{d \Omega} {\cal P} = - 2 \imag (f^* g) . 
\eea

\subsection{Photoproduction}

In terms of the helicity amplitudes $H_i$ of Eq.\ (\ref{hviafpipho}),
the differential cross sections and the three single-polarization
observables can be written as \cite{bmz95}:
\be
\renewcommand{\arraystretch}{2}
\begin{array}{lccll}
\displaystyle \frac{d \sigma}{d \Omega} &=& &
\displaystyle \frac{\mathrm q}{2 \mathrm k} 
\sum\limits_1^4 {\abs {\widetilde H_i}}^2 , \nonumber \\
\displaystyle \frac{d \sigma}{d \Omega} \Sigma &=& &
\displaystyle \frac{\mathrm q}{\mathrm k} 
\real 
( \widetilde H_1 \widetilde H_4^* - \widetilde H_2 \widetilde H_3^*)
 & \qquad \mbox{Photon asymmetry} , \nonumber \\
\displaystyle \frac{d \sigma}{d \Omega} {\cal P} &=& - & 
\displaystyle \frac{\mathrm q}{\mathrm k} 
\imag 
( \widetilde H_1 \widetilde H_3^* + \widetilde H_2 \widetilde H_4^*)
 & \qquad \mbox{Recoil polarization} , \nonumber \\
\displaystyle \frac{d \sigma}{d \Omega} {\cal T} &=& &
\displaystyle \frac{\mathrm q}{\mathrm k} 
\imag 
( \widetilde H_1 \widetilde H_2^* + \widetilde H_3 \widetilde H_4^*)
 & \qquad \mbox{Target asymmetry} . 
\end{array}
\renewcommand{\arraystretch}{1.0}
\ee
In complete analogy to the hadronic case, the total cross sections are
given in terms of the partial waves by:
\be
\sigma = \frac{2 \pi}{{\mathrm q}^2}
\sum\limits_l 
\left [ 
(l+1)
({\abs {\widetilde {\cal T}^M_{l+}}}^2 + 
{\abs {\widetilde {\cal T}^E_{l+}}}^2) +
l
({\abs {\widetilde {\cal T}^E_{l-}}}^2 + 
{\abs {\widetilde {\cal T}^M_{l-}}}^2)
\right ] .
\ee
From wich the reduced cross section can be calculated via:
\be
\sigma_{red} = \sqrt{\frac{\sigma {\mathrm k}}{4 \pi {\mathrm q}}} .
\ee

\subsection{Compton scattering}

Again, as in the case of photproduction, the observables are best 
expressed in terms of the helicity amplitudes (\ref{hviafcomp})
\cite{prs74}:
\be
\renewcommand{\arraystretch}{2}
\begin{array}{lccl}
\displaystyle \frac{d \sigma}{d \Omega} &=& &
\displaystyle \frac{1}{2} 
\left [
{\abs {\widetilde \Phi_1}}^2 + {\abs {\widetilde \Phi_2}}^2 +
2 ({\abs {\widetilde \Phi_3}}^2 + {\abs {\widetilde \Phi_4}}^2 ) +
{\abs {\widetilde \Phi_5}}^2 + {\abs {\widetilde \Phi_6}}^2
\right ] , \nonumber \\
\displaystyle \frac{d \sigma}{d \Omega} \Sigma &=& - &
\real \left (
(\widetilde \Phi_1 + \widetilde \Phi_5) \widetilde \Phi_3^*
+ (\widetilde \Phi_2 - \widetilde \Phi_6) \widetilde \Phi_4^*
\right )
, \nonumber \\
\displaystyle \frac{d \sigma}{d \Omega} {\cal P} &=& - & 
\imag \left (
(\widetilde \Phi_1 + \widetilde \Phi_5) \widetilde \Phi_4^*
- (\widetilde \Phi_2 - \widetilde \Phi_6) \widetilde \Phi_3^*
\right ) 
\qquad = \displaystyle \frac{d \sigma}{d \Omega} {\cal T}
.
\end{array}
\renewcommand{\arraystretch}{1.0}
\ee
Finally, the total Compton cross section is given by:
\be
\sigma = \frac{2 \pi}{{\mathrm k}^2}
\sum\limits_l 
\left [ 
(l + 1)
({\abs {\widetilde T^{MM}_{l+}}}^2 + 
{\abs {\widetilde T^{EE}_{l+}}}^2 +
2({\abs {\widetilde T^{ME}_{l+}}}^2 + 
  {\abs {\widetilde T^{EM}_{l+}}}^2)) \nonumber \\
+ l
({\abs {\widetilde T^{EM}_{l-}}}^2 + 
{\abs {\widetilde T^{ME}_{l-}}}^2)
\right ] .
\ee
%

%
%

\newpage

%
%



\pagestyle{empty}

\setcounter{table}{0} \setcounter{figure}{0}

\begin{table}[!ht]
\begin{center}
\renewcommand{\arraystretch}{1.2}
\begin{tabular}{c|c|c|c|c|c|c|c}
 & Partial- &  & & \multicolumn{4}{c}{$l_{\varphi}$, CGLN-Notation} \\
\raisebox{1.5ex}[-1.5ex]{Multipol}  &
welle &
\raisebox{1.5ex}[-1.5ex]{$J$} &
\raisebox{1.5ex}[-1.5ex]{$P$} &
\multicolumn{2}{c|}{$P_{\varphi} = -$} &
\multicolumn{2}{c}{$P_{\varphi} = +$} \\
\hline\hline
${\cal M}_{l+}$ & $-\alpha {\cal T}_{l+}^M$ &
$l+\einh$ & $(-)^{l+1}$ &   $l$ &     $M_{l+}$ & $l+1$ & $M_{(l+1)-}$ \\
${\cal M}_{l-}$ & $-\alpha {\cal T}_{l-}^M$ &
$l-\einh$ & $(-)^{l+1}$ &   $l$ &     $M_{l-}$ & $l-1$ & $M_{(l-1)+}$ \\
${\cal E}_{l+}$ & $\alpha {\cal T}_{l+}^E$ &
$l+\einh$ &     $(-)^l$ & $l+1$ & $E_{(l+1)-}$ &   $l$ & $E_{l+}$ \\
${\cal E}_{l-}$ & $-\alpha {\cal T}_{l-}^E$ &
$l-\einh$ &     $(-)^l$ & $l-1$ & $E_{(l-1)+}$ &   $l$ & $E_{l-}$ \\
\end{tabular}
\renewcommand{\arraystretch}{1.0}
\end{center}
\caption{Connection of the multipoles ${\cal M}_{l \pm}$ and ${\cal
    E}_{l \pm}$ to the partial waves and usual CGLN multipoles. $l$
  denotes the photon, $J$ the total and $l_{\varphi}$ the meson angular
  momentum; $\alpha = \frac{1}{\sqrt {\mathrm k \mathrm {q} \; l(l+1)}}$.}
\label{multinot}
\end{table}

\begin{table}[!ht]
\begin{center}
\begin{tabular}{l|c|c|c|c|c|c|c|c|c}
        & 
$\chi^2/{\rm DF}$ & 
$\chi^2_{\gamma \gamma}/{\rm DF}$ & 
$\chi^2_{\gamma \pi}/{\rm DF}$ & 
$\chi^2_{\gamma \eta}/{\rm DF}$ & 
$\chi^2_{\gamma K}/{\rm DF}$ & 
$\chi^2_{\pi \pi}/{\rm DF}$ & 
$\chi^2_{\pi \zeta}/{\rm DF}$ & 
$\chi^2_{\pi \eta}/{\rm DF}$ & 
$\chi^2_{\pi K}/{\rm DF}$ \\
\hline\hline
SM95-pt-1 &  9.61 & 7.15 & 13.08 &  6.09 & 5.17 & 3.13 & 5.86 & 1.73 & 3.28 \\
SM95-pt-2 &  7.76 & 5.20 &  9.62 &  3.00 & 3.91 & 5.78 & 9.43 & 1.95 & 3.77 \\
SM95-pt-3 &  5.58 & 3.40 &  6.69 &  2.78 & 4.09 & 3.88 & 8.10 & 1.86 & 4.21 \\
\end{tabular} 
\end{center}
\caption{$\chi^2$ per datapoint values for the different fits. First
  line: Ohta's method with fixed hadronic parameters, second: Ohta's
  method with a refit of the hadronic parameters, third: Haberzettl's
  method, all parameters refitted. Also the $\chi^2/{\rm DF}$-values
  for the different reaction channels are given separately.}
\label{chi2comp}
\end{table}

\begin{table}[ht]
\begin{center}
\begin{tabular}{c|r||c|r|c|r||c|r}
$g$                 & Value & $g$                 & Value & $\kappa$                 & Value & $g$                            & Value \\
\hline\hline                                       
$g_{\pi NN}$        & 13.05 & $g_{\rho NN}$       &  2.35 & $\kappa_{\rho NN}$       &  2.26 & $g_{\gamma \eta \rho^0}$       &  1.12 \\
                    & 13.05 &                     &  1.94 &                          &  3.44 &                                &  1.08 \\
                    & 13.09 &                     &  2.07 &                          &  3.01 &                                &  1.05 \\
\hline                                             
$g_{\eta NN}$       &  1.13 & $g_{a_0 NN}$        &  0.18 & -- & -- & $g_{\gamma \eta \omega}$       &  0.20 \\
                    &  1.01 &                     &  0.52 & -- & -- &                                &  0.23 \\
                    &  0.36 &                     &  0.41 & -- & -- &                                &  0.36 \\
\hline                                             
$g_{K N \Lambda}$   & -6.12 & $g_{K^* N \Lambda}$ & -6.52 & $\kappa_{K^* N \Lambda}$ & -0.43 & -- & -- \\
                    & -6.25 &                     & -5.82 &                          & -0.42 & -- & -- \\
                    & -8.65 &                     & -5.99 &                          & -0.45 & -- & -- \\
\end{tabular} 
\end{center}
\caption{Couplings of the mesons to the nucleon and $g_{\gamma \eta
    \rho^0, \omega}$ as obtained in the fits. First line: SM95-pt-1,
  second: SM95-pt-2, third: SM95-pt-3.}
\label{mesparmSM95}
\end{table}

\begin{table}[ht]
\begin{center}
\begin{tabular}{c|r|r|r|r|r|r|r|r|r|r}
                 & $M$ & $\Gamma_{tot}$ &
\multicolumn{2}{c|}{$\Gamma_{\pi N}$}  & \multicolumn{2}{c|}{$\Gamma_{\zeta N}$} & \multicolumn{2}{c|}{$\Gamma_{\eta N}$} & \multicolumn{2}{c}{$\Gamma_{K \Lambda}$} \\
$L_{2I,2S}$       &[GeV]  & [MeV]&  [MeV] & \% &  [MeV] & \% &  [MeV] & \% &  [MeV] & \% \\
\hline\hline
$\nres S11{1535}$ & 1.543 &  151 &  56(+) &  37 &   5(+) &   3 &  90(+) &  60 &     -- &  -- \\
                  & 1.553 &  213 &  67(+) &  31 &   6(+) &   3 & 140(+) &  66 &     -- &  -- \\
                  & 1.549 &  215 &  67(+) &  31 &  13(+) &   6 & 135(+) &  63 &     -- &  -- \\[0.5ex]
$\nres S11{1650}$ & 1.692 &  209 & 155(+) &  74 &  41(+) &  20 &   0(-) &   0 &  13(+) &   6 \\
                  & 1.689 &  216 & 167(+) &  77 &  37(+) &  17 &   0(-) &   0 &  12(+) &   6 \\
                  & 1.684 &  194 & 141(+) &  73 &  43(+) &  22 &   1(-) &   1 &   9(+) &   5 \\
\hline
$\nres P11{1440}$ & 1.448 &  334 & 202(+) &  60 & 132(+) &  40 &  0.95$^a$ &   0 &     -- &  -- \\
                  & 1.438 &  300 & 178(+) &  59 & 122(+) &  41 & -1.00$^a$ &   0 &     -- &  -- \\
                  & 1.479 &  513 & 316(+) &  62 & 197(+) &  38 &  2.79$^a$ &   0 &     -- &  -- \\[0.5ex]
$\nres P11{1710}$ & 1.727 &  266 &   1(+) &   0 & 138(-) &  52 &  89(+) &  33 &  38(+) &  14 \\
                  & 1.724 &  272 &   0(+) &   0 & 134(-) &  49 & 115(+) &  42 &  23(+) &   8 \\
                  & 1.709 &  284 &   0(+) &   0 & 145(-) &  51 &  90(+) &  32 &  49(+) &  17 \\
\hline
$\nres P13{1720}$ & 1.771 &  344 &  74(+) &  22 & 241(+) &  70 &  24(+) &   7 &   5(+) &   1 \\
                  & 1.776 &  396 &  89(+) &  22 & 270(+) &  68 &  30(+) &   8 &   7(+) &   2 \\
                  & 1.801 &  637 & 135(+) &  21 & 475(+) &  75 &  23(+) &   4 &   4(+) &   1 \\
\hline
$\nres D13{1520}$ & 1.510 &  101 &  58(+) &  57 &  43(-) &  43 &  10$^b$(+) &   0 &     -- &  -- \\
                  & 1.512 &  110 &  58(+) &  53 &  52(-) &  47 &  49$^b$(+) &   0 &     -- &  -- \\
                  & 1.512 &   93 &  52(+) &  56 &  41(-) &  44 &  43$^b$(+) &   0 &     -- &  -- \\[0.5ex]
$\nres D13{1700}$ & 1.901 &  359 &  35(+) &  10 & 300(+) &  84 &  24(-) &   7 &   0(+) &   0 \\
                  & 1.910 &  222 &  34(+) &  15 & 175(+) &  79 &  13(-) &   6 &   0(+) &   0 \\
                  & 1.940 &  412 &  43(+) &  10 & 309(+) &  75 &  59(-) &  14 &   1(+) &   0 \\
\hline\hline
$\nres S31{1620}$ & 1.598 &  150 &  44(+) &  29 & 106(-) &  71 &     -- &  -- &     -- &  -- \\
                  & 1.604 &  173 &  57(+) &  33 & 116(-) &  67 &     -- &  -- &     -- &  -- \\
                  & 1.579 &  153 &  32(+) &  21 & 121(-) &  79 &     -- &  -- &     -- &  -- \\
\hline
$\nres P33{1232}$ & 1.230 &  110 & 110(+) & 100 &     -- &  -- &     -- &  -- &     -- &  -- \\
                  & 1.229 &  110 & 110(+) & 100 &     -- &  -- &     -- &  -- &     -- &  -- \\
                  & 1.228 &  110 & 110(+) & 100 &     -- &  -- &     -- &  -- &     -- &  -- \\[0.5ex]
$\nres P33{1600}$ & 1.686 &  405 &  59(+) &  15 & 346(+) &  85 &     -- &  -- &     -- &  -- \\
                  & 1.743 &  590 &  96(+) &  16 & 494(+) &  84 &     -- &  -- &     -- &  -- \\
                  & 1.721 &  485 &  74(+) &  15 & 411(+) &  85 &     -- &  -- &     -- &  -- \\
\hline
$\nres D33{1700}$ & 1.675 &  547 &  70(+) &  13 & 477(+) &  87 &     -- &  -- &     -- &  -- \\
                  & 1.668 &  408 &  71(+) &  17 & 337(+) &  83 &     -- &  -- &     -- &  -- \\
                  & 1.677 &  387 &  55(+) &  14 & 332(+) &  86 &     -- &  -- &     -- &  -- \\
\end{tabular} 
\end{center}
\caption{Extracted resonance parameters. Notation as in Table\
  \protect\ref{mesparmSM95}. $^a$: the coupling $g_{\eta
    N R}$ is given instead of the partial width, $^b$: width in keV.
  The signs of the couplings are given in brackets.}
\label{rescouplSM95}
\end{table}

\begin{table}[ht]
\begin{center}
\begin{tabular}{c|r|r|r|r}
                  & $z_{\pi N}$ & $z_{\zeta N}$ & $z_{\eta N}$ & $z_{K \Lambda}$ \\
\hline\hline
$\nres P13{1720}$ &  -2.200 &  -0.210 &  -1.993 &  -0.421 \\
                  &  -2.134 &  -0.218 &  -1.938 &  -0.448 \\
                  &  -0.242 &   0.226 &  -2.453 &  -0.553 \\
\hline
$\nres D13{1520}$ &   0.352 &  -0.171 &   0.823 &      -- \\
                  &   0.311 &  -0.317 &   0.574 &      -- \\
                  &   0.319 &  -0.658 &   0.646 &      -- \\[0.5ex]
$\nres D13{1700}$ &  -1.281 &  -0.990 &   0.195 &  -2.240 \\
                  &  -1.434 &  -0.777 &   0.582 &   1.383 \\
                  &   0.424 &   0.887 &   0.516 &   0.616 \\
\hline\hline
$\nres P33{1232}$ &  -0.306 &      -- &      -- &      -- \\
                  &  -0.339 &      -- &      -- &      -- \\
                  &  -0.352 &      -- &      -- &      -- \\[0.5ex]
$\nres P33{1600}$ &   1.587 &   0.094 &      -- &      -- \\
                  &   1.532 &   0.086 &      -- &      -- \\
                  &  -0.100 &  -0.753 &      -- &      -- \\
\hline
$\nres D33{1700}$ &   0.628 &  -0.212 &      -- &      -- \\
                  &   0.606 &  -0.222 &      -- &      -- \\
                  &  -0.681 &   0.367 &      -- &      -- \\
\end{tabular} 
\end{center}
\caption{Fitted hadronic $z$-parameters of the spin-$\dreih$ resonances.
  Notation as in Table\ \protect\ref{mesparmSM95}.}
\label{reszparmSM95}
\end{table}

\begin{table}[!ht]
\begin{center}
\begin{tabular}{l|rr|rr|r|r}
$L_{2I,2S}$ & \multicolumn{2}{c|}{$A_{1/2}$} &
\multicolumn{2}{c|}{$A_{3/2}$} & $z_1$ & $z_2$ \\
\hline\hline
$\nres S11{1535}$ &   70 $\pm$ 12 &  -46 $\pm$ 27 & \multicolumn{2}{c|}{--} & -- & -- \\
                  &   60 $\pm$ 15 &  -20 $\pm$ 35 & \multicolumn{2}{c|}{--} & -- & -- \\
                  &   93 &  -43 & \multicolumn{2}{c|}{--} & -- & -- \\
                  &  101 &  -60 & \multicolumn{2}{c|}{--} & -- & -- \\
                  &  106 &  -63 & \multicolumn{2}{c|}{--} & -- & -- \\[0.5ex]
$\nres S11{1650}$ &   53 $\pm$ 16 &  -15 $\pm$ 21 & \multicolumn{2}{c|}{--} & -- & -- \\
                  &   69 $\pm$  5 &  -15 $\pm$  5 & \multicolumn{2}{c|}{--} & -- & -- \\
                  &   31 &  -28 & \multicolumn{2}{c|}{--} & -- & -- \\
                  &   33 &  -23 & \multicolumn{2}{c|}{--} & -- & -- \\
                  &   45 &  -26 & \multicolumn{2}{c|}{--} & -- & -- \\
\hline
$\nres P11{1440}$ &  -65 $\pm$  4 &   40 $\pm$ 10 & \multicolumn{2}{c|}{--} & -- & -- \\
                  &  -63 $\pm$  5 &   45 $\pm$ 15 & \multicolumn{2}{c|}{--} & -- & -- \\
                  &  -73 &   51 & \multicolumn{2}{c|}{--} & -- & -- \\
                  &  -66 &   55 & \multicolumn{2}{c|}{--} & -- & -- \\
                  &  -84 &   47 & \multicolumn{2}{c|}{--} & -- & -- \\[0.5ex]
$\nres P11{1710}$ &    9 $\pm$ 22 &   -2 $\pm$ 14 & \multicolumn{2}{c|}{--} & -- & -- \\
                  &    7 $\pm$ 15 &   -2 $\pm$ 15 & \multicolumn{2}{c|}{--} & -- & -- \\
                  &    8 &   -4 & \multicolumn{2}{c|}{--} & -- & -- \\
                  &    4 &    4 & \multicolumn{2}{c|}{--} & -- & -- \\
                  &   19 &  -19 & \multicolumn{2}{c|}{--} & -- & -- \\
\hline
$\nres P13{1720}$ &   18 $\pm$ 30 &    1 $\pm$ 15 &  -19 $\pm$ 20 &  -29 $\pm$ 61 & -- & -- \\
                  &  -15 $\pm$ 15 &    7 $\pm$ 15 &    7 $\pm$ 10 &   -5 $\pm$ 25 & -- & -- \\
                  &   36 &   20 &   23 &   32 &   0.028 &  -2.840 \\
                  &   30 &   23 &   51 &   28 &  -0.282 &  -2.760 \\
                  &   23 &    2 &   75 &  -17 &  -0.852 &   1.086 \\
\hline
$\nres D13{1520}$ &  -24 $\pm$  9 &  -59 $\pm$  9 &  166 $\pm$  5 & -139 $\pm$ 11 & -- & -- \\
                  &  -20 $\pm$  7 &  -48 $\pm$  8 &  167 $\pm$  5 & -140 $\pm$ 10 & -- & -- \\
                  &   -6 &  -46 &  140 & -150 &  -0.323 &  -1.361 \\
                  &   -9 &  -47 &  152 & -157 &  -0.256 &  -1.244 \\
                  &    3 &  -47 &  136 &  -98 &  -0.265 &  -0.475 \\[0.5ex]
$\nres D13{1700}$ &  -18 $\pm$ 13 &    0 $\pm$ 50 &   -2 $\pm$ 24 &   -3 $\pm$ 44 & -- & -- \\
                  &  -16 $\pm$ 14 &    6 $\pm$ 24 &   -9 $\pm$ 12 &  -33 $\pm$ 17 & -- & -- \\
                  &   20 &   -6 &  -20 &   24 &  -1.734 &   1.372 \\
                  &  -34 &   34 &  -11 &   18 &   2.015 &  -0.614 \\
                  &    5 &   47 &   41 &  -55 &  -1.171 &  -2.322 \\
\end{tabular}
\end{center}
\caption{Extracted helicity couplings (in units of
  $10^{-3}$ GeV$^{-1/2}$) and $z$-parameters for the $I\!=\!\einh$
  resonances. The first value denotes the $A^p$ and the second
  $A^n$. First line: PDG-Values \protect\cite{pdg96}, second: Arndt et
  al.\ \protect\cite{sm95}, third: SM95-pt-1, fourth: SM95-pt-2, fifth:
  SM95-pt-3.}
\label{resheli12SM95}
\end{table}

\begin{table}[!ht]
\begin{center}
\begin{tabular}{l|r|r|r|r}
$L_{2I,2S}$ & $A_{1/2}$ & $A_{3/2}$ & $z_1$ & $z_2$ \\
\hline\hline
$\nres S31{1620}$ &   27 $\pm$ 11 & -- & -- & -- \\
                  &   35 $\pm$ 20 & -- & -- & -- \\
                  &    0 & -- & -- & -- \\
                  &    0 & -- & -- & -- \\
                  &   -4 & -- & -- & -- \\
\hline              
$\nres P33{1232}$ & -140 $\pm$  5 & -258 $\pm$  6 & -- & -- \\
                  & -141 $\pm$  5 & -261 $\pm$  5 & -- & -- \\
                  & -132 & -253 &  -0.534 &   1.372 \\
                  & -129 & -244 &  -0.512 &   1.351 \\
                  & -126 & -233 &  -0.267 &  -0.658 \\[0.5ex]
$\nres P33{1600}$ &  -23 $\pm$ 20 &   -9 $\pm$ 21 & -- & -- \\
                  &  -18 $\pm$ 15 &  -25 $\pm$ 15 & -- & -- \\
                  &  -12 &  -35 &   0.456 &  -2.345 \\
                  &  -14 &  -44 &  -0.202 &  -4.493 \\
                  &  -26 &  -52 &   2.782 &  -4.479 \\
\hline              
$\nres D33{1700}$ &  104 $\pm$ 15 &   85 $\pm$ 22 & -- & -- \\
                  &   90 $\pm$ 25 &   97 $\pm$ 20 & -- & -- \\
                  &  102 &  172 &  -0.630 &   0.532 \\
                  &   83 &  139 &  -2.446 &   0.664 \\
                  &   75 &   98 &   0.462 &  -0.862 \\
\end{tabular}
\end{center}
\caption{Extracted helicity couplings (in units of
  $10^{-3}$ GeV$^{-1/2}$) and $z$-parameters for the $I\!=\!\dreih$
  resonances. Notation as in Table\ \protect\ref{resheli12SM95}.}
\label{resheli32SM95}
\end{table}

\begin{table}[ht]
\begin{center}
\renewcommand{\arraystretch}{1.2}
\begin{tabular}{c|r|c|r|c|r|c|r|c|r|c|r}
                  & Value &                   & Value &
                  & Value &                   & Value &
                  & Value &                   & Value \\
                  & [GeV] &                   & [GeV] &
                  & [GeV] &                   & [GeV] &
                  & [GeV] &                   & [GeV] \\
\hline\hline
     $\Lambda_N$  &  1.23 & $\Lambda_{1/2}^h$ &  1.24 &
$\Lambda_{1/2}^e$ &  0.92 & $\Lambda_{3/2}^h$ &  1.06 & 
$\Lambda_{3/2}^e$ &  3.98 &       $\Lambda_t$ &  0.70 \\
                  &  1.19 &                   &  1.24 &
                  &  0.93 &                   &  1.10 &
                  &  3.00 &                   &  0.72 \\
                  &  1.24 &                   &  1.72 &
                  &  1.06 &                   &  1.13 &
                  &  3.59 &                   &  0.70 \\
\end{tabular} 
\renewcommand{\arraystretch}{1.0}
\end{center}
\caption{Values of the fitted cutoff-parameters $\Lambda$. Notation as
  in Table\ \protect\ref{mesparmSM95}. For the nucleon resonances the
  indices $h$ and $e$ name the Cutoffs at the hadronic and
  electromagnetic vertices.}
\label{cutoffsSM95}
\end{table}

\begin{table}[ht]
\begin{center}
\begin{tabular}{l|c|c|c|c|c|c}
 & 
$\chi^2/{\rm DF}$ & 
$\chi^2_{\gamma K}/{\rm DF}$ & 
$\chi^2_{\pi K}/{\rm DF}$ & 
$g_{KN \Lambda}$ & 
$g_{K^*N \Lambda}$ & 
$\kappa_{K^*N \Lambda}$ \\
\hline\hline
PV & 7.76 & 3.91 & 3.77 & -6.25 & -5.82 & -0.42 \\
PS & 7.74 & 3.70 & 3.41 & -6.82 & -6.20 & -0.43 \\
\end{tabular} 
\end{center}
\caption{Comparison of SM95-pt-2 and a fit using PS-coupling at the $K
  N \Lambda$-vertex. Shown are only the various $\chi^2$-values and
  the coupling constants $g_{KN \Lambda}$, $g_{K^*N \Lambda}$ and
  $\kappa_{K^*N \Lambda}$.}
\label{PSvsPV}
\end{table}

\clearpage


\begin{figure}[ht]
  \centerline{ \includegraphics[width=14cm]{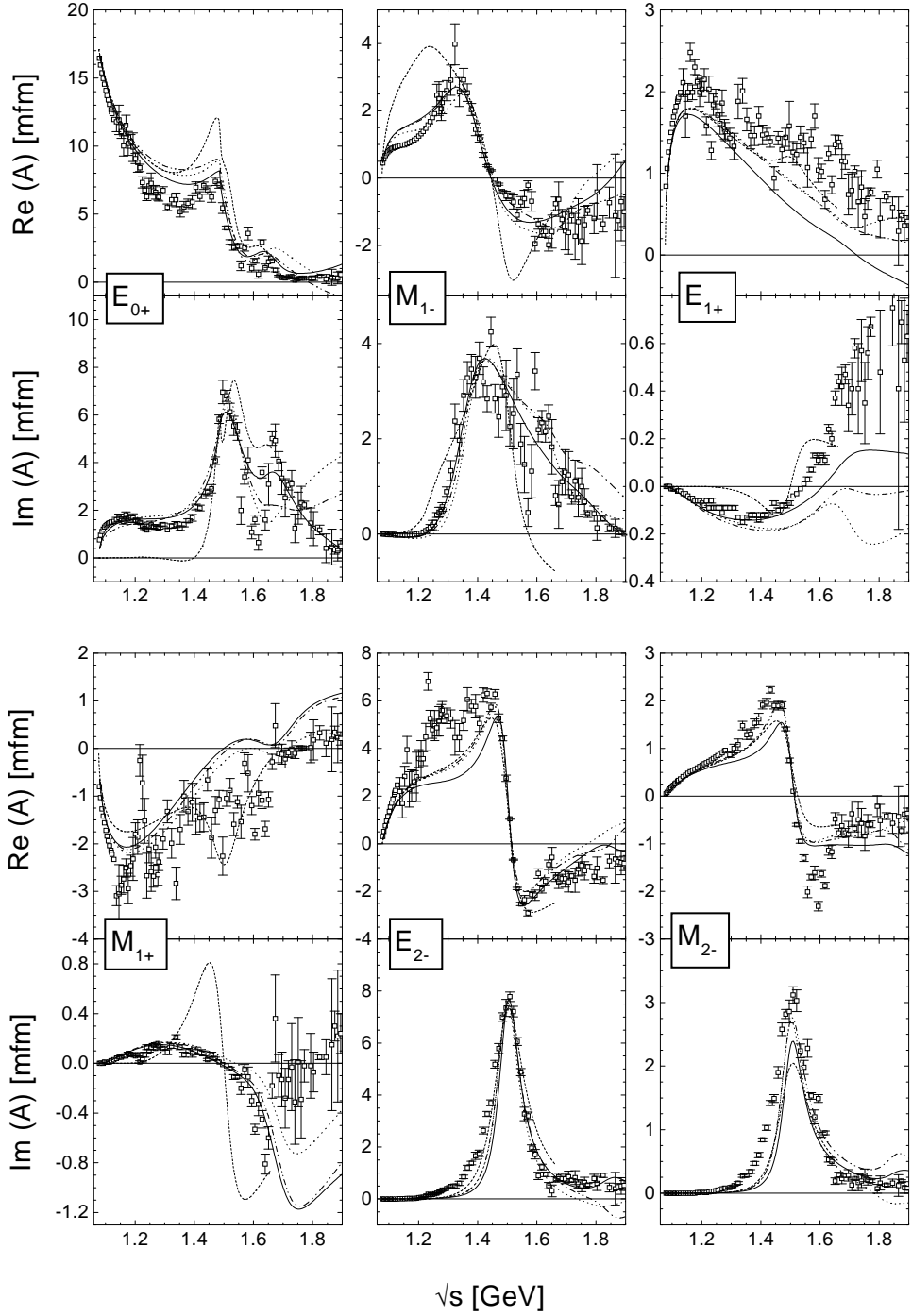} }
\caption{Results for the proton multipoles of pion photoproduction
  together with the data from SP97 \protect\cite{sm95}. Shown are all
  three fits: (\dotdot) SM95-pt-1, (\dashdot) SM95-pt-2, (\solid)
  SM95-pt-3. For comparison we also show the results of a $T$-matrix
  calculation \protect\cite{fm97} (\dash).}
\label{gpipSM}
\end{figure}

\begin{figure}[ht]
  \centerline{ \includegraphics[width=14cm]{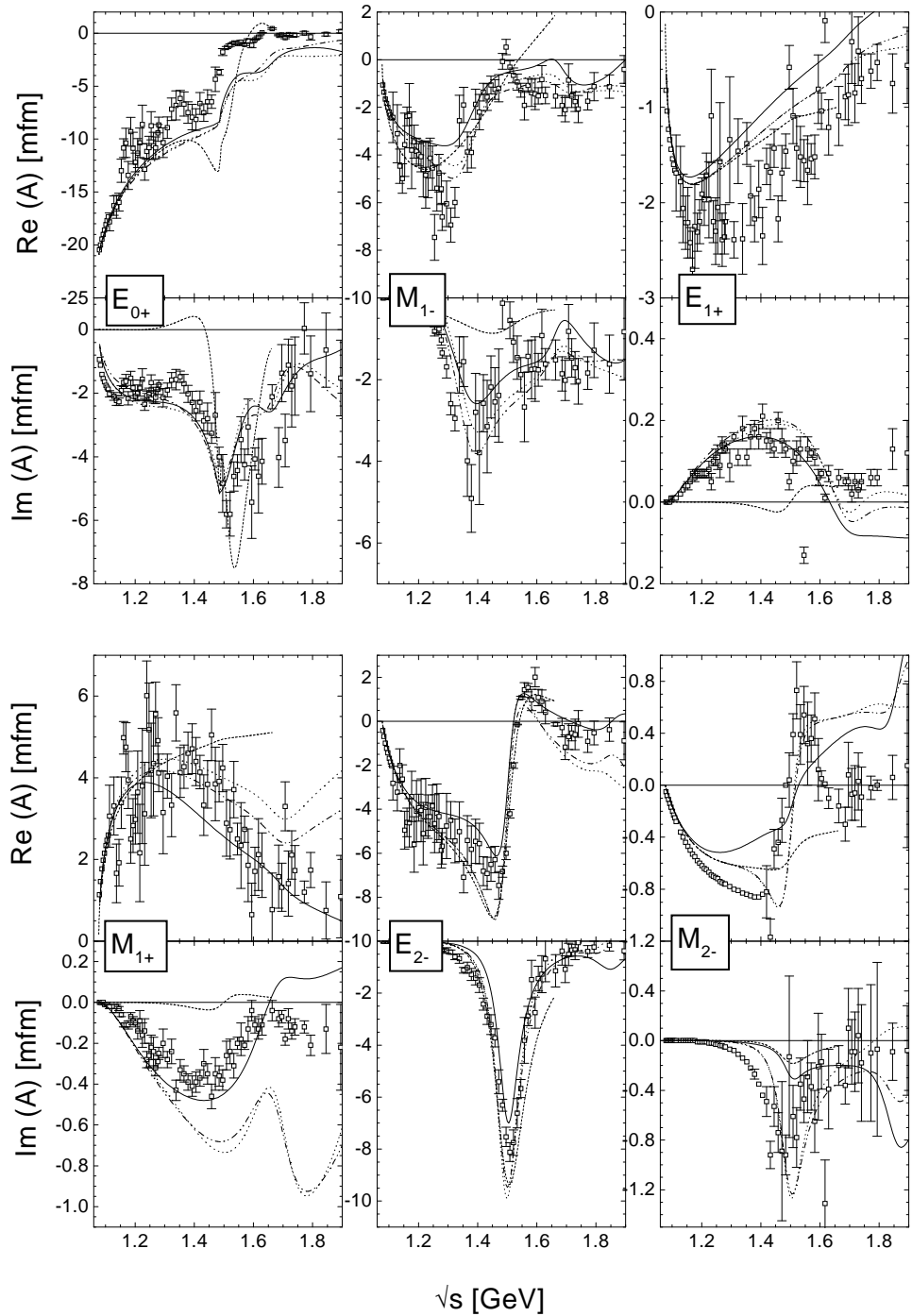} }
\caption{Same as Fig.\ \ref{gpipSM}, but for the neutron multipoles.}
\label{gpinSM}
\end{figure}

\begin{figure}[ht]
  \centerline{ \includegraphics[width=14cm]{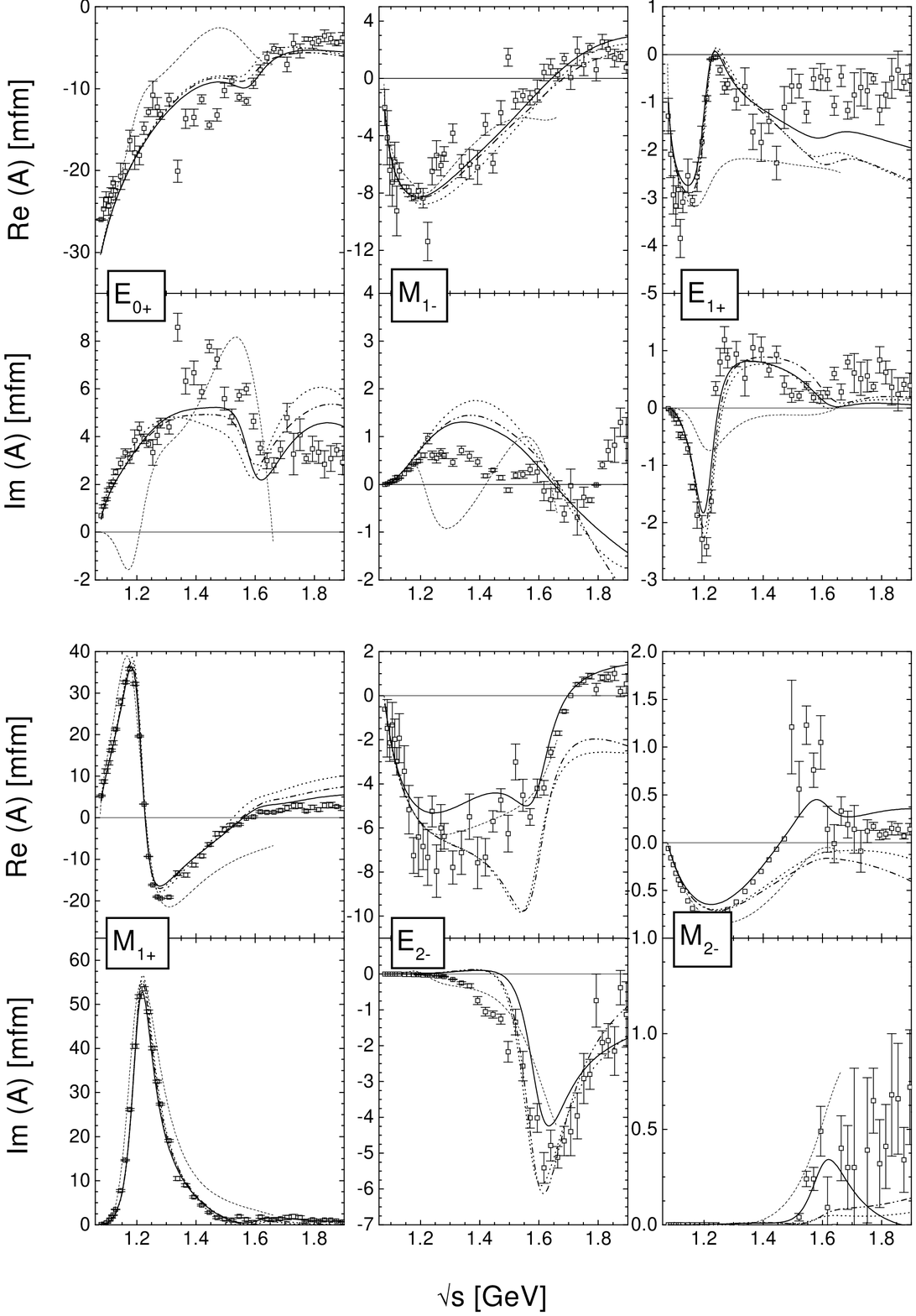} }
\caption{Same as Fig.\ \ref{gpipSM}, but for the $I\!=\!\dreih$
  multipoles.}
\label{gpi32SM}
\end{figure}

\begin{figure}[ht]
  \centerline{ \includegraphics[width=14cm]{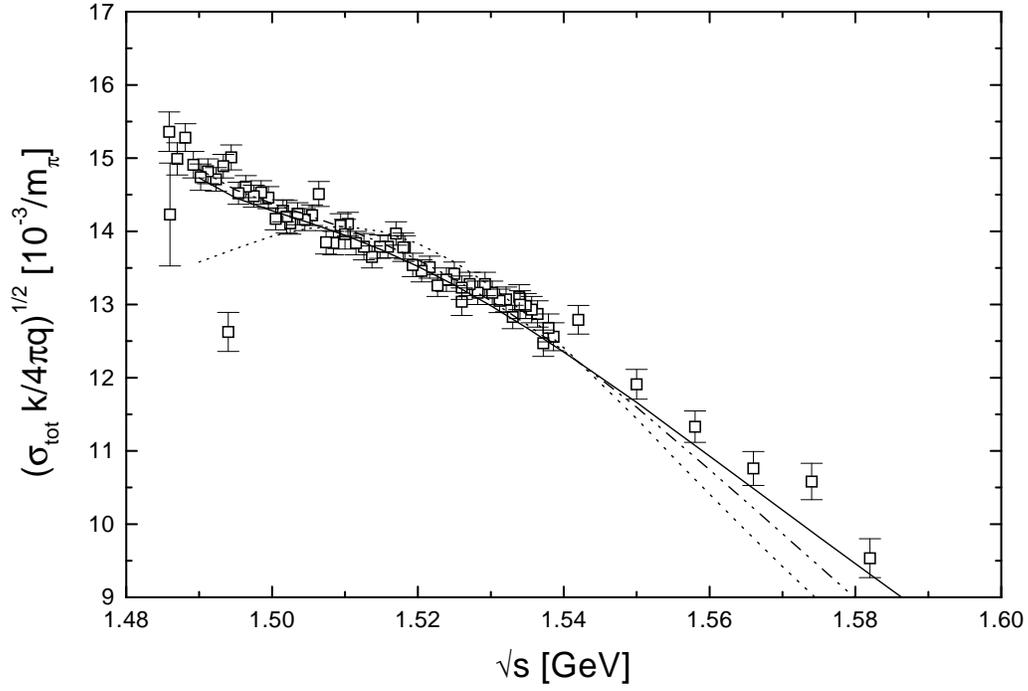} }
  \centerline{ \includegraphics[width=14cm]{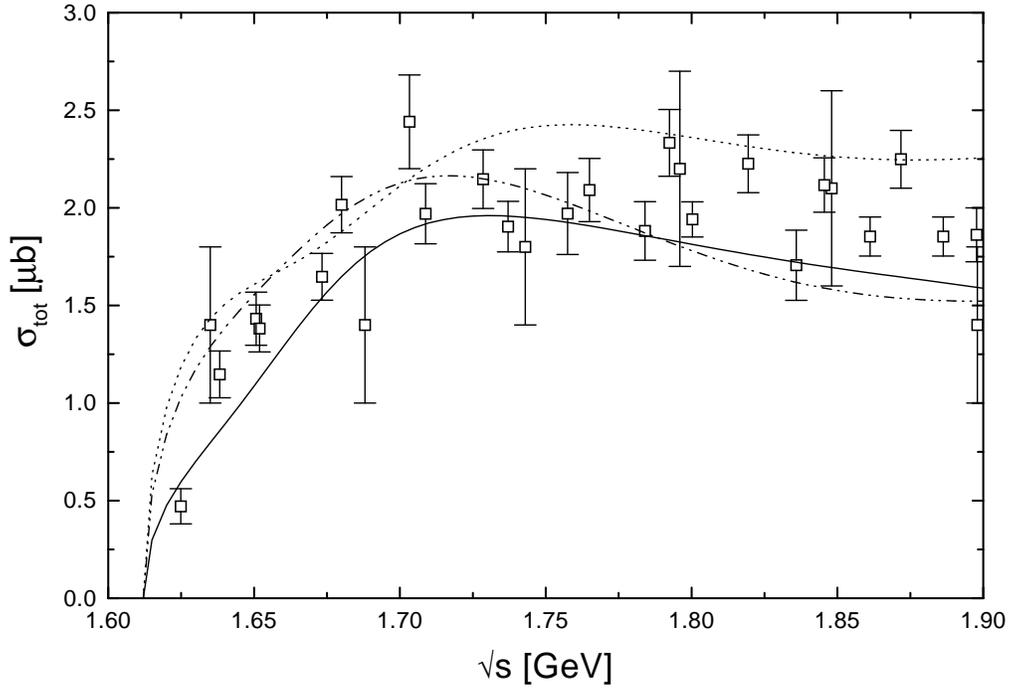} }
\caption{Results for the total $\gamma p \to \eta p$ (upper plot) and
  $\gamma p \to K^+ \Lambda$ (lower) cross sections as compared to
  data from \protect\cite{gptoep,gptokl}. In the case of eta
  photoproduction the reduced cross section is shown. The linecodes
  are those of Fig.\ \ref{gpipSM}.}
\label{gekSM95}
\end{figure}

\begin{figure}[ht]
  \centerline{ \includegraphics[width=14cm]{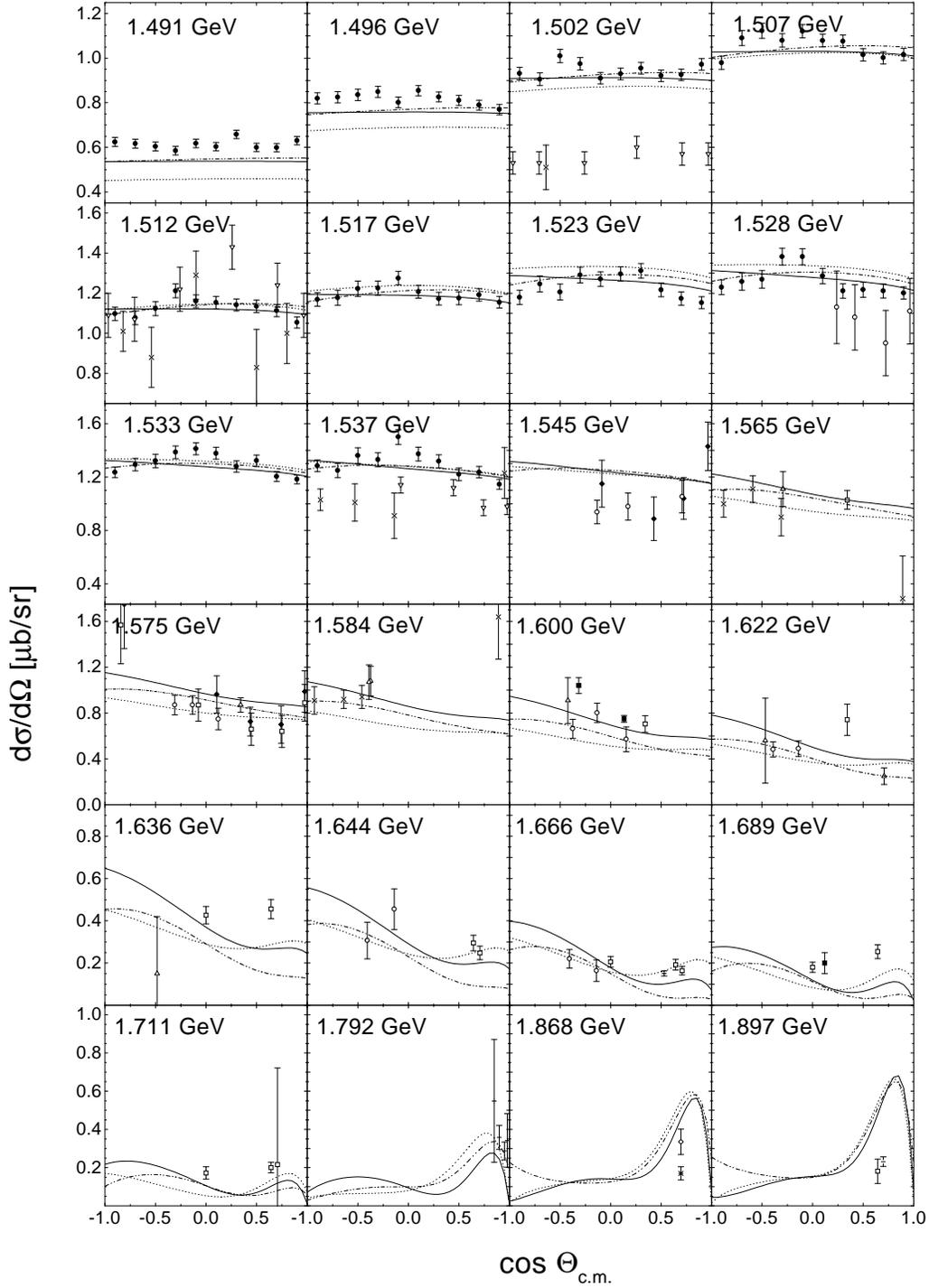} }
\caption{Comparison with data for the calculated differential
  $\gamma p \to \eta p$ cross sections at different energies. The
  datapoints are taken from: \protect\cite{k95} ({\large $\bullet$}),
  \protect\cite{gptoep}, the linecodes are those of Fig.\ 
  \ref{gpipSM}.}
\label{geSM95}
\end{figure}

\begin{figure}[ht]
  \centerline{ \includegraphics[width=14cm]{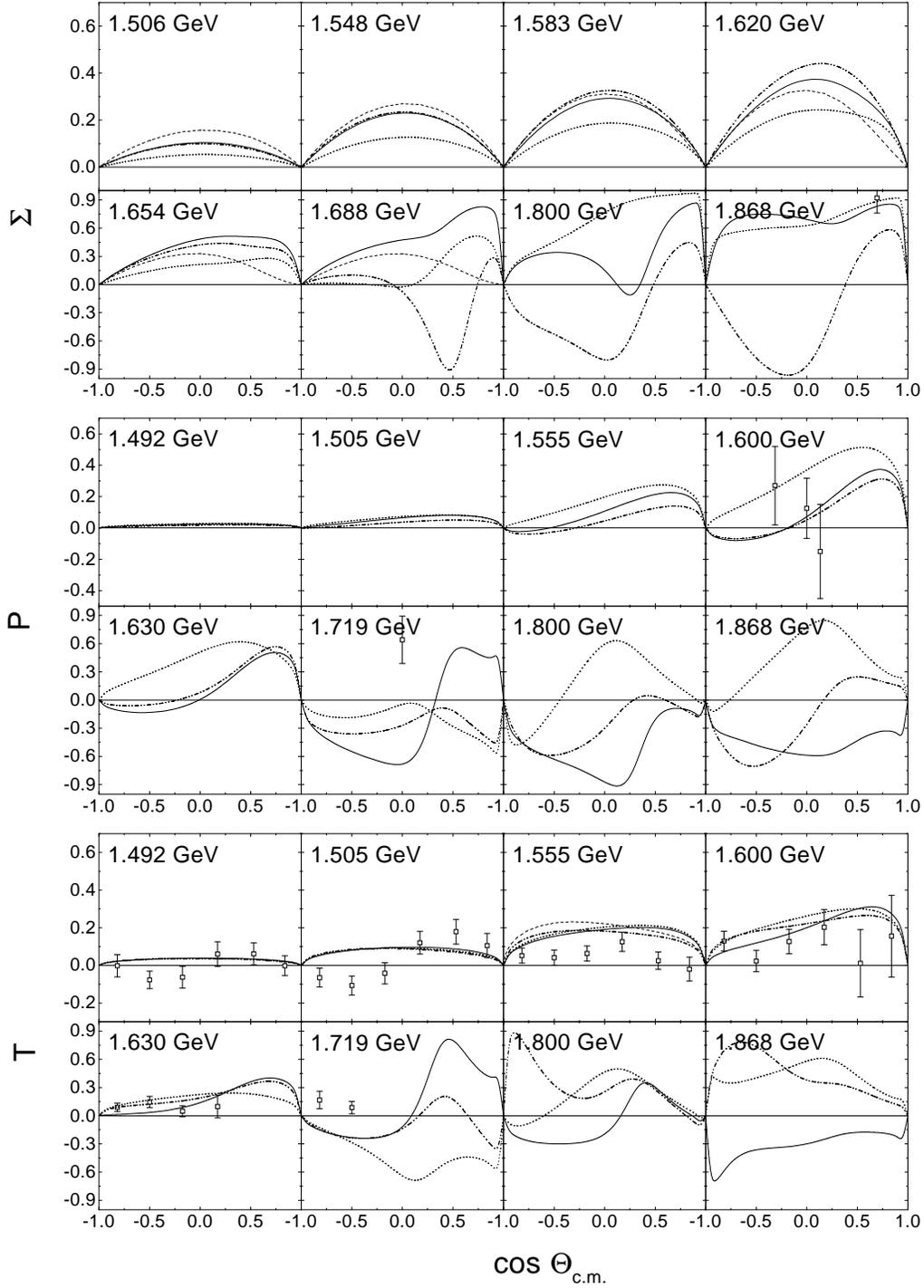} }
\caption{Polarization observables $\Sigma$, ${\cal P}$ and ${\cal T}$
  for $\gamma p \to \eta p$ as compared to the data for different
  energies. For the photon asymmetry $\Sigma$ also the calculation of
  Kn\"ochlein et al.\ \protect\cite{kdt95} is shown (\dash). The
  datapoints are taken from \protect\cite{gptoep,bock97}, the
  linecodes are those of Fig.\ \ref{gpipSM}.}
\label{gepolSM95}
\end{figure}

\begin{figure}[ht]
  \centerline{ \includegraphics[width=14cm]{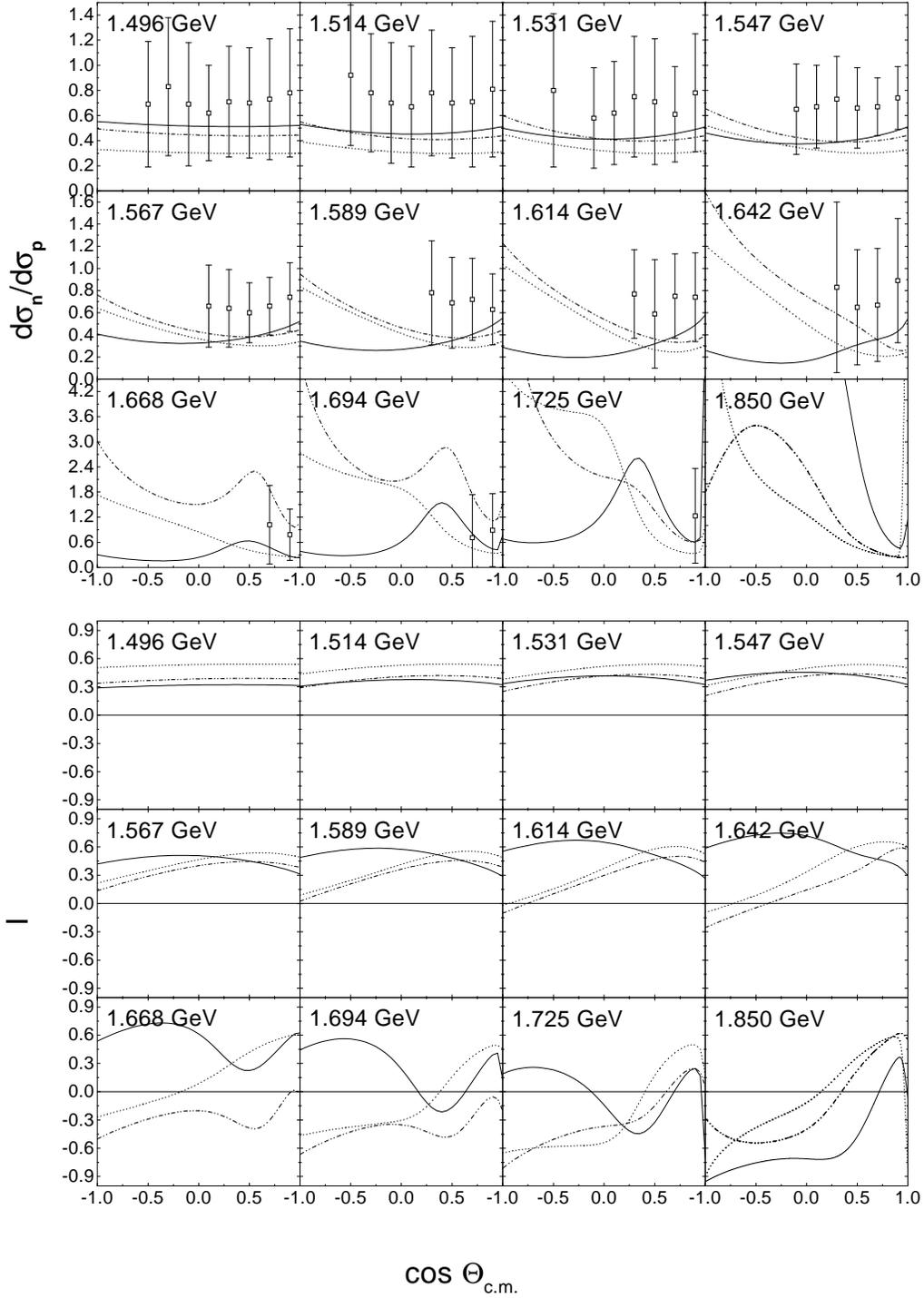} }
\caption{Neutron/Proton ratios and the isospin asymmetry ${\cal I}$
  for $\gamma p \to \eta p$ as compared to the data for different
  energies. The datapoints are taken from \protect\cite{hr97}, the
  linecodes are those of Fig.\ \ref{gpipSM}.}
\label{genprSM95}
\end{figure}

\begin{figure}[ht]
  \centerline{ \includegraphics[width=14cm]{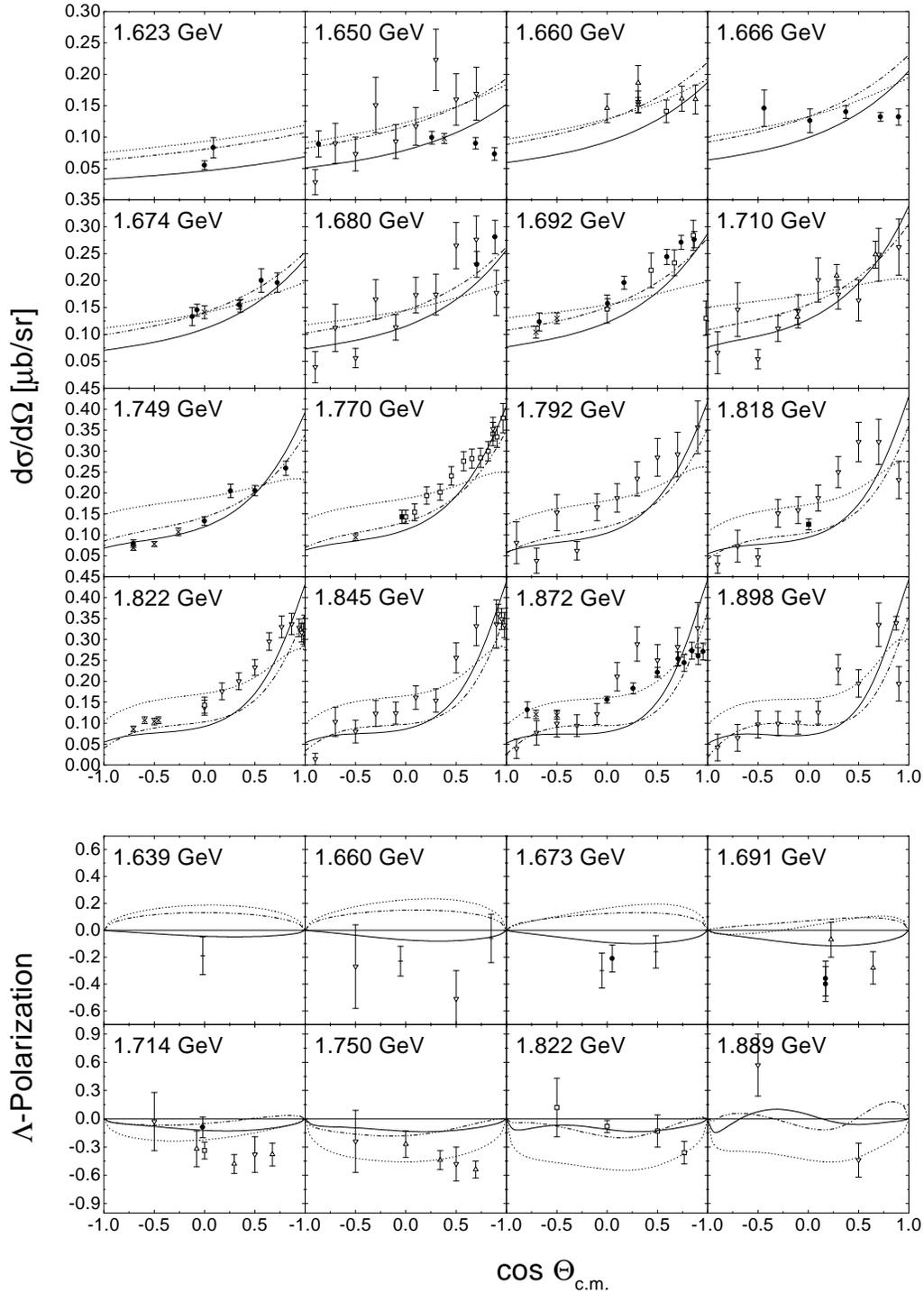} }
\caption{Comparison with data for the calculated differential
  $\gamma p \to K^+ \Lambda$ cross sections and
  $\Lambda$-polarizations for different energies. The datapoints are
  taken from: \protect\cite{barth97} ({\large $\Box$}),
  \protect\cite{gptokl}, the linecodes are those of Fig.\ 
  \ref{gpipSM}.}
\label{gkSM95}
\end{figure}

\begin{figure}[ht]
  \centerline{ \includegraphics[width=12cm]{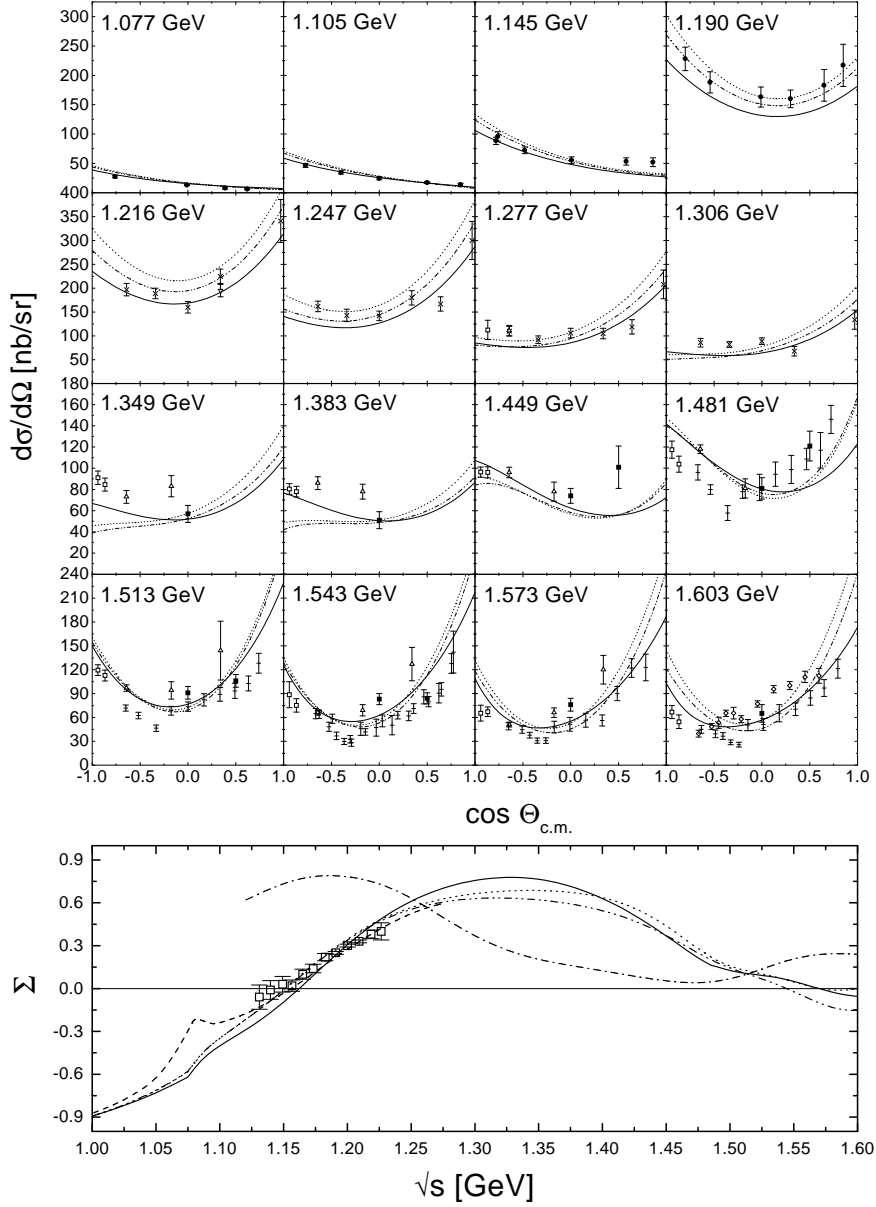} }
\caption{Comparison with data from \protect\cite{gptogp,b96} for the
  calculated $\gamma p \to \gamma p$ differential cross sections (top)
  and photon asymmetries (bottom). The linecodes are those of Fig.\ 
  \ref{gpipSM}. Also shown are the result of the isobar-model of Wada
  et al.\ \protect\cite{gptogp} (\dashsdot) and the dispersion relation
  calculation from L'vov \protect\cite{lvov81} (\dash).}
\label{ggSM95}
\end{figure}

\begin{figure}[ht]
  \centerline{ \includegraphics[height=10cm]{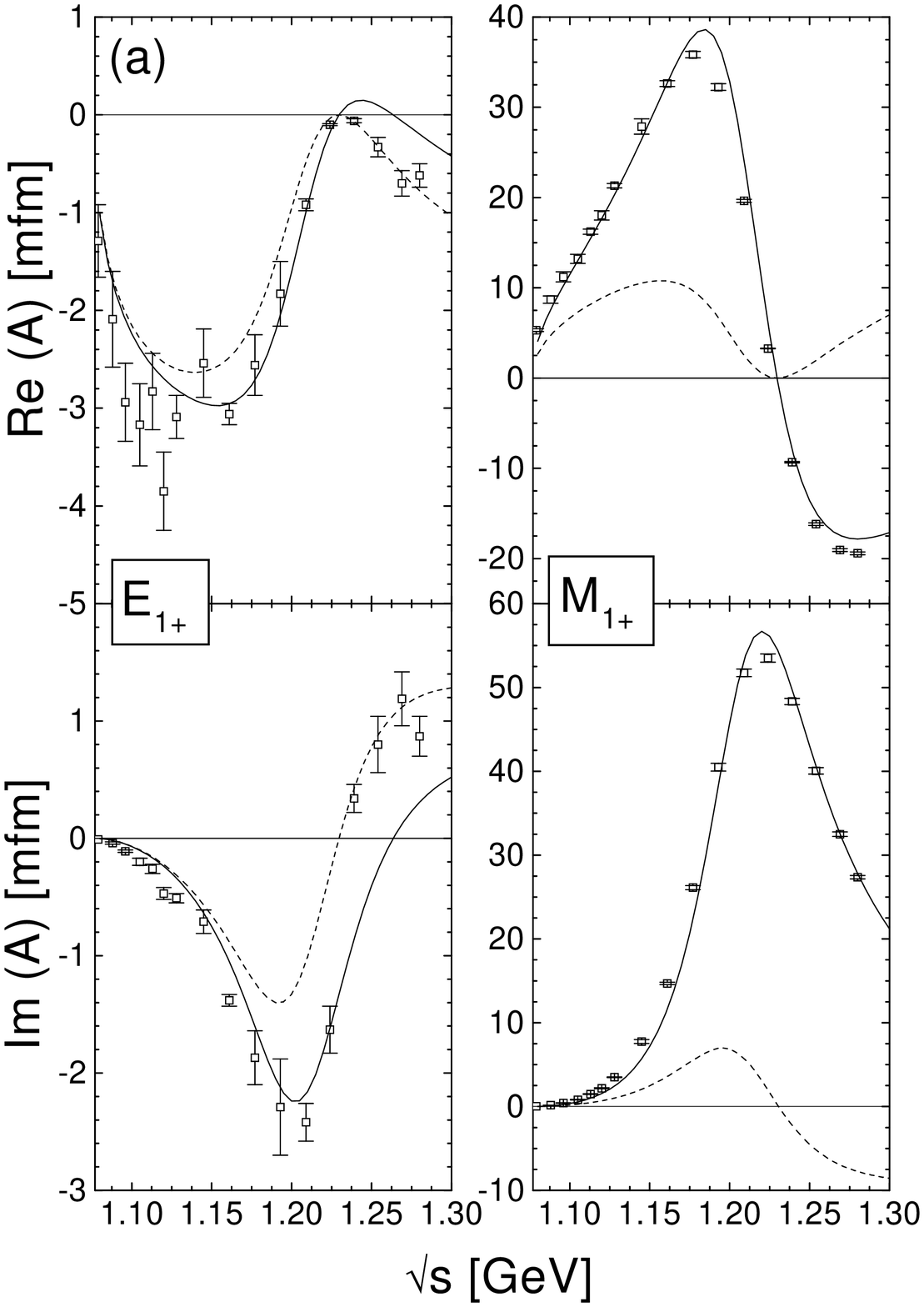}
               \includegraphics[height=10cm]{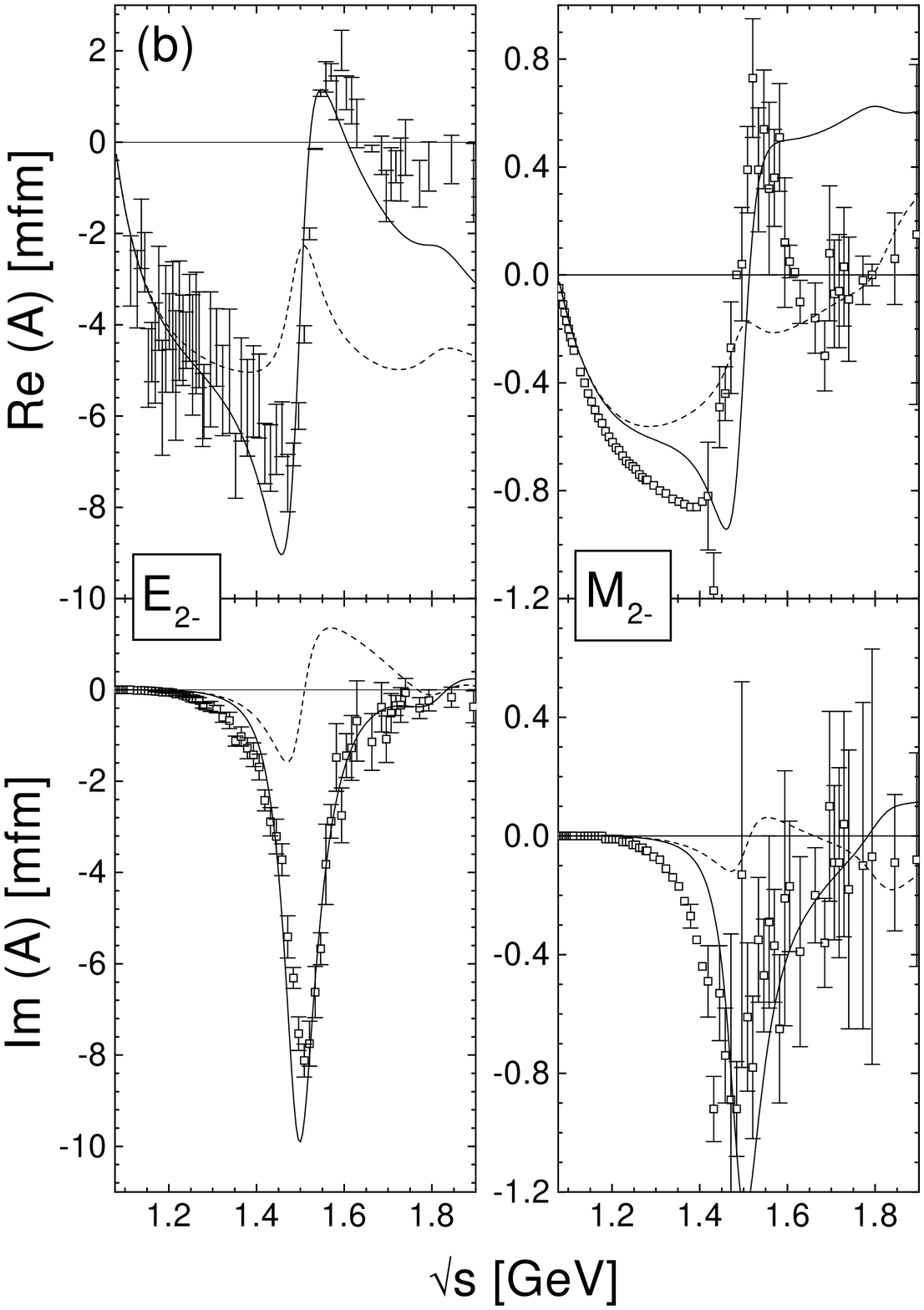} }
\caption{Influence of the rescattering on some of the multipoles.
  Shown are the results for the fit with fixed hadronic parameters
  with (\solid) and without (\dash) direct resonance coupling. {\bf
    (a):} $\nres P33{1232}$ and $E_{1+}^{3/2}$- and
  $M_{1+}^{3/2}$-multipoles. {\bf (b):} $\nres D13{1520}$ and
  $E_{2-}^{n}$- and $M_{2-}^{n}$-multipoles.}
\label{p33d13unitar}
\end{figure}

\begin{figure}[ht]
  \centerline{ \includegraphics[width=12cm]{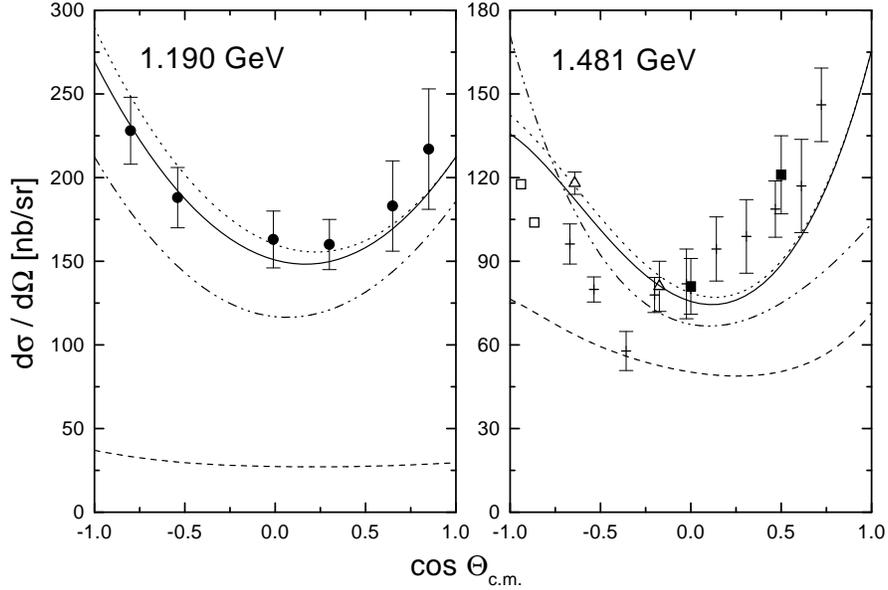} }
\caption{Comparison of contributions to the differential $\gamma p \to
  \gamma p$ cross section (data from \protect\cite{gptogp}). Full
  calculation using SM95-pt-2 (\solid), without the $\pi^0$ and $\eta$
  contributions in the $t$-channel (\dotdot), Born $s$- und
  $u$-channel diagrams only (\dash). {\bf Left:} $\nres P33{1232}$
  contributions only (\dashdot). {\bf Right:} $\nres D13{1520}$
  contribution only (\dashdot).}
\label{ggcontrib}
\end{figure}

\begin{figure}[ht]
  \centerline{ \includegraphics[width=12cm]{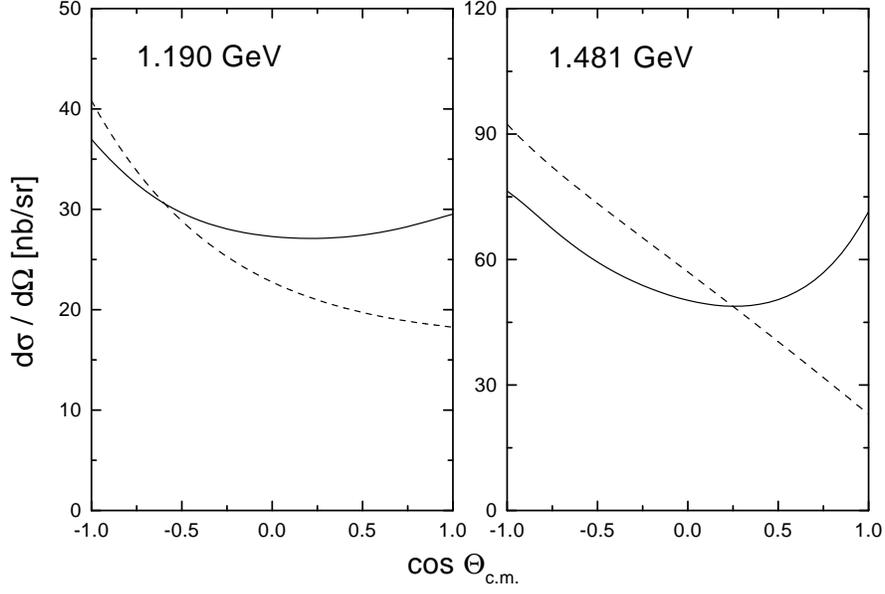} }
\caption{Influence of the rescattering on the nucleon $s$- and
  $u$-channel contributions to Compton scattering. Shown are the
  differential cross sections for two energies for the $K$-matrix
  calculation (\solid) and using the $T$-matrix approximation
  (\dash).}
\label{ggborn}
\end{figure}

\begin{figure}[ht]
  \centerline{ \includegraphics[width=12cm]{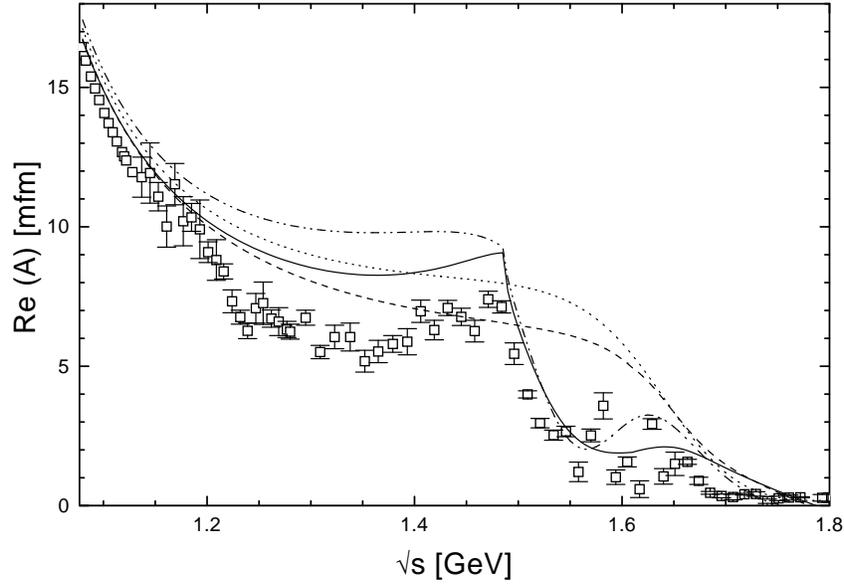} }
\caption{Different contributions to $\real (E_{0+}^p)$. Full
  calculation using SM95-pt-2 (\solid), SM95-pt-2, but without the
  $\nres S11{1535}$ (\dash), (\dashdot) and (\dotdot) show the results
  of \protect\cite{sau96}, both with and without the $\nres
  S11{1535}$.}
\label{s11contrib}
\end{figure}

\begin{figure}[ht]
  \centerline{ \includegraphics[width=12cm]{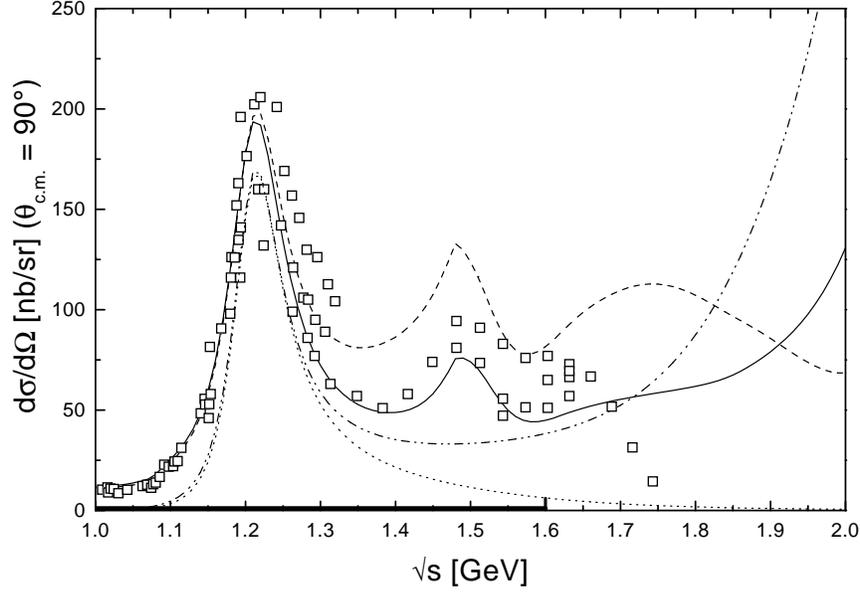} }
\caption{Sensitivity of the Compton cross section under 90$^\circ$ on
  the helicity couplings of the $\nres D13{1520}$. Full calculation
  using SM95-pt-2 (\solid), result using SM95-pt-2 and PDG-values for
  $A_{1/2,3/2}^p$ of the $\nres D13{1520}$ (\dash) and the $\nres
  P33{1232}$ contribution only for $\Lambda^e_{3/2}$ = 4.0 GeV
  (\dashdot) and 1.1 GeV (\dotdot). The data from \protect\cite{gptogp}
  are shown without their errorbars. The marking on the energy axis
  indicates the range of data used in the fits.}
\label{ggd13}
\end{figure}

\end{document}